\documentclass[preprints,article,accept,preprint, moreauthors,pdftex,10pt,a4paper]{mdpi/mdpi} 

\firstpage{1} 
\pubvolume{4}
\issuenum{4}
\articlenumber{0}
\pubyear{2019}
\copyrightyear{2019}
\history{Received: 1 October 2019; Accepted: 28 November 2019; Published: 6 December 2019}

\pdfoutput=1

\Title{Certified Randomness From Steering Using Sequential Measurements}

\usepackage{algorithm}
\usepackage{algorithmic}

\usepackage{dsfont}

\usepackage{qcircuit}

\usepackage{wrapfig}
\usepackage{subcaption}
\usepackage{amsmath,amsthm,amsfonts,amssymb} 
\usepackage{fontenc}
\usepackage{graphicx,multirow}
\usepackage{tikz}
\usetikzlibrary{shapes, arrows, shadows}
\usepackage[scaled]{helvet}
\usepackage[toc, page]{appendix}
\usepackage{graphicx}				
\usepackage{changes}
\usepackage{lipsum}

\newcommand{\secref}[1]{Section \ref{#1}}
\newcommand{\appref}[1]{Appendix \ref{#1}}

\newcommand{\figref}[1]{Figure \ref{#1}}
\renewcommand{\eqref}[1]{(\ref{#1})}

\newcommand{\bra}[1]{\langle #1|}
\newcommand{\ket}[1]{|#1\rangle}

\newcommand{\comment}[1]{\footnote{}}

\newcommand{\tr}{\textsf{tr}}
\newcommand{\Tr}{\textsf{Tr}}

\newtheorem{theorem_fix}{Theorem}
\newtheorem{lemma_fix}{Lemma}

\let\Algorithm\algorithm

\renewcommand\algorithm[1][]{\Algorithm[#1]\setstretch{1.6}}

\Author{Brian Coyle $^{1,}$*\orcidA{},  Elham Kashefi $^{1, 2}$ and Matty J. Hoban $^{3}$}

\AuthorNames{Brian Coyle,  Elham Kashefi and Matty J. Hoban}

\address{%
$^{1}$ \quad School of Informatics, University of Edinburgh, 10 Crichton Street, Edinburgh EH8 9AB, United Kingdom.\\
$^{2}$ \quad  Laboratoire  d'Informatique  de  Paris  6,  CNRS, Sorbonne  Universit\'{e},  4  Place  Jussieu,  75005  Paris, France; ekashefi@gmail.com.\\
$^{3}$ \quad Department of Computing, Goldsmiths, University of London, New Cross, London SE14 6NW, United Kingdom; matty.hoban@googlemail.com.}

\corres{Correspondence: brian.coyle@ed.ac.uk}

\conference{the Proceedings of the 9th International Workshop on Physics and Computation} 

\abstract{The generation of certifiable randomness is one of the most promising applications of quantum technologies. Furthermore, the intrinsic non-locality of quantum correlations allow us to certify randomness in a device-independent way, i.e. one need not make assumptions about the devices used. Due to the work of Curchod et. al.,  a single entangled two-qubit pure state can be used to produce arbitrary amounts of certified randomness. However, the obtaining of this randomness is experimentally challenging as it requires a large number of measurements, both projective and general. Motivated by these difficulties in the device-independent setting, we instead consider the scenario of one-sided device independence where certain devices are trusted, and others not; a scenario motivated by asymmetric experimental set-ups such as ion-photon networks. We show how certain aspects of previous work can be adapted to this scenario and provide theoretical bounds on the amount of randomness which can be certified. Furthermore, we give a protocol for unbounded randomness certification in this scenario, and provide numerical results demonstrating the protocol in the ideal case. Finally, we numerically test the possibility of implementing this scheme on near-term quantum technologies, by considering the performance of the protocol on several physical platforms.}

\keyword{one-sided device independence; randomness generation; randomness certification; quantum cryptography; semi-definite programming, self testing.}

\begin{document}

\section{Introduction}\label{sec:intro}

Quantum physics has the potential to make a great impact upon information technology, especially through the development of universal quantum computers. However, near-term quantum devices will not be capable of fault-tolerant, universal quantum computation. Luckily these devices will still be of use for information processing tasks, in particular as genuine random number generators. Certifiable (private) random numbers can then be used for cryptography, the simulation of physical systems, or other randomised algorithms. By certifiable we mean that there is a certificate guaranteeing that the randomness is private and unpredictable from any external agent (who is not directly using the device). This certificate may be predicated on certain assumptions, which could be computational or physical in nature, depending on the degree of security desired. 

It is now well established that quantum systems are capable of producing data that is unpredictable, and thus random to any external agent, even when one has perfect knowledge of the quantum system. Unfortunately, in practice it can be difficult to have perfect knowledge of quantum systems, especially if they are somewhat noisy, as near-term quantum devices will be. These (often classical) sources of noise can appear as unpredictable as the randomness resulting from the quantum systems, so one must have an excellent characterisation of the sources of noise to extract the true quantum randomness. Indeed, if the noise is just classical data then it could have been generated by some external process and thus an external agent could, in principle, keep a copy of this data and use it to predict the output data of a quantum device.

There does exist a convenient approach to certifiable quantum random number generation, which is \textit{device-independent randomness certification}. In this scenario one does not need a complete characterisation of a device; genuine randomness is certified by the violation of a Bell inequality \cite{bell_einstein_1964} between two, or more, devices. That is, certification is achieved via the statistics produced in a Bell test, without any specific assumptions made on the devices producing the statistics.The kind of assumption made in this approach is to assume that devices are quantum mechanical or that there multiple, non-communicating devices that might share some resource. Furthermore, there are no computational assumptions made about the device producing the randomness. The downside of this approach is that a genuine violation of a Bell inequality is experimentally daunting, with the first loophole-free demonstrations emerging very recently \cite{hensen_loophole-free_2015}.

Given the experimental challenges of device-independent random number generation,  \cite{pironio_random_2010, colbeck_quantum_2009} a promising and practical route to certifiable randomness generation is within the scope of \textit{one-sided} device-independent quantum information \cite{cavalcanti_quantum_2017}. In this setting, certain devices are assumed to be perfectly characterised (through trusted and characterised measurement devices) while others are not. Randomness can be certified based on the violation of a steering inequality \cite{wiseman_steering_2007}, which is the analogue of a Bell inequality for this new setting.

Within the framework of device-independent randomness certification it was shown that a single entangled pair of qubits (in a pure state) can be a source of ``unbounded" random numbers, one qubit for each wing of the Bell experiment \cite{curchod_unbounded_2017}. That is, one can fix a value $N$ of random bits that one would like to obtain, and then construct a scheme with sequences of measurements on the two-qubit state that will produce $N$ bits of randomness. Thus by using a sequence of measurements, one can exceed the randomness possible from a single general measurement, which for a qubit is $2$ bits \cite{optimal}. One issue is that this randomness certification scheme involves a large number of measurements (exponential in the size of the output random string) for one of the parties and limits its utility for various protocols. 

In this work, we study the adaptation of the above sequential measurement scenario to the one-sided device independent scenario. In doing so, we develop a more robust scheme no longer requiring exponentially many measurements for one of the parties. We present an analytical bound on the min entropy of our randomness generation scheme. We then go on to given numerical results to derive more optimal rates of randomness generation. Furthermore, we discuss how the scheme could be implemented in current architectures for networked quantum information processing. This work is an extended version of the following conference paper \cite{coyle_one-sided_2018}.

\begin{table}[h!]
\centering
\begin{tabular}{ |c|c|c| } 
 \hline
 & Our work & Ref. \cite{curchod_unbounded_2017} \\ \hline
Alice & Untrusted & Trusted \\ \hline
Bob & Untrusted  & Untrusted\\ \hline
Randomness certified & $\Omega(n)$ & $\Omega(n)$ \\ \hline
Number of measurements required & $O(n)$ & $O(2^{n})$ \\ \hline
Method & Steering inequality violation & Bell inequality violation \\ \hline
Relevant Quantity & Steerable assemblage & Non-local probability distribution\\
&$\{\sigma_{\vec{b}|\vec{y}}\}$& $\{P(a\vec{b}|x\vec{y})\}$\\
\hline 
\end{tabular}
\caption{Comparison between the device-independent and one-sided device-independent sequential randomness generation between our work and the work in Ref. \cite{curchod_unbounded_2017}. The positive integer $n$ is the number of measurements made in a sequence of measurements. Here we see that there is an exponential improvement in the number of measurements required.}
\label{table:1}
\end{table}

\subsection{Related work}
In Table \ref{table:1} we compare our work with that of \cite{curchod_unbounded_2017}, showing how, by trusting one party's measurements, we exponentially reduce the number of measurements required. In work by Skrzypczyk and Cavalcanti, it was shown how by increasing the local Hilbert space dimension of the quantum state held by Alice and Bob, more randomness can be certified in the one-sided device-independent scenario \cite{skrzypczyk18}. In particular, for a local dimension $d$, then $\Omega(\log d)$ bits can be certified. This work is built on a series of works in one-sided device-independent randomness certification, with \cite{passaro_optimal_2015} establishing tools based on semi-definite programming. 

Our cryptographic scenario is intermediate between the device-independent and the device-dependent scenarios. Another such example of an intermediate scenario is that of \textit{semi-device-independent quantum information} \cite{brunner, semi}, where one bounds the dimension (or energy) of the Hilbert space of the systems involved. Randomness certification has been shown in this scenario, with experimental implementations of various protocols \cite{Rusca, lunghi}. This scenario is not comparable with that of one-sided device-independence due to the different assumptions, but it demonstrates that such intermediate scenarios are of broad interest.

\section{One-Sided Device Independence and Randomness Certification}\label{sec:one_sided_randomness}

Before introducing the scenario it is worthwhile briefly motivating it first from an experimental point-of-view. One particular kind of experimental set-up we have in mind is an atom-photon hybrid experiment, where one system is an atom in a cavity, and the other system is a photon, which is emitted from the atom. Instead of an atom in a cavity, an ion in a trap is another possibility. Photons are convenient for long-range communication, and ion trap technology is associated with high fidelity operations and excellent system control. As a result the detection efficiency in an ion trap is very close to perfect, but in spite of recent advances, photo-detectors are not. In a device-independent scheme, a lower detection efficiency can compromise the security of a protocol, so to circumvent these issues we can resort to the one-sided device-independent setting (1sDI). In this setting, the photonic system is taken to be trusted and well characterised thus ruling out detector-based attacks, and the atomic system is treated as a black box. 

An extra motivation for this 1sDI scenario will be when one wants to consider sequences of measurements on the same system, as we will do. We need our technology to allow for the possibility of returning a quantum system after a measurement (thus being a non-trivial quantum instrument). This is experimentally challenging for photonic systems, but feasible within ion trap technology. Ideally we would thus like our trusted system to make very simple operations, such as a single measurement that does not return a quantum state as an output. In this way, we can see one-sided device independence as exploiting the best features of a hybrid quantum information experiment. This will be pertinent when we come to discuss implementations of our randomness certification scheme. 

The idea of producing certifiable randomness using steering was first studied by Law \textit{et al.} \cite{law_quantum_2014}, and then by Passaro \textit{et al.} \cite{passaro_optimal_2015}, which utilises the techniques of semi-definite programming. The broad scenario considered in 1sDI information processing for randomness generation is the following. There are two parties, Alice (A), and Bob (B), who can share some resource. We allow for the possibility of a third party, Eve (E), having prepared the shared quantum resource. Alice's share of the resource is assumed to be a quantum system with a known Hilbert space upon which Alice can perform arbitrary (characterised) quantum operations. In particular, Alice can perform tomographically complete measurements. Bob's share of the resource is contained within a black box and he can only input classical data into the box and retrieve more classical data; he does not have any knowledge of the inner workings of the black box, only that it has a quantum description. Bob can only collect statistics of the input and output data.

\begin{wrapfigure}{r}{0.5\textwidth}
  \begin{center}
    \includegraphics[width=0.45\textwidth]{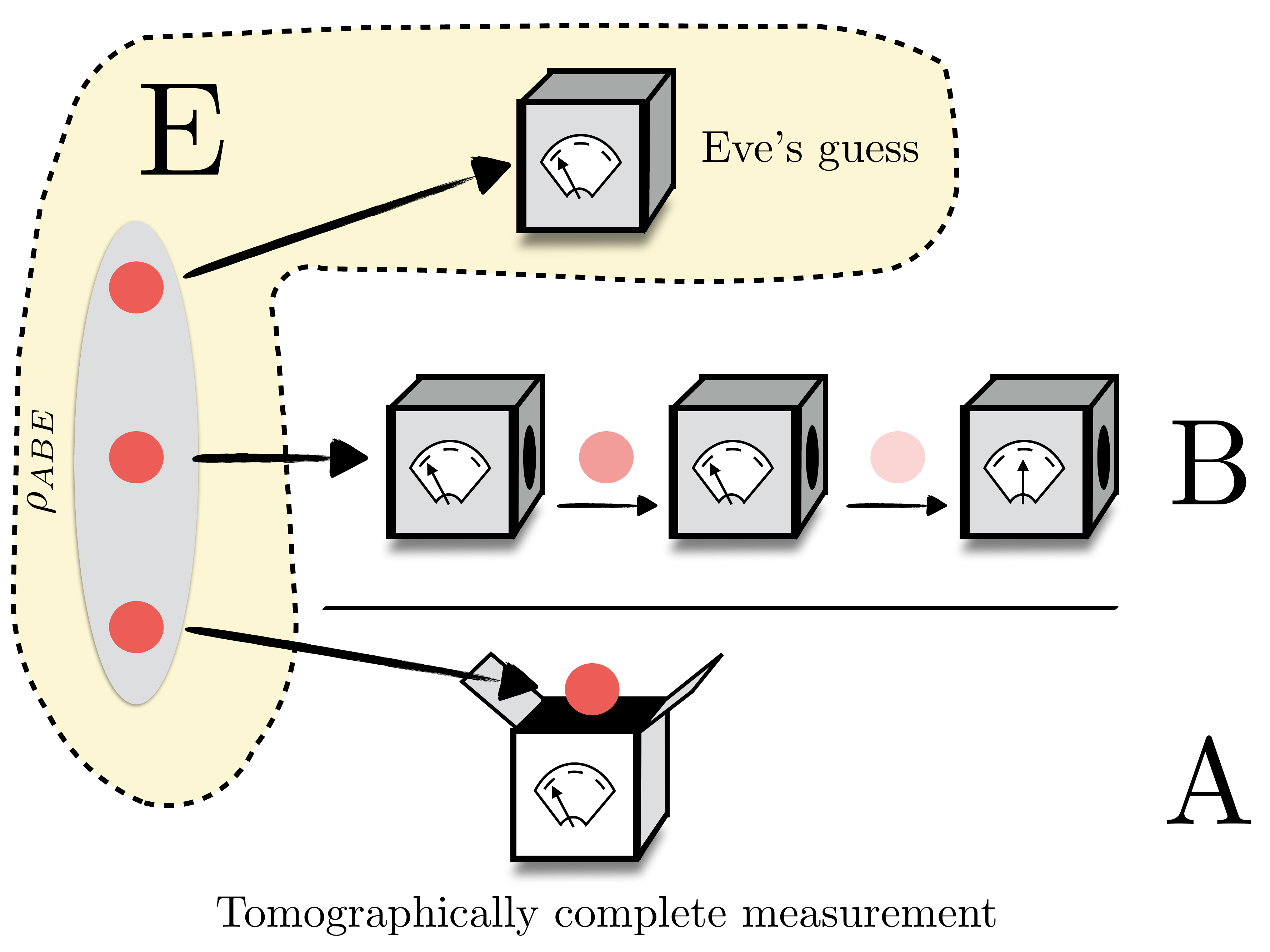}
  \end{center}
  \caption{Illustration of the tripartite scenario between Alice, Bob and Eve, in which Bob also makes a sequence of measurements and Alice can make trusted measurements. Eve tries to guess the outcomes of Bob's measurements.}
    \label{fig:1sdi_protocol_scenario}
\end{wrapfigure}

Given this scenario, the way in which we certify the randomness generated is through a (slightly modified) \textit{non-local guessing game} \cite{Silleras, passaro_optimal_2015}. We give a schematic of this guessing game in Figure \ref{fig:1sdi_protocol_scenario}. In this game, in each round, Eve prepares a quantum state $\vert\psi\rangle_{ABE}$, which we can assume to be pure through the Stinespring dilation (we could dilate the Hilbert spaces of Bob and Eve, for example). Then one subsystem is each distributed to Alice and Bob so that they share the joint state $\rho_{AB}=\textrm{tr}_{E}\vert\psi\rangle\langle\psi|_{ABE}$. Since Alice has access to her respective subsystem she is able to characterise $\rho_{A}=\textrm{B}\rho_{AB}$, but Bob does not have direct access to his subsystem. Inside Bob's device if he inputs the classical variable $y$, which is his choice of measurement, and gets the output $b$ then a measurement  is made on Bob's subsystem, which is described by the positive operator $M^{B}_{b|y}$ such that $\sum_{b}M^{B}_{b|y}=\mathbb{I}_{B}$ for all $y$. Eve will then in each round perform a measurement that will generate an outcome $z$, which will be her guess of Bob's outcome $b$; this measurement will be described by a positive operator $N_z^E$ such that $\sum_{z}N^{E}_{z}=\mathbb{I}_{E}$. 

In this setting, Eve's goal is to optimise over the state $\vert\psi\rangle_{ABE}$ and measurements $N_z^E$ that will give her the best chance to guess the outcome of Bob's measurement. Importantly, Eve's strategy has to be compatible with the observed statistics of what Alice and Bob observe. Note that in this game, the most compact way of describing what Alice and Bob observe (assuming Alice performs tomography on her system) is described by the \textit{assemblage} $\{\sigma_{b|y}\}_{b,y}$, which is a set where each element can be described as
\begin{equation}
\sigma_{b|y}=\textrm{tr}_{B}\left(\mathbb{I}_{A}\otimes M^{B}_{b|y}\rho_{AB}\right),
\end{equation}
which can be viewed as a sub-normalised density matrix describing the state of Alice's system after the measurement $M^{B}_{b|y}$ is made such that $\sum_{b}\sigma_{b|y}=\rho_{A}$ for all $y$. This assemblage is merely Alice's and Bob's observed assemblage, but really every element is obtained in the following way:
\begin{eqnarray}
\sigma_{b|y}&=&\textrm{tr}_{BE}\left(\sum_{z}\mathbb{I}_{A}\otimes M^{B}_{b|y}\otimes N^{E}_{z}\vert\psi\rangle\langle\psi|_{ABE}\right)\\
&=&\sum_{z}\sigma_{z,b|y}
\end{eqnarray}
where we have course-grained over all of Eve's measurement outcomes, or guesses, and introduced the identity
\begin{equation}
\sigma_{z,b|y}=\textrm{tr}_{BE}\left(\mathbb{I}_{A}\otimes M^{B}_{b|y}\otimes N^{E}_{z}\vert\psi\rangle\langle\psi|_{ABE}\right),
\end{equation}
which can be seen as the sub-normalised state of Alice's system conditioned on Bob's and Eve's particular measurement outcomes. 

Returning to the game, we quantify Eve's ability to guess Bob's outcome with the \textit{guessing probability}. We first assume that that Bob will aim to generate randomness from only one particular input, denoted by $y^{*}$, and Eve knows $y^{*}$. The guessing probability for Eve's output $z$ to correctly guess Bob's output $b$ for choice $y^{*}$ is then
\begin{equation}
p_{\textrm{guess}}(y^{*})=\sum_{z}\delta_{b,z}\textrm{tr}_{A}\sigma_{z,b|y^{*}}.
\end{equation}
This can be seen as the sum over $z$ of the probabilities $p(z,b|y^{*})$ when $b=z$ \cite{passaro_optimal_2015}.

We will now expand upon this set-up to allow for Bob's measurement to be a sequence of measurements. That is, we describe Bob's input $y$ and output $b$ to be tuples of length $n$, so that $y:=(y_{1},y_{2},y_{3},...,y_{n})$ and $b:=(b_{1},b_{2},b_{3},...,b_{n})$. That is, Bob makes a sequence of measurements where each $i$th measurement in the sequence corresponds to the measurement choice $y_{i}$ with output $b_{i}$. We assume that the output $b_{i}$ is obtained before the choice $y_{i+1}$ is made, and thus we impose a constraint of causality: measurement outcomes in the past are independent of future measurement choices. A consequence of this, for example, is that $p(b_{1}|y_{1},y_{2})=p(b_{1}|y_{1})$, i.e. the probability of observing $b_{1}$ given $y_{1}$ is independent of the future choice of $y_2$. Since $\textrm{tr}_{A}\sigma_{b|y}=p(b|y)$ for $y:=(y_{1},y_{2},y_{3},...,y_{n})$ and $b:=(b_{1},b_{2},b_{3},...,b_{n})$, this then has consequences for the assemblage. For example, for $n=2$,
\begin{equation}
\sum_{b_{2}}\sigma_{b_{1},b_{2}|y_{1},y_{2}}=\sigma_{b_{1}|y_{1},y_{2}}=\sigma_{b_{1}|y_{1}},
\end{equation}
and likewise for larger $n$. At this point it is worthwhile pointing out that any assemblage that satisfies these causality constraints in addition to non-signalling constraints, i.e. $\sum_{b}\sigma_{b|y}=\rho_{A}$ for all $y$, can be realised by Alice and Bob sharing a quantum state and Bob making an appropriate sequence of measurements, as proven in \cite{Sainz2019}.

These are all of the constraints in the scenario that we are considering when allowing for sequences of measurements on a state. The goal is given all of these constraints, to give bounds on the guessing probability $p_{\textrm{guess}}(y^{*})$ given an observed assemblage $\{\sigma_{b|y}\}_{b,y}$. One method for doing this is through semi-definite programming \cite{passaro_optimal_2015}, and we will return to this technique when it comes to presenting numerical results. We will also give analytical results based on \textit{self-testing} in the steering scenario \cite{supic}. One unifying aspect to our results is that instead of certifying randomness given the observed assemblages, we can certify randomness based on the violation of \textit{steering inequalities}, which are analogous to Bell inequalities. More generally, a steering inequality violation results directly from some observed statistics for Alice. Therefore we can certify randomness based on statistical tests given particular (known) measurements made by Alice. Within this work, it will be made clear how $p_{\textrm{guess}}(y^{*})$ is being calculated. 

Given the guessing probability $p_{\textrm{guess}}(y^{*})$, we can compute a related quantity, which is the certifiable \textit{min entropy} of Bob's outcomes:
\begin{align}
    H_{min}(b|y^{*},z) := -\log_2 p_{\textrm{guess}}(y^{*}) \label{minentropy}
\end{align}
As we can see this is directly related to the guessing probability. That is, if the set of possible outcomes $b$ has cardinality $2^{m}$ and $p_{\textrm{guess}}(y^{*})=2^{-m}$ then the min entropy associated with Bob's outcomes is $m$ bits. In this way, Bob's device is a source of $m$ bits of certifiable randomness.

\section{A Scheme for Unbounded Randomness Generation}\label{sec:scheme}

In this section we will describe an honest strategy in which a sequence of measurements made upon half of a two-qubit entangled state can result in a large amount of observed randomness. In the subsequent sections we will give methods to certify that this is genuine randomness, but for now we will not concern ourselves with certification. 

The scheme is similar to that of \cite{curchod_unbounded_2017}. We will call this scheme the \textit{Two-Qubit Sequential Measurement} (TQSM) scheme. We have that Bob can implement non-projective measurements in "rotated versions" of the Pauli-$X$ and $Z$ bases, and Alice has the functionality implement a tomographically complete set of measurements, for example to measure the Pauli observables, $X, Y, Z$ since this is sufficient for her to do quantum state tomography to certify Bob's random outcomes. 

First, for simplicity, we will consider Bob just making one sequence, i.e. a sequence of $n$ measurements for $n=1$ so that $y:=y_{1}$ and $b:=b_{1}$. We have that Bob can make a choice between two dichomotic measurements, so that $y$, $b\in\{0,1\}$. When Bob makes choice $y=0$ ($y=1$), he will make a (possibly non-projective) rotated version of a measurement in the Pauli-Z (Pauli-X) basis. 

We will now describe these "rotated" measurements in terms of their associated Kraus operators. These operators are of the form $\Pi^{\omega}_{b|y}$ where $\omega$ is an angle and $b$, $y$ are the bits as defined above. Consider the following operators:
\begin{equation}\label{1sdikrausoperators} 
\begin{split}
\Pi^{\phi}_{0|0} = \cos(\phi)\ket{0}\bra{0}+\sin(\phi)\ket{1}\bra{1}, \qquad
\Pi^{\phi}_{1|0} =  \cos(\phi)\ket{1}\bra{1}+\sin(\phi)\ket{0}\bra{0}\\   
\Pi^{\theta}_{0|1} = \cos(\theta)\ket{+}\bra{+}+\sin(\theta)\ket{-}\bra{-}, \qquad
\Pi^{\theta}_{1|1} = \cos(\theta)\ket{-}\bra{-}+\sin(\theta)\ket{+}\bra{+}
\end{split}
\end{equation}
The positive-operator valued measure (POVM) constructed from these Kraus operators that Bob implements on his half of the shared state will be of the form
*
\begin{equation*}
M^{\omega}_{b|y} = \left(\Pi^{\omega}_{b|y}\right)^\dagger\left(\Pi^{\omega}_{b|y}\right).
\end{equation*}
*
These Kraus operators reduce to the usual projective Pauli-X and Pauli-Z basis projectors for $\theta = \phi = 0$. Therefore, if Alice and Bob share the pure quantum state $|\psi\rangle_{AB}$ and Bob makes a measurement in, say, the rotated Pauli-X basis, and gets the outcome $b=1$, the post-measurement state will be 
\begin{equation*}
    \rho_{AB}=\frac{\mathbb{I}_{A}\otimes\Pi_{1|1}^{\phi}|\psi\rangle\langle\psi|\mathbb{I}_{A}\otimes\Pi_{1|1}^{\phi}}{|\mathbb{I}_{A}\otimes\Pi_{1|1}^{\phi}|\psi\rangle|^{2}}
\end{equation*}
Very similar expressions are then obtained for the other Kraus operators. It should be noted that for all pure states  $|\psi\rangle_{AB}=\alpha|00\rangle+\beta|11\rangle$ the post-measurement state $\rho_{AB}$ will also be pure \cite{curchod_unbounded_2017}. The post-measurement pure state shared by Alice and Bob after outcome $b$ for input $y$ will be
\begin{equation}\label{postmeas_single}
    |\psi_{b|y}\rangle = U_{A}^{b|y}\otimes U_{B}^{b|y}\left(\cos(\zeta_{b|y})|00\rangle+\sin(\zeta_{b|y})|11\rangle\right)
\end{equation}
where unitaries $U_{A}^{b|y}$ and $U_{B}^{b|y}$, and angle $\zeta_{b|y}$ depend on the initial quantum state and the angle of the rotated measurement. We point out that such an angle and unitaries exist (and can be calculated).

What is the probability of getting the outcome $b$ given $y$? This will be $p(b|y)=|\mathbb{I}_{A}\otimes\Pi_{b|y}^{\phi}|\psi\rangle|^{2}$. We will only care about the case where $y=1$, since for this case if $|\psi\rangle=\alpha|00\rangle+\beta|11\rangle$ we have that
\begin{equation*}
    p(b|y)=|\mathbb{I}_{A}\otimes\Pi_{b|y=1}^{\phi}|\psi\rangle|^{2}=\frac{1}{2}.
\end{equation*}
Therefore, assuming that Alice and Bob share that state and Bob makes that measurement (in the honest setting) then Bob's outcome for $y=1$ will be perfectly random. This will then be the basis of the certified randomness in this scheme.

The above is what happens for a sequence consisting of one measurement. For sequences of measurements of length $n$ for $n\geq 2$, the post-measurement state $|\psi_{b}\rangle$ as described in \ref{postmeas_single} will be relevant. Note that up to the unitaries $U_{A}^{b|y}$ and $U_{B}^{b|y}$, the state $|\psi_{b|y}\rangle$ is of the form $\alpha|00\rangle+\beta|11\rangle$. Therefore, if after his first measurement, Bob applies the unitary $(U_{B}^{b|y})^{\dagger}$ to his share of the state, the joint state will be
\begin{equation*}
    |\psi_{b|y}\rangle = U_{A}^{b|y}\otimes \mathbb{I}\left(\cos(\zeta_{b|y})|00\rangle+\sin(\zeta_{b|y})|11\rangle\right).
\end{equation*}
Now after applying this unitary, Bob can make another measurement that is a rotated Pauli measurement. Now Bob's input $y$ will be a tuple of length $2$, i.e. $y=(y_{1},y_{2})$. For the second round, Bob's choices of measurements are again between two rotated Pauli basis measurements, where $y_2=0$ is for the Z basis and $y_2=1$ is for the X basis.

If $y_{1}=y_{2}=1$, then Bob performs the rotated X measurement, followed by a correcting unitary, then another rotated X measurement and another corrective unitary. The post-measurement state after this second measurement (and unitary) will be 
\begin{equation*}
    |\psi_{b_{1},b_{2}|1,1}\rangle = U_{A}^{b_{1},b_{2}|y_{1}=1,y_{2}=1}\otimes \mathbb{I}_{B}\left(\cos(\zeta_{b_{1},b_{2}|1,1})|00\rangle+\sin(\zeta_{b_{1},b_{2}|1,1})|11\rangle\right),
\end{equation*}
again for appropriately chosen unitaries and angles. 

The probability of getting the outcomes $b:=(b_1,b_2)$ for inputs $y=(1,1)$ is straightforwardly calculated to be $p(b_1,b_2|1,1)=\frac{1}{4}$. Thus for a sequence of two measurements with each being the rotated $X$ basis measurement, we have two perfectly random outcomes $(b_1,b_2)$. In general, for this sequence of rotated measurement, correcting unitary, rotated measurement and so on, if there $n$ measurements, then the probability $p(b_{1},b_{2},...,b_{n}|y=(1,1,...,1))=2^{-n}$. 

This TQSM scheme thus gives us randomness assuming a particular state and sequence of measurements made by Bob. In subsequent sections the goal will be to remove the assumptions of the state and measurements but certify (almost) the same amount of randomness in the 1sDI scenario. It turns out that the randomness in the TQSM scheme can be certified. That is, to reproduce the observed assemblage (or statistics) between Alice and Bob, Eve would have to prepare devices that implement something equivalent to, or extremely close to, the TQSM scheme. Since this scheme produces lots of randomness, so will the certified version.

Before moving on, it is worthwhile to point out how this scheme differs from that presented in \cite{curchod_unbounded_2017}. The important distinction is that in the scheme of \cite{curchod_unbounded_2017}, in addition to Bob making a sequence of measurements, Alice had to choose from a number of measurements that increased with the length of the sequence. This is because the certification was done in the device-independent setting, and not the 1sDI setting. The number of measurements Alice makes will never depend on the number of measurements in the sequence; it will only depend on the dimension of Alice's Hilbert space since she only needs to do at most a tomographically complete measurement.

\section{Certifiable Unbounded Randomness Generation}\label{sec:unbounded_randomness}

In this section we will give an analytical method for certifying the randomness in a sequential scenario that is suited to the TQSM scheme. In particular, we will show that the TQSM scheme can produce an unbounded amount of certifiable randomness: for an arbitrary integer $N$, there is a sequence of measurements that produces $\Omega(N)$ bits of certifiable randomness. 

In order to certify randomness in the 1sDI setting, we cannot assume the initial state shared by Alice, Bob and Eve nor the measurement sequence made by Bob; we can only assume the Hilbert space of Alice's system, which from now on will be assumed to be two, i.e. Alice holds a qubit system. As mentioned earlier, e can assume that the state $|\psi\rangle_{ABE}$ shared by Alice, Bob and Eve is pure. We can additionally assume for cryptographic purposes that the measurements in Bob's sequence are all projective. For example, the non-projective measurements in the TQSM scheme can be simulated by projective measurements on a potentially larger Hilbert space (we outline such an approach in \appref{app:min-entropy-proofs}).

We introduce notation to refer to Bob's measurements. In particular we will introduce observables for each of Bob's measurements in the sequence. For the first measurement in the sequence, the choice of measurement corresponding to $y_1=0$ and $y_1=1$ will have the observable $Z_{B}=M^{B}_{0|y_1=0}-M^{B}_{1|y_1=0}$ and $X_{B}=M^{B}_{0|y_1=1}-M^{B}_{1|y_1=1}$ respectively, where $M^{B}_{b_{1}|y_{1}}$ being Bob's POVM corresponding to the outcome $b_{1}$ for input $y_{1}$. For subsequent measurements we will introduce a piece of notation that $b^{i}:=(b_{1},b_{2},...,b_{i})$ and $y^{i}:=(y_{1},y_{2},...,y_{i})$ will be the tuple of all values of $b_{i}$ and $y_{i}$ from $1$ to $i$ consecutively (and inclusive). The observable corresponding to the $(i+1)$th measurement in the sequence after obtaining the outcomes $b^{i}$ for choices $y^{i}$ will be denoted as $Z_{B}^{b^{j}|y^{i}}=M^{B}_{0|b^{i},y^{i},y_{i+1}=0}-M^{B}_{1|b^{i},y^{i},y_{i+1}=0}$ and $X_{B}^{b^{j}|y^{i}}=M^{B}_{0|b^{i},y^{i},y_{i+1}=1}-M^{B}_{1|b^{i},y^{i},y_{i+1}=1}$.

The method for certifying this randomness is for Alice to choose between three of the Pauli measurements. Note that Alice does not have to randomly choose between measurements. In each round of the guessing game, Alice can choose a different Pauli basis, but this can be chosen deterministically. Based on the statistics gathered from these three Pauli measurements and Bob's sequence of measurements. Note that every single-qubit observable can be written as a linear combination of Pauli matrices so it is sufficient to make Pauli measurements and calculate the statistics for an arbitrary observable a posteriori. 
As part of the certification we have a \textit{statistical criteria} that the statistics obtained by Alice and Bob need to satisfy. If the statistics satisfy the criteria then this is the certificate that the outcomes of Bob's sequence of measurements is random. To wit, Eve will not be able to perfectly predict the outcomes of Bob's measurements. The statistical criteria will be based on the TQSM scheme.

From the TQSM scheme we have that after the $i$th measurement and correctly unitary, the state of Alice and Bob's two-qubit state will be 
\begin{equation}\label{postmeas_ith}
    |\psi_{b^{i}|y^{i}}\rangle = U_{A}^{b^{i}|y^{i}}\otimes \mathbb{I}_{B}\left(\cos(\zeta_{b^{i}|y^{i}})|00\rangle+\sin(\zeta_{b^{i}|y^{i}})|11\rangle\right).
\end{equation}
We will use the unitaries and angles in this post-measurement state to outline the statistical criteria. For each measurement in a sequence, there will be statistical criteria that should be satisfied. For simplicity we will start with the first measurement in the sequence. 

The statistical criteria we will use can be derived from considering Alice and Bob both making Pauli-X and Pauli-Z measurements on a two-qubit pure entangled state of the form $\alpha|00\rangle+\beta|11\rangle$. The criteria essentially compares the observed statistics with those that would be obtained from perfect Pauli measurements on such an entangled state. These criteria will be then be used for self-testing the devices by showing that their behaviour will not deviate from Pauli measurements on an entangled state. For future work, it would be of interest to use a steering inequality instead of these three separate criteria. Recall that the TQSM scheme is very similar to pure Pauli measurements on a two-qubit pure entangled state, except for some rotation in the typically non-projective measurements. Hence we wish to leverage this fact to produce certifiable randomness. The statistical criteria is 
\begin{eqnarray}\label{crit1}
\vert\langle \tau^{A}_{Z}\otimes Z_{B}\rangle-1\vert&\leq&\epsilon_{1}\nonumber\\
\vert\langle \tau^{A}_{X}\otimes X_{B}\rangle-\sin(2\zeta)\vert&\leq&\epsilon_{2}\nonumber\\
\vert\langle \tau^{A}_{Z}\rangle-\cos(2\zeta)\vert&\leq&\epsilon_{1},
\end{eqnarray} 
where $\tau_{Z}$ and $\tau_{X}$ are the Pauli-Z and Pauli-X observables respectively and $\epsilon_1$, $\epsilon_2$ are real, positive numbers. The angle $\zeta$ just comes from the target pure state between Alice and Bob $|\psi\rangle=\cos(\zeta)|00\rangle+\sin(\zeta)|11\rangle$. For subsequent measurements in the sequence, after the $i$th measurement, we have the following criteria for the $(i+1)$th measurement in the sequence:
\begin{eqnarray}\label{critnew1}
\vert\langle U_{A}^{b^{i}|y^{i}}\tau^{A}_{Z}\left(U_{A}^{b^{i}|y^{i}}\right)^{\dagger}\otimes Z_{B}^{b^{i}|y^{i}}\rangle-1\vert&\leq&\epsilon_{1}^{i+1}\nonumber\\
\vert\langle U_{A}^{b^{i}|y^{i}}\tau^{A}_{X}\left(U_{A}^{b^{i}|y^{i}}\right)^{\dagger}\otimes X_{B}^{b^{i}|y^{i}}\rangle-\sin(2\zeta_{b^{i}|y^{i}})\vert&\leq&\epsilon_{2}^{i+1}\nonumber\\
\vert\langle U_{A}^{b^{i}|y^{i}}\tau^{A}_{Z}\left(U_{A}^{b^{i}|y^{i}}\right)^{\dagger}\rangle-\cos(2\zeta_{b^{i}|y^{i}})\vert&\leq&\epsilon_{1}^{i+1},
\end{eqnarray} 
where the unitary $U_{A}^{b^{i}|y^{i}}$ and $\zeta_{b^{i}|y^{i}}$ are the same as in \ref{crit1}. Just as with \ref{crit1}, $\epsilon_{1}^{1+1}$ and $\epsilon_{2}^{1+1}$ are real, positive numbers. We will call the conjunction of the critera in \ref{crit1} and all criteria \ref{critnew1} for all $i$ the sequential steering criteria (SSC). 

It should be emphasised again that in the SCC, Alice does not need to make a measurement corresponding to the observable $U_{A}^{b^{i}|y^{i}}\tau^{A}_{Z}\left(U_{A}^{b^{i}|y^{i}}\right)^{\dagger}$, say, since for a known unitary $U_{A}^{b^{i}|y^{i}}$, this observable can be written as a real linear combination of Pauli matrices. Thus Alice only needs to measure the Pauli observables to recover the relevant expectation values.

If we take the TQSM scheme and start introducing parameters for the rotated measurements, then we can adjust the SSC parameters to suit the TQSM scheme. For each measurement in the sequence for the measurement in the rotated Pauli $Z$ basis, we will fix the angle $\phi$ to be equal to zero so that the POVM is for the outcome $0$ ($1$) is $|0\rangle\langle 0|$ ($|1\rangle\langle1|$). For the rotated Pauli $X$ measurement we fix the angle to be $\theta_{i}$ for the $i$th measurement in the sequence, which we can fix later but it will be in the range $\theta_{i}\in]0,\frac{\pi}{4}[$. Therefore the POVM for $y_{i}=1$, we have $M^{\theta_i}_{0|y_{i}=1}=\cos(\theta_{i})|+\rangle\langle+|+\sin(\theta_{i})|-\rangle\langle-|$ and $M^{\theta_i}_{1|y_{i}=1}=\cos(\theta_{i})|-\rangle\langle-|+\sin(\theta_{i})|+\rangle\langle+|$. 

One point to make at this point is for the choice parameters to give the criteria in \ref{schemecrit}, after Bob makes the measurement choice $y_{1}=0$ in any round then he makes a projective measurement. The problem with this is that the post-measurement state will be a product state, and no longer entangled; entanglement is necessary to certify randomness in the 1sDI scenario we have here. To get around this issue, we alter the scheme, as is suggested in \cite{curchod_unbounded_2017}, such that after any time Bob makes a projective measurement, he does not make any more measurements in the sequence. That is, a measurement in the $(i+1)$th round will only follow the measurement choice $y_i=1$. Therefore, the only bit-strings $y$ that will be produced by Bob will be consist of a bit-value ($0$ or $1$) prefixed by all ones. When we look at numerical approaches to randomness certification we will relax this constraint to look for optimal amounts of randomness.

When we put these details and values for the measurements into the SSC we obtain the following bounds:
\begin{eqnarray}\label{schemecrit}
\vert\langle U_{A}^{b^{i}|y^{i}}\tau^{A}_{Z}\left(U_{A}^{b^{i}|y^{i}}\right)^{\dagger}\otimes Z_{B}^{b^{i}|y^{i}}\rangle-1\vert&=&0\nonumber\\
\vert\langle U_{A}^{b^{i}|y^{i}}\tau^{A}_{X}\left(U_{A}^{b^{i}|y^{i}}\right)^{\dagger}\otimes X_{B}^{b^{i}|y^{i}}\rangle-\sin(2\zeta_{b^{i}|y^{i}})\vert&\leq&2\sin^{2}(\theta_{i})\nonumber\\
\vert\langle U_{A}^{b^{i}|y^{i}}\tau^{A}_{Z}\left(U_{A}^{b^{i}|y^{i}}\right)^{\dagger}\rangle-\cos(2\zeta_{b^{i}|y^{i}})\vert&=&0.
\end{eqnarray} 
We will use these values to certify the randomness produced by the TQSM scheme.

Coming back to randomness certification, for a sequence of measurements, the sequence of inputs $y^{*}$ from which we will obtain a string of $n$ random bits will be the all-ones string, i.e. $y^{*}=(1,1,...,1)$. The following (informally stated) result gives an upper bound on the guessing probability for Eve to correctly guess Bob's sequence of measurement outcomes. 

\begin{theorem_fix}\label{thm1}
For Bob making a sequence of $n$ measurements yielding the outcome bit-string $b$ of length $n$, if Alice, Bob and Eve share some initial state $\ket{\psi}_{ABE}$, if Eve makes a measurement associated with operators $\{M_z\}_{z}$, where $z$ is Eve's guess of Bob's outcome $b$, Eve's guessing probability is
\begin{equation}
p_{\textrm{guess}}(y^{*})=\sum_{z}\delta_{b,z}\textrm{tr}_{A}\sigma_{z,b|y^{*}}.
\end{equation}
and if for each $i$, if the SSC is satisfied and for all $\epsilon_{1}^{i}$ and $\epsilon_{2}^{i}$ (with $\epsilon_{1}^{1}=\epsilon_{2}$ and $\epsilon_{1}^{1}=\epsilon_{2}$) from the statements of the SSC, then
\begin{equation}
p_{\textrm{guess}}\leq \prod_{i=1}^{n}\left(\frac{1}{2}+\sqrt{\epsilon^{i}_{1}}\left(3\sqrt{2}+2+\frac{5}{2\sin(\zeta_{b^{i-1}|y^{i-1}})}\right)+3\sqrt{\epsilon^{i}_{1}+\epsilon^{i}_{2}}\left(\frac{1}{\sqrt{2}\sin(\zeta_{b^{i-1}|y^{i-1}})}+1\right)\right),
\end{equation}
and if $\epsilon^{i}_{1}=0$ for all $i$, then
\begin{equation}
P_{\textrm{guess}}\leq \prod_{i=1}^{n}\left(\frac{1}{2}+3\sqrt{\epsilon^{i}_{2}}\left(\frac{1}{\sqrt{2}\sin(\zeta_{b^{i-1}|y^{i-1}})}+1\right)\right).
\end{equation}
\end{theorem_fix}
The proof of this theorem can be found in the Appendix, section \ref{app:min-entropy-proofs}. This theorem uses techniques from self-testing in the 1sDI setting as developed in \cite{supic}. Of independent interest we present a method to self-test all partially entangled two-qubit states in a robust manner.

Given Theorem \ref{thm1} we can certify an unbounded amount of randomness assuming all of the SSC is satisfied. In particular, for the TQSM scheme we can give bounds on the amount of bits that will be certified, as indicated in the following the result.
\begin{theorem_fix}
\textit{If all statistics satisfy the SSC with $\epsilon^{i}_{1}=0$ and $\epsilon^{i}_{2}=2\sin^{2}(\theta_{i})$ for all $i$ and $\theta_{i}$ as a free choice of angle that is assumed to be small, then the certifiable randomness $H_{min}(b|y^{*},z)$ for Bob's sequence of $n$ measurements is
\begin{equation*}
H_{min}(b|y^{*},z)\geq (1-c)n,
\end{equation*}
where $c\in]0,1[$. Furthermore, the TQSM scheme achieves this asymptotic behaviour as its resulting statistics will satisfy the SSC for the chosen values $\epsilon^{i}_{1}=0$ and $\epsilon^{i}_{2}=2\sin^{2}(\theta_{i})$ for all $i$. 
}
\end{theorem_fix}

\begin{proof} If we take the result of Theorem \ref{thm1} and convert the probability into a min entropy we have
\begin{eqnarray*}
H_{min}(b|y^{*},z)&\geq&-\sum_{i=1}^{n}\log_{2}\left(\frac{1}{2}+3\sqrt{\epsilon^{i}_{2}}\left(\frac{1}{\sqrt{2}\sin(\zeta_{b^{i-1}|y^{i-1}})}+1\right)\right)\\
&=&n-\sum_{i=1}^{n}\log_{2}\left(1+6\sqrt{\epsilon^{i}_{2}}\left(\frac{1}{\sqrt{2}\sin(\zeta_{b^{i-1}|y^{i-1}})}+1\right)\right)\\
&\geq&n-\frac{6}{\textrm{ln}2}\sum_{i=1}^{n}\sqrt{\epsilon^{i}_{2}}\left(\frac{1}{\sqrt{2}\sin(\zeta_{b^{i-1}|y^{i-1}})}+1\right)\\
&=&n-\frac{6}{\textrm{ln}2}\sum_{i=1}^{n}\sin(\theta_{i})\left(\frac{1}{\sin(\zeta_{b^{i-1}|y^{i-1}})}+\sqrt{2}\right)\\
&\geq&n-\frac{6\sqrt{2}}{\textrm{ln}2}\sum_{i=1}^{n}\theta_{i}\left(\frac{1}{\zeta_{b^{i-1}|y^{i-1}}}+1\right)\\
&\geq&n-\frac{12\sqrt{2}}{\textrm{ln}2}\sum_{i=1}^{n}\theta_{i}\left(\frac{1}{\zeta_{b^{i-1}|y^{i-1}}}\right),
\end{eqnarray*}
where in the third line we have that $\textrm{ln}(1+\alpha)\leq\alpha$, and in the fourth line we use the value of $\epsilon^{i}_{2}=2\sin^{2}(\theta_{i})$ from the statement of the theorem. In the fifth line we have that $\sin(\alpha)\geq\frac{\alpha}{\sqrt{2}}$ and $\sin(\alpha)\leq\alpha$ for $\alpha\in]0,\frac{\pi}{4}[$, which will always be the case by construction. Then in the sixth line, we used the fact that $(\frac{1}{\alpha}+1)\leq\frac{2}{\alpha}$ for $\alpha\in]0,\frac{\pi}{4}[$, which will always be the case by construction. In the conditions, we can choose $\theta_{i}=d\zeta_{b^{i-1}|y^{i-1}}$ for constant $d=\frac{c\textrm{ln}2}{12\sqrt{2}}$, such that $H_{\textrm{min}}(\textbf{B}|\textbf{Y}=\textbf{0},E)\geq (1-c)N$, thus completing the proof. 
\end{proof}

Note that by appropriate choice of the measurement parameters for the rotated Pauli-X basis measurement we can get arbitrarily close to the $n$ bits of randomness by reducing the constant $c$ in the statement of the theorem. We cannot reduce this constant to $0$ since this would involve one of the rotated Pauli-X measurements would become projective, and we would not be able to certify randomness.

\section{Numerical Results} \label{sec:numericalresults}

The previous analytical results indicate that unbounded randomness is possible, but the methods employed are perhaps sub-optimal in extracting the most randomness from the TQSM scheme. In this section we will employ numerical techniques, similar to those developed in \cite{passaro_optimal_2015}, to give an indication of how robust the scheme is for randomness generation. 

The methods employed in this section are based in semi-definite programming (SDP). We will take the approach that given the violation of a steering inequality, can we certify the randomness. A violation of a steering inequality implies that there must be certifiable randomness present. In this way the violation of the steering inequality is the certificate for the randomness. First we will outline how to derive a steering inequality from assemblages.

Given an assemblage, a method was derived to determine the steerability of the assemblage via an SDP by Skrzypczyk et. al., \cite{skrzypczyk_quantifying_2014}. The steering weight (SW) is given to be the solution to the following SDP, (\ref{steeringweightsdp}):
\begin{align}
        \begin{array}{lllll}
        SW(\{\sigma_{b|y}\})=& \min        & 1- \tr\sum\limits_\lambda\sigma_\lambda &\\ 
          & \text{s.t. }&\sigma_{b|y} - \sum\limits_{\lambda}D(b|y,\lambda)\sigma_\lambda \geq 0 & \forall b,y\\ 
          &      & \sigma_{\lambda} \geq 0,    &\forall \lambda 
        \end{array} \label{steeringweightsdp}
\end{align}
where $\{\sigma_\lambda\}$ is an assemblage that Eve could produce for Alice using hidden variables $\lambda$. This SDP has a corresponding dual program given by:
\begin{align}
        \begin{array}{lllll}
            SW(\{\sigma_{b|y}\}) =& \max        & 1 - \tr\sum\limits_{by}F_{b|y}\sigma_{b|y}                &\\ 
                &\text{s.t. } &  \sum\limits_{by}D(b|y,\lambda)F_{b|y}-\mathds{1} \geq 0  &\forall \lambda\\ 
                &             & F_{b|y}  \geq 0,                                          & \forall b,y
        \end{array} \label{steeringweightdual}
\end{align}
The dual program, (\ref{steeringweightdual}), is the most relevant for this work, as shown in \cite{skrzypczyk_quantifying_2014}, the dual variables of the SDP, (\ref{steeringweightdual}), in fact define a steering inequality, $\{F_{b|y}\}$, for which the assemblage, $\{\sigma_{b|y}\}$, produces an optimal violation, if one exists. We will use these steering inequalities as the fundamental building block for our sequential certification scheme.

We now return calculating the certifiable randomness in terms of the guessing probability for Eve to guess Bob's measurement outcomes. For simplicity, we will first study the case of a single measurement before giving the results for a sequence of measurements. With just a single measurement, the maximum guessing probability is given as the solution to the following SDP:

\begin{align}
\begin{array}{ccll}
p_{\mathrm{guess}} =& \max\limits_{\{\sigma^{E}_b|y\}_{b,y}}&\tr_A\sigma^{E}_{b|y^*} &\\ 
&\text{s.t. } &\sum\limits_{b,y}F_{b|y}\sigma^{E}_{b|y} = v& \label{gpsdprelax}\\ 
&&\sum\limits_{b}\sigma^{E}_{b|y} = \sum\limits_{b}\sigma^{E}_{b|y'} & \forall z, y \neq y'\\ 
&&\sigma^{E}_{b|y}\succeq 0 & \forall y,b
\end{array}
\end{align}
The steering inequality $\{F_{b|y}\}$ is the one determined by the SDP \eqref{steeringweightdual}, which is optimally violated by the observed assemblage. The SDP \eqref{gpsdprelax} allows Eve to create, for Alice, any assemblage, $\{\sigma_{b|y}^E\}$, as long as this assemblage obeys the constraints in the SDP. 

The first constraint enforces the fact that this assemblage should produce the observed violation of the steering inequality, $\{F_{b|y}\}$,  which is found as a result of Alice computing the optimal values for the steering weight SDP \eqref{steeringweightdual}. Of course, if the assemblage that Alice observes is \textit{not steerable}, i.e.\@ it produces a steering weight of $0$, then this will be reflected in the observed violation of a steering inequality, i.e.\@ there will not be one for any steering inequality. The second constraint enforces that Alice and Bob cannot communicate faster than the speed of light (no-signalling condition), while the last constraint enforces that Eve must produce a valid assemblage for Alice i.e. it must be a positive semidefinite matrix.

We can now extend this scenario to one in which Bob implements a sequence of measurements on his half of the shared state. Defining the protocol for $n$ rounds is therefore straightforward. The idea will be that for each measurement in the sequence there will be a steering inequality, and an observed violation. The steering inequalities and violations will be obtained from the assemblages produced by the TQSM scheme, where the SW is calculated and a steering inequality generated for each measurement round in the sequence of measurements. Once we have this set of steering inequalities, she can determine the guessing probability for Eve, as the solution of the following SDP:
\begin{align}
\begin{array}{|l|l|}
    \hline
    p_{\text{guess}}= \max\limits_{b,y} \tr_A\sigma^E_{b|y = y^*}\qquad \text{s.t.\@}&\\ \hline
\begin{array}{lll}
 \sum\limits_{b,y} F_{b|y}\sigma^E_{b|y} &= v_n,   \\
\sum\limits_{b^{n-1},y^{n-1}} F_{b^{n-1}|y^{n-1}}\sigma^E_{b^{n-1}|y^{n-1}}   &= v_{n-1}  \\ \nonumber
                            \vdots                      & \vdots\\
 \sum\limits_{b_1,y_1} F_{b_1|y_1}\sigma^E_{b_1|y_1}    &  = v_1, 
\end{array} & \begin{array}{lll}
\sum\limits_{b_n} \sigma^E_{b|y} &= \sigma^E_{b^{n-1}|y^{n-1}} , &  \forall y_n \\
 \sum\limits_{b_{n-1}} \sigma^E_{b^{n-1}|y^{n-1}} &= \sigma^E_{b^{n-2}|y^{n-2}} & \forall y^{n-1} \\
                                    &\vdots     & \vdots\\
\sum\limits_{b_1}\sigma^E_{b_1|y_1} &= \rho_A   & \forall y_1 
\end{array}\\ \hline
\begin{array}{lll}
\sum\limits_{b} \sigma^E_{b|y} &= \sum\limits_{b} \sigma^E_{b|y'} & \forall y,y' \\
 \sum\limits_{b^{n-1}} \sigma^E_{b^{n-1}|y^{n-1}}  &= \sum\limits_{b^{n-1}} \sigma^E_{b^{n-1}|{y^{n-1}}'} & \forall y^{n-1},{y^{n-1}}' \\
                                        &\vdots                                 & \vdots\\
 \sum\limits_{b_1}\sigma^E_{b_1|y_1}    &= \sum\limits_{b_1}\sigma^E_{b_1|y_1'} & \forall y_1,y_1' 
\end{array} 
&  
\begin{array}{lll}
 \sigma^E_{b|y}         &\succeq 0  &  \forall y,b \\
 \sigma^E_{b^{n-1}|y^{n-1}} &\succeq 0  &  \forall y^{n-1},b^{n-1} \\
                                        &\vdots     &  \vdots \\
 \sigma^E_{b_1|y_1}                     &\succeq 0  &  \forall y_1, b_1
\end{array} \\ \hline
\end{array} 
\end{align}

The solution of this SDP is the guessing probability and the maximum over the trace of all the assemblages that Eve can create for Alice at the end of the protocol, $\sigma^E_{b|y = y^*}$, for a particular input string, $y^*$. Again, Eve knows from which measurement settings, $y^*$, Bob wants to extract randomness. The steering inequality violations can be calculated by Alice for the assemblage she observes. The constraints of the SDP are similar to the single measurement case except for the addition of one new set of constraints which are required for a sequence. These particular constraints enforce causality in the measurement sequence, as mentioned earlier. Recall that, as mentioned earlier, any assemblage satisfying these constraints can be implemented by Alice and Bob sharing a quantum state and by Bob making appropriate measurements \cite{Sainz2019}.

To obtain the most amount of randomness, for the final measurement round, the measurement operators will become projective, i.e. $\theta_n = \phi_n = 0$ and the state at round $n-1$ should be a pure entangled state. In this case, it is possible to define the steering inequality explicitly, as done in \cite{skrzypczyk_quantifying_2014}:
\begin{align}
F_{b|y} &= \alpha\left(\mathds{1} - \frac{\sigma_{b|y}}{\tr(\sigma_{b|y})}\right) \label{projectivesteeringfunctional}
\end{align}
where $\alpha$ is chosen sufficiently large. A choice of $\alpha = 100$ was chosen for all numerical results in this paper. Clearly, this choice of a steering inequality automatically gives a violation value of $v_n = 0$.

\subsection{Ideal Case} \label{ssec:ideal}

In this section, we present numerical results to illustrate the performance of the TQSM scheme assuming ideal functionality of devices. As a convention, it will be assumed that Bob always measures in the noisy X basis in the first round, with the final measurement round in the protocol being projective, $\theta_n = 0$ or $\phi_n = 0$, depending on whether $n$ is odd or even. We will also allow for the possibility that both of Bob's possible measurements for each measurement in the sequence can be non-projective.

\begin{figure}[H]
        \centering
            \includegraphics[width =0.6\linewidth, height= 0.5\linewidth]{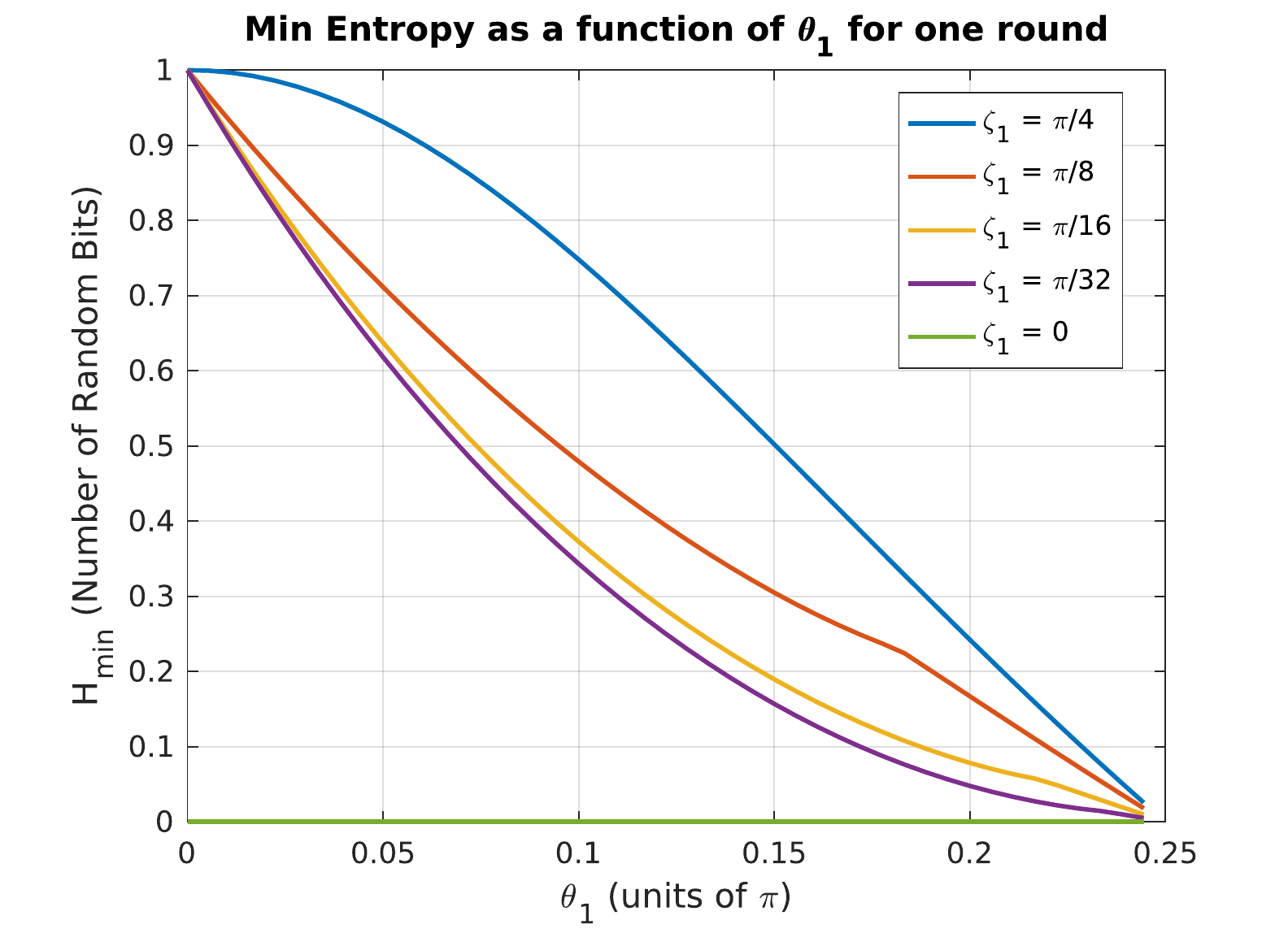}
            \caption{$H_{min}$ for one round of measurements, using a range of initial states, $\zeta_1$, as a function of initial measurement angle, $\theta_1$}
            \label{fig:OneRound}
\end{figure}

For completeness, the min entropy for one round of measurement is plotted as a function of measurement angles used for the first round, with the rotated X measurements for a range of values of $\theta_1$, as seen in \figref{fig:OneRound}. All measurements are applied on the following initial pure state:

\begin{align}
    \ket{\Psi(\zeta_1)} &= \cos(\zeta_1)\ket{00}+\sin(\zeta_1)\ket{11} \label{initialstate}
\end{align}

$\ket{\Psi(\zeta_1)}$ was measured for values of: $\zeta_1 \in \{0, \frac{\pi}{32}, \frac{\pi}{16}, \frac{\pi}{8}, \frac{\pi}{4} \}$. The solution of this SDP clearly reproduce the already known results for a single measurement round, as is done in \cite{passaro_optimal_2015}, \cite{skrzypczyk_paulskrzypczyk/steeringreview_2018}, but using our SDP which is slightly different than the one derived in those works. As expected, when $\zeta_1 = 0$, no randomness can be certified as the state becomes a product state. In the opposite end of the spectrum, for $\zeta_1 = \pi/4$, the maximal amount of randomness can be certified, since this state is maximally entangled between Alice and Bob.

\begin{figure}[H]
\centering
\begin{subfigure}{.5\textwidth}
  \centering
  \includegraphics[width=\linewidth]{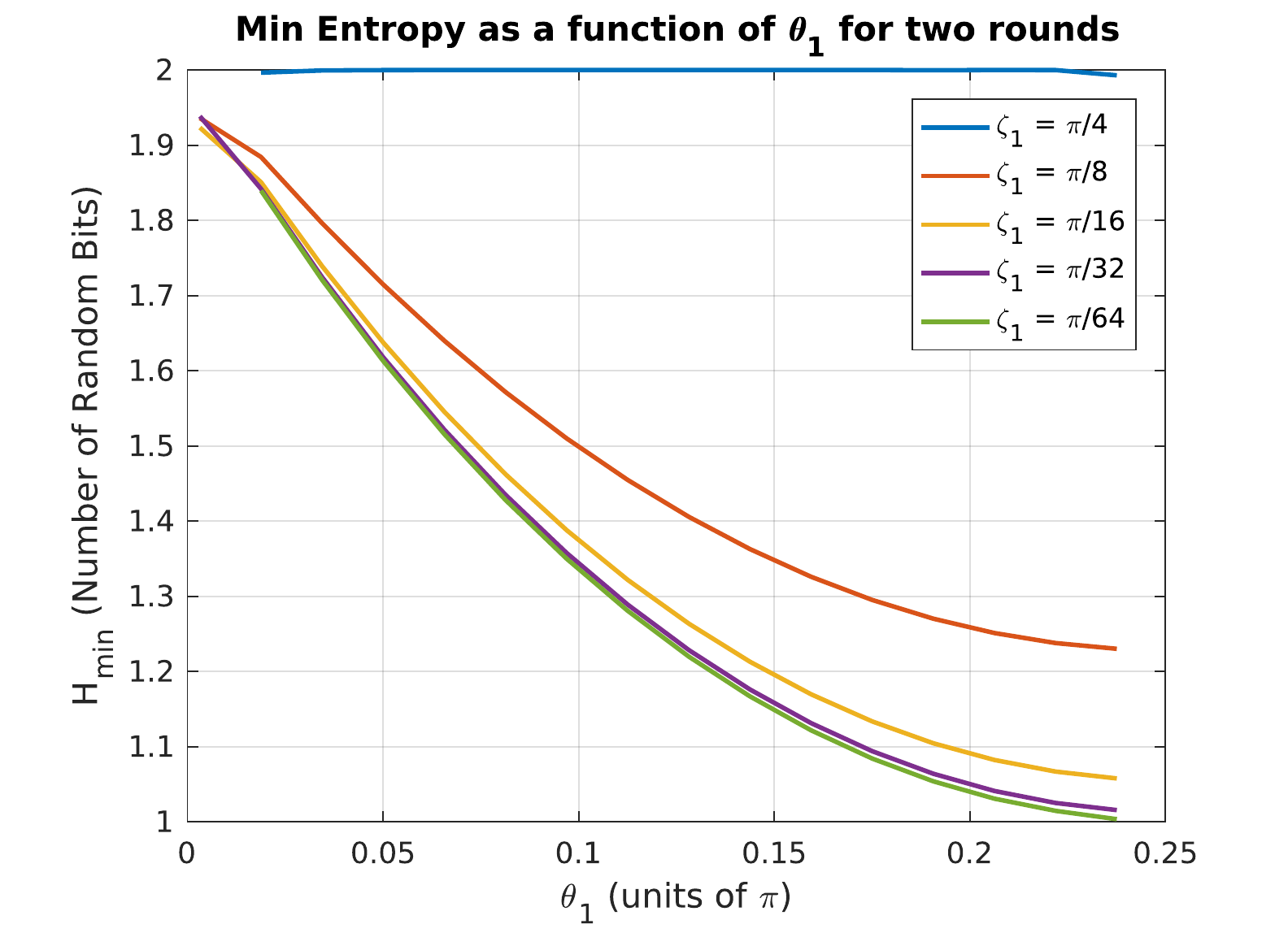}
   \caption{}
 \label{fig:TwoRounds1}
\end{subfigure}%
\begin{subfigure}{.5\textwidth}
  \centering
  \includegraphics[width=\linewidth]{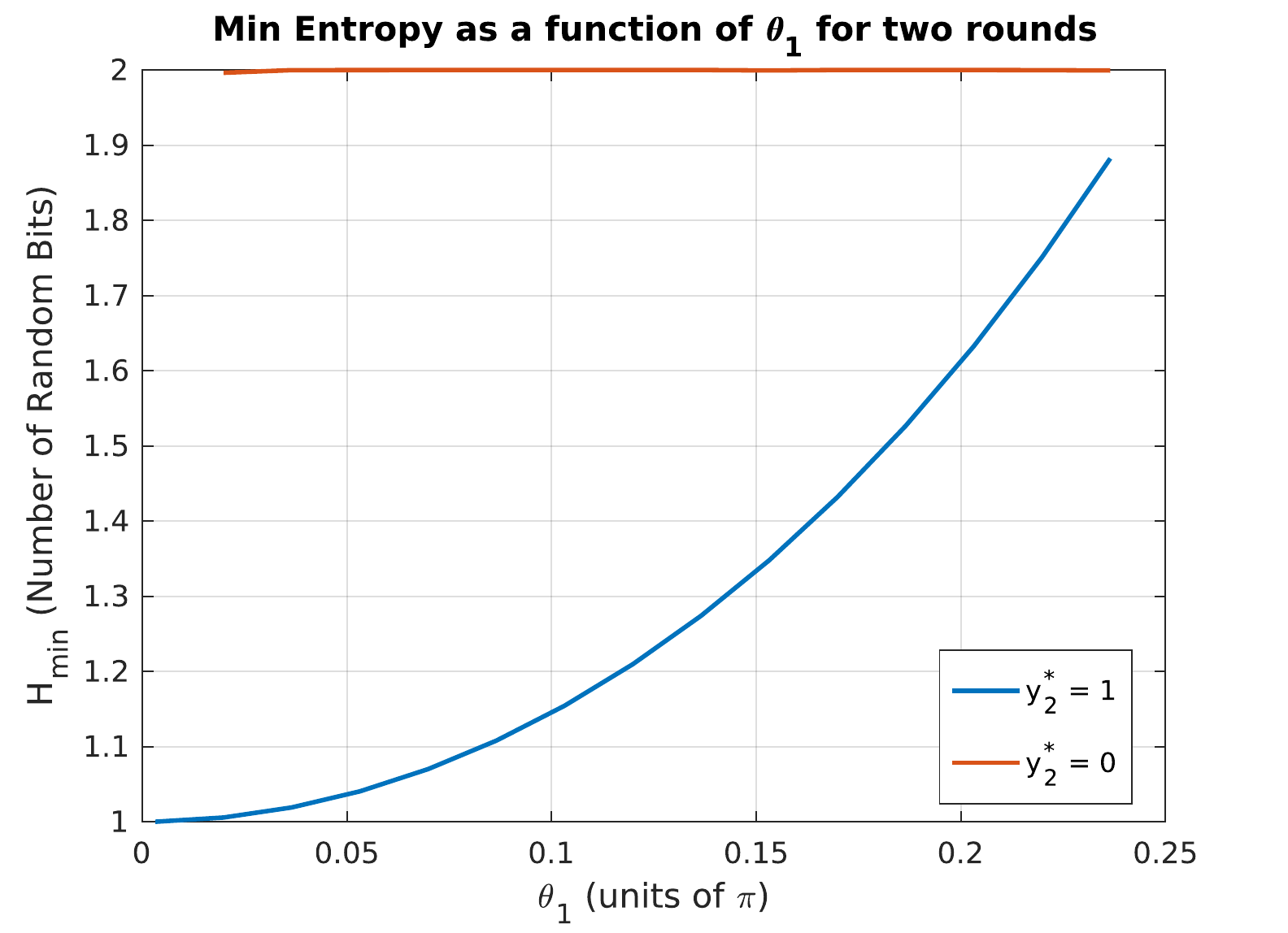}
   \caption{}
\label{fig:TwoRounds2}
\end{subfigure}
\caption{\textbf{(a)} $H_{min}$ for two rounds of measurements, with a range of initial states, $\zeta_1 \in (0,\frac{\pi}{4}]$ and $\phi_1 = \theta_2 = \phi_2 =0$. \textbf{(b)} Difference in certified randomness when choosing between measurement settings $y^*_2 = 1 \text{ or } y^*_2  = 0$ in the second measurement round}
\label{fig:TwoRounds}
\end{figure}

\figref{fig:TwoRounds1} and \figref{fig:TwoRounds2} show the results after two measurement rounds. In \figref{fig:TwoRounds1}, the measurement in round one was taken to be in the noisy $X$ basis, with a range of initial angles $\zeta_1$, and the measurement in round two was taken to be in the usual computational basis, $\phi_2 = 0$. \figref{fig:TwoRounds2} illustrates the difference in choosing different measurement choices for the second round, i.e.\@ between $y_2^* = 0$, or $y_2^* = 1$, with maximal randomness certified after sequential measurements in alternating bases,  $y_1^* = 1, y_2^* = 0$. We cut the graphs at the extremes of the measurement angles ($\theta_1 =\{0, \pi/4\})$ in order to avoid the discontinuity that occurs as soon as the first round measurement undergoes the transition from projective to non-projective.

An interesting feature of the protocol can be seen in \figref{fig:TwoRounds1}, for the case of $\zeta_1 = \pi/4$. It turns out that in this case a maximal amount of randomness can be certified, \textit{for all} initial measurement angles, $\theta_1$. This behaviour illustrates the fundamental difference between the steering, and fully device independent scenario and the more robust nature of quantum steering. In the latter, one observes the amount of certifiable randomness \textit{decreases} monotonically as ($\theta_1\rightarrow 0$), corresponding to the first round measurement becoming non-interactive. We leave a further analysis of this phenomenon to future work.

\begin{figure}[H]
\centering
\begin{subfigure}{.34\textwidth}
  \centering
  \includegraphics[width=\linewidth, height=0.9\linewidth]{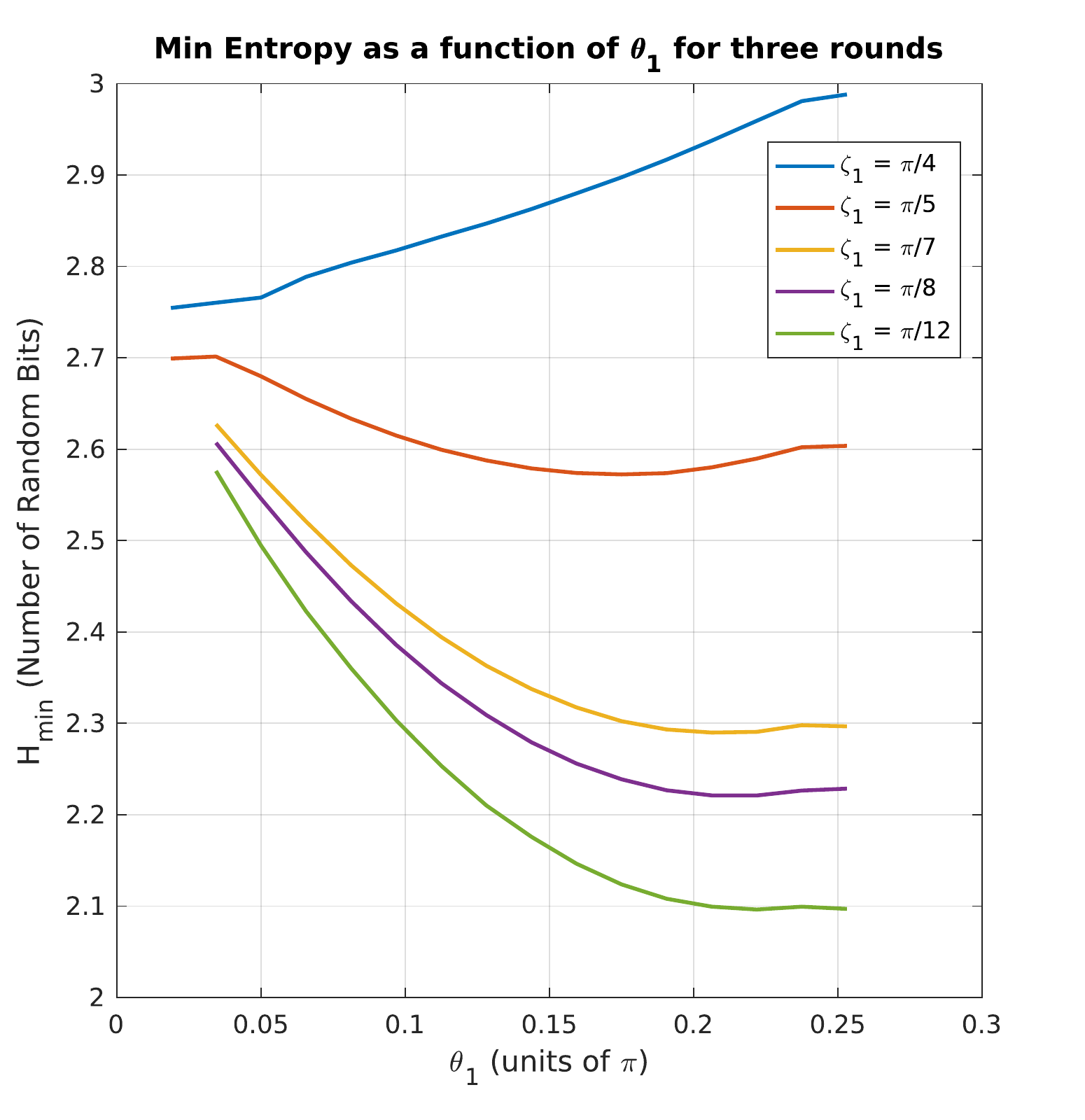}
   \caption{}
\label{fig:ThreeRounds1}
\end{subfigure}%
\begin{subfigure}{.34\textwidth}
  \centering
  \includegraphics[width=\linewidth, height=0.9\linewidth]{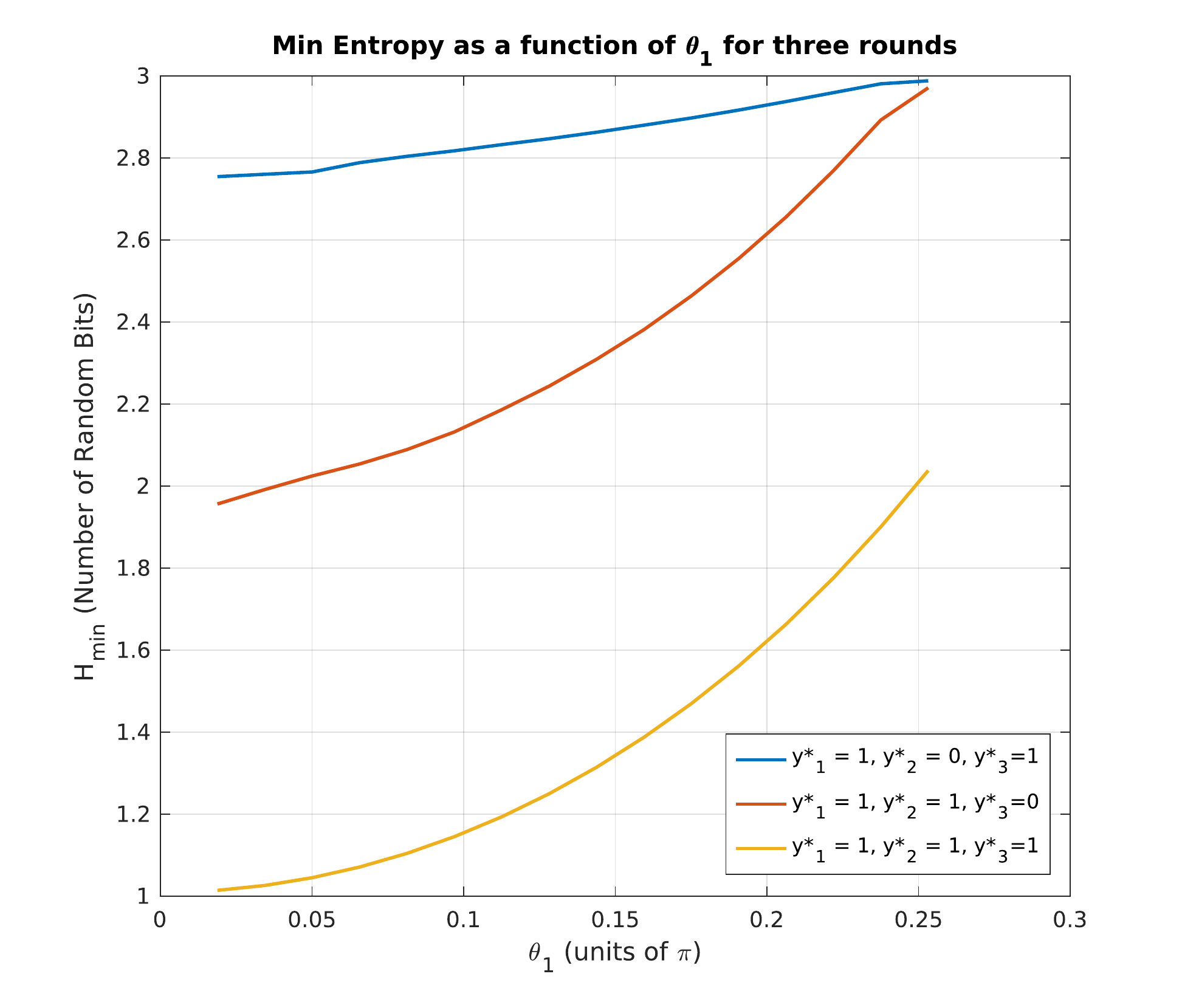}
   \caption{}
    \label{fig:ThreeRounds2}
\end{subfigure}%
\begin{subfigure}{.34\textwidth}
  \centering
  \includegraphics[width=\linewidth, height=0.9\linewidth]{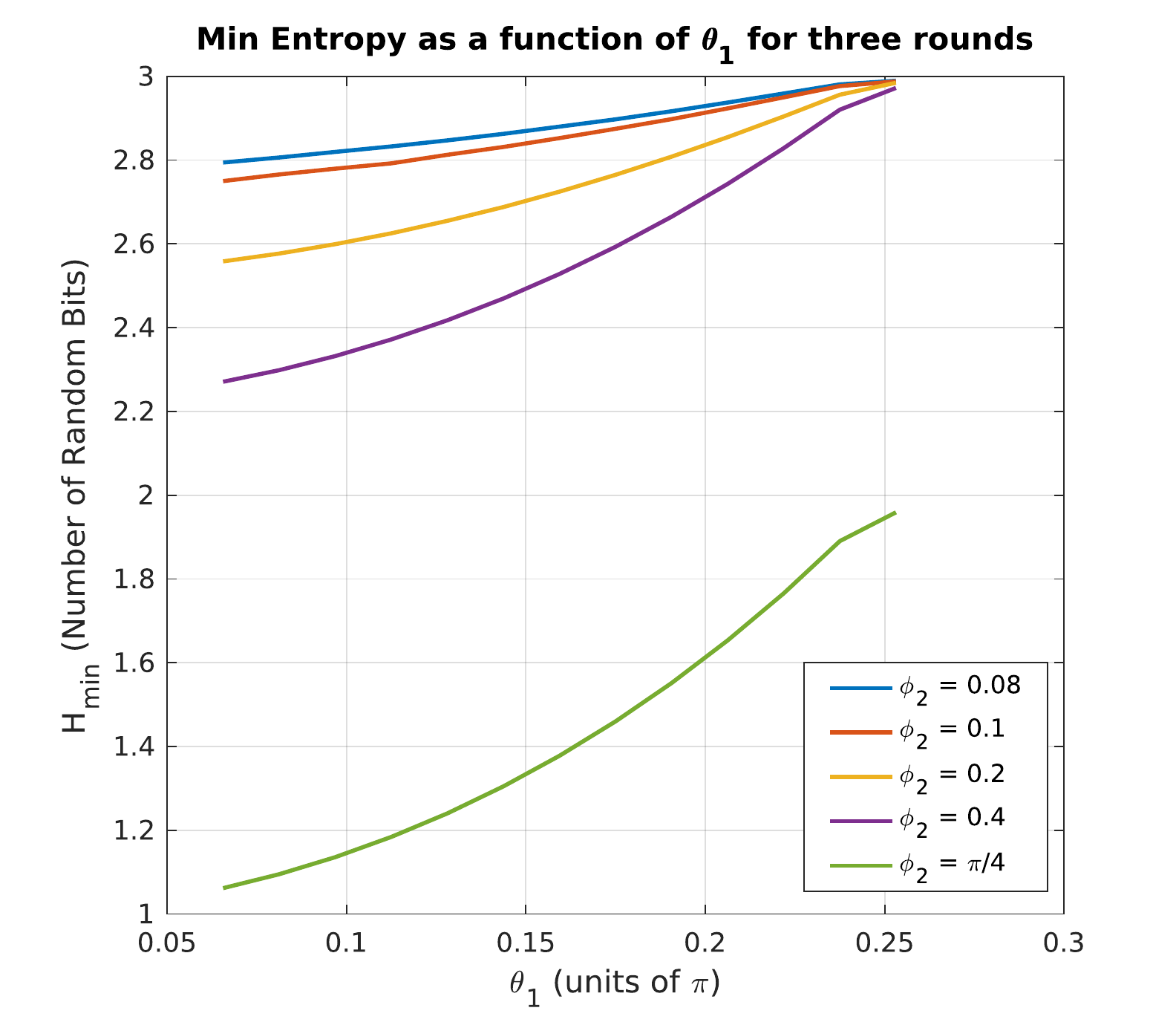}
   \caption{}
\label{fig:ThreeRounds3}
\end{subfigure}
\caption{\textbf{(a)} $H_{min}$ using various initial states, with initial angles, $\zeta_1 \in \{\frac{\pi}{4}, \frac{\pi}{5}, \frac{\pi}{7}, \frac{\pi}{8}, \frac{\pi}{12}\}$. \textbf{(b)} $H_{min}$ using various measurement settings, $y_1^*, y_2^*, y_3^*$. \textbf{(c)} $H_{min}$ using various angles in the second round, $\phi_2 \in \{0.08 , 0.1 , 0.2, 0.4, \frac{\pi}{4}\}$ rad.}
\label{fig:ThreeRounds}
\end{figure}

Finally, \figref{fig:ThreeRounds} illustrates numerical results for the protocol with three measurement rounds. The protocol proceeds in exactly the same manner as for one and two rounds. In particular, in the first round, Bob can choose between a non-projective measurement in the noisy $X_{\theta_1}$ basis, or if the particular run of the protocol is a test, he will measure in the projective $Z_{0}$ basis. In the second round, he will choose to measure in the noisy $Z_{\phi_2}$ basis, or the $X_{0}$ basis for a test run. In the final round, he will choose to measure in the projective ($\theta_3 = 0$) $X_{0}$ basis, or the  projective ($\phi_3 = 0$) $Z_{0}$ basis for a test. Again, \figref{fig:ThreeRounds2} reiterates the optimality of using an alternating sequence of non-projective measurements, with the most randomness produced with the setting $y_1^* = 1, y_2^* = 0, y_3^* = 1$ in this example. \figref{fig:ThreeRounds3} shows the results for various second round measurement angles, and the amount of randomness that can be certified increases as the measurement angle, $\phi_2\rightarrow 0$.

In these results we see that the amount of randomness that can be certified using the numerics is quite robust. This then could make this scheme amenable to experiment. In the next section we will adapt these numerical techniques to look at experimental feasibility of this randomness certification scheme.

\section{Towards Experimental Implementations}\label{sec:exp_implementation}

\subsection{Networked Ion Trap Implementation}\label{ssec:ion_trap_implementation}
The framework in which we have designed this protocol, assuming a \textit{malicious} adversary, Eve, is general enough to include the scenario in which she is not intentionally trying to interfere with our randomness generation, but instead we can imagine that Eve simply made some error in building the devices. This would correspond to introducing some noise, for example, in our state preparation and/or measurement apparatus. This noise assumption is clearly a subcase of the malicious adversary scenario. This mentality allows us to use our protocol to evaluate the usefulness of some current available technologies for randomness generation purposes, in some simple cases. In particular, we will restrict to assuming we only have some noise in our state preparation, but all other parts of the device works perfectly. To do so, we test the state introduced in \cite{nigmatullin_minimally_2016}, which can be produced between two parties in a networked architecture of ion traps:

\begin{align}
    \rho_\epsilon &= (1-\epsilon)\Phi^+ + \frac{\epsilon}{3}\Phi^-+\frac{\epsilon}{3}\Psi^+ +\frac{\epsilon}{3}\Psi^- \label{rawstate1}
\end{align}

where $\phi^+$, $\phi^-$, $\psi^+$, and $\psi^-$ are the standard 2-qubit Bell states. The state, (\ref{rawstate1}), is  a mixed state assuming uniform depolarising noise. In \cite{nigmatullin_minimally_2016}, this  state is assumed to be one produced by two ion traps entangled by a photonic link. The simple noise model is chosen to allow use of a technique to purify the state. In particular, after 3 rounds of this purification protocol, the resulting states are given by:

\begin{align}
    \rho^{(0)}_\epsilon &= (1-\epsilon)\Phi^+ + \frac{\epsilon}{3}\Phi^-+\frac{\epsilon}{3}\Psi^+ +\frac{\epsilon}{3}\Psi^- \label{rawstate2}\\
    \rho^{(1)}_\epsilon &= \left(1-\frac{2}{3}\epsilon-\frac{2}{3}\epsilon^2\right)\Phi^+ 
    +\left(\frac{2}{9}\epsilon+\frac{2}{9}\epsilon^2\right)\Phi^-  + \frac{2}{9}\epsilon^2\Psi^+ + \frac{2}{9}\epsilon^2\Psi^- + O(\epsilon^3)  \label{1roundpure}\\ 
    \rho^{(2)}_\epsilon &= \left(1- \frac{8}{9}\epsilon^2- \frac{8}{27} \epsilon^3\right)\Phi^+    \frac{4}{9}\epsilon^2\Phi^-
    + \frac{4}{9}\epsilon^2\Psi^+ + \frac{8}{27}\epsilon^3\Psi^- + O(\epsilon^4)
   \label{tworoundpure}\\ 
    \rho^{(3)}_\epsilon &= \left(1- \frac{2}{9}\epsilon^2- \frac{16}{27}\epsilon^3\right)\Phi^+\frac{2}{9}\epsilon^2\Phi^- + \frac{8}{27}\epsilon^3\Psi^+ + \frac{8}{27}\epsilon^3\Psi^- + O(\epsilon^4) \label{threeroundpure}
\end{align}

where $\rho^{(i)}_\epsilon$ is the state produced after $i$ rounds of the purification protocol. 

Currently, raw entanglement between two ion traps, connected with an entangling photon, has been achieved with a fidelity of about $85\% \implies \epsilon \approx 0.15$, \cite{hucul_modular_2014}. Starting with this level of raw infidelity, the purification protocol produces states of infidelity $ \epsilon \approx 0.1, 0.02, 0.005$ after one, two and three rounds respectively. The fidelity is given by \eqref{fidelity}, \cite{nielsen_quantum_2011}, and taken to be between the actual state $\rho^{(i)}$, and the pure Bell state, $\Phi^+$:

\begin{align}
    F(\rho^{(i)}_{\epsilon},\Phi^+) = \Tr\left(\sqrt{\sqrt{\rho^{(i)}_\epsilon}\Phi^+\sqrt{\rho^{(i)}_\epsilon}}\right)  \label{fidelity}
\end{align}

\begin{figure}[H]
\centering
\begin{subfigure}{.5\textwidth}
  \centering
  \includegraphics[width=\linewidth]{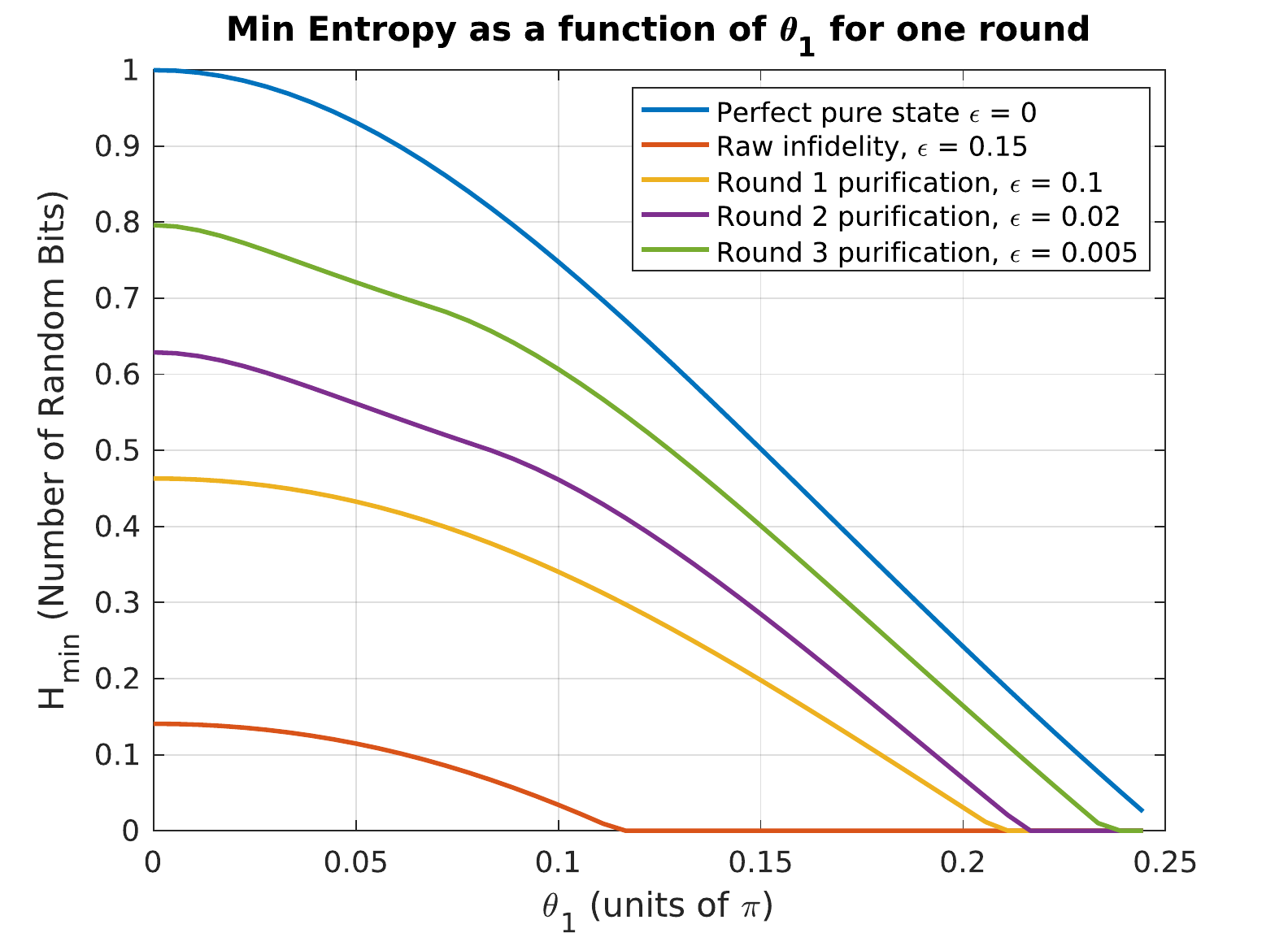}
   \caption{}
        \label{fig:oneroundpure}
\end{subfigure}%
\begin{subfigure}{.5\textwidth}
  \centering
  \includegraphics[width=\linewidth]{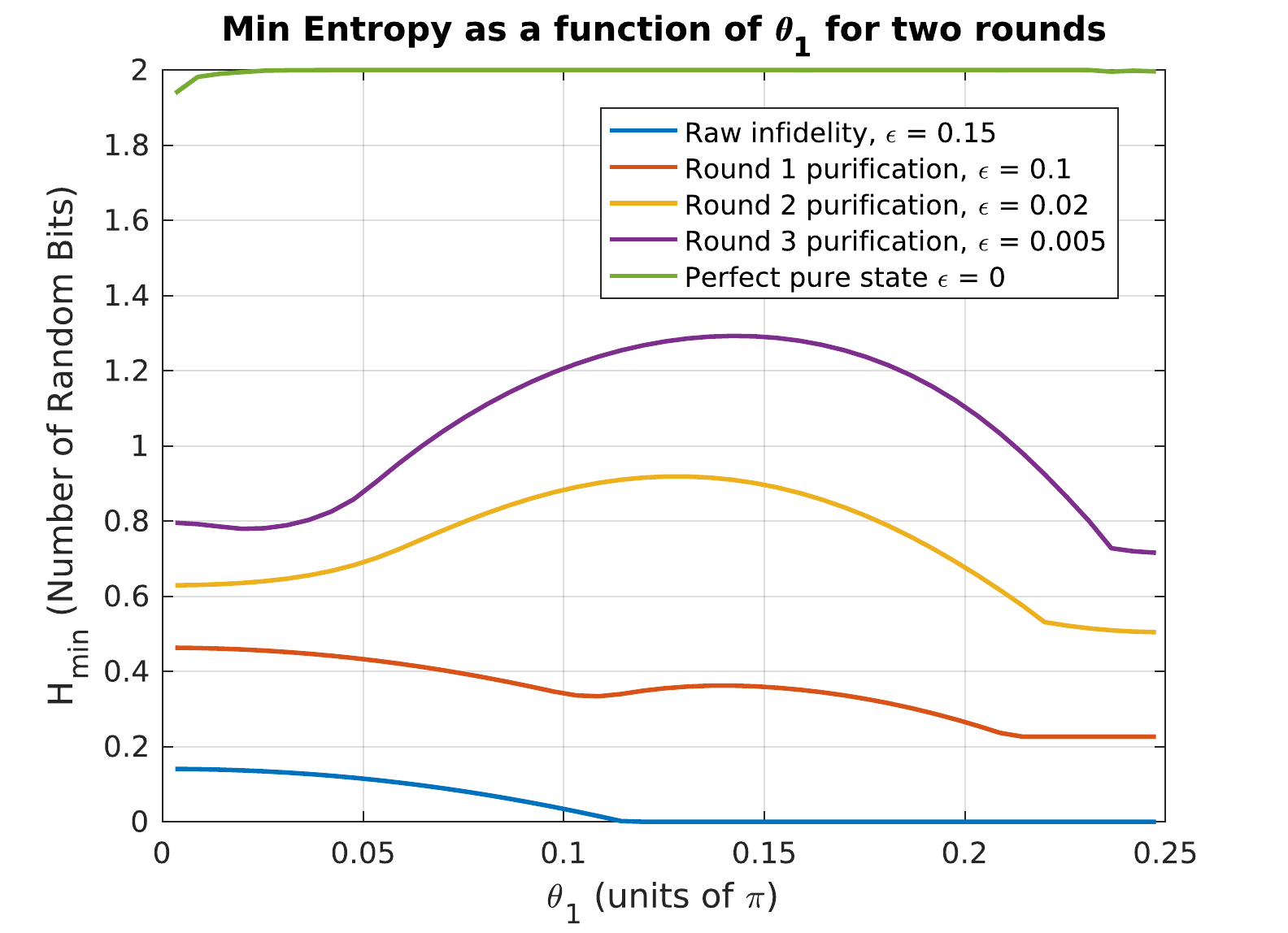}
   \caption{}
        \label{fig:tworoundspure}
\end{subfigure}
\caption{\textbf{(a)} Single measurement on the raw entangled state \eqref{rawstate1} ($\epsilon = 0.15$), the states produced after three rounds of the purification protocol, (\ref{1roundpure}, \ref{tworoundpure}, \ref{threeroundpure}), with $\epsilon = 0.1, 0.02, 0.005$ respectively and a perfect pure state with $\epsilon =0 $. \textbf{(b)} Two rounds of measurement on the raw entangled state (\ref{rawstate1}) ($\epsilon = 0.15$), the states produced after three rounds of the purification protocol, with the same parameters as \textbf{(a)}}
\label{fig:onetworoundspure}
\end{figure}

Given the levels of entanglement present in the states above, we test the advantage of using a sequence of measurements vs. a single measurement on a noisy entangled state. \figref{fig:oneroundpure}) shows the result after a single X measurement on the states (choosing $y^*_1 = 1$) (\ref{rawstate2}, \ref{1roundpure}, \ref{tworoundpure}, \ref{threeroundpure}). Clearly, maximal randomness can be certified in the case where the measurement is projective, as expected. It can also be seen that by using the raw entangled state, \eqref{rawstate2}, very little randomness can be certified, with a maximum of approximately 0.15 bits.

\figref{fig:tworoundspure} illustrates the results after two rounds of measurements, where the second round measurements are projective, $\theta_2 = \phi_2 = 0$. The case of $\theta_1 = 0$ gives the same result as the single measurement scenario, since in this case the first measurement is projective and hence no randomness can be certified in the second round.

Unfortunately, it can be seen that no extra randomness can be certified in two measurement rounds on the raw entangled state, \eqref{rawstate1}. However, after two or more rounds of the purification protocol, indeed more randomness can be certified by using a sequence vs. a single measurement, as indicated by the peaks in \figref{fig:tworoundspure}. The infidelity for which the sequence becomes more useful than a single measurement can be seen to be approximately in the interval $\epsilon \in (0.06, 0.07)$.

\begin{figure}[H]
\centering
\begin{subfigure}{.5\textwidth}
  \centering
  \includegraphics[width=\linewidth]{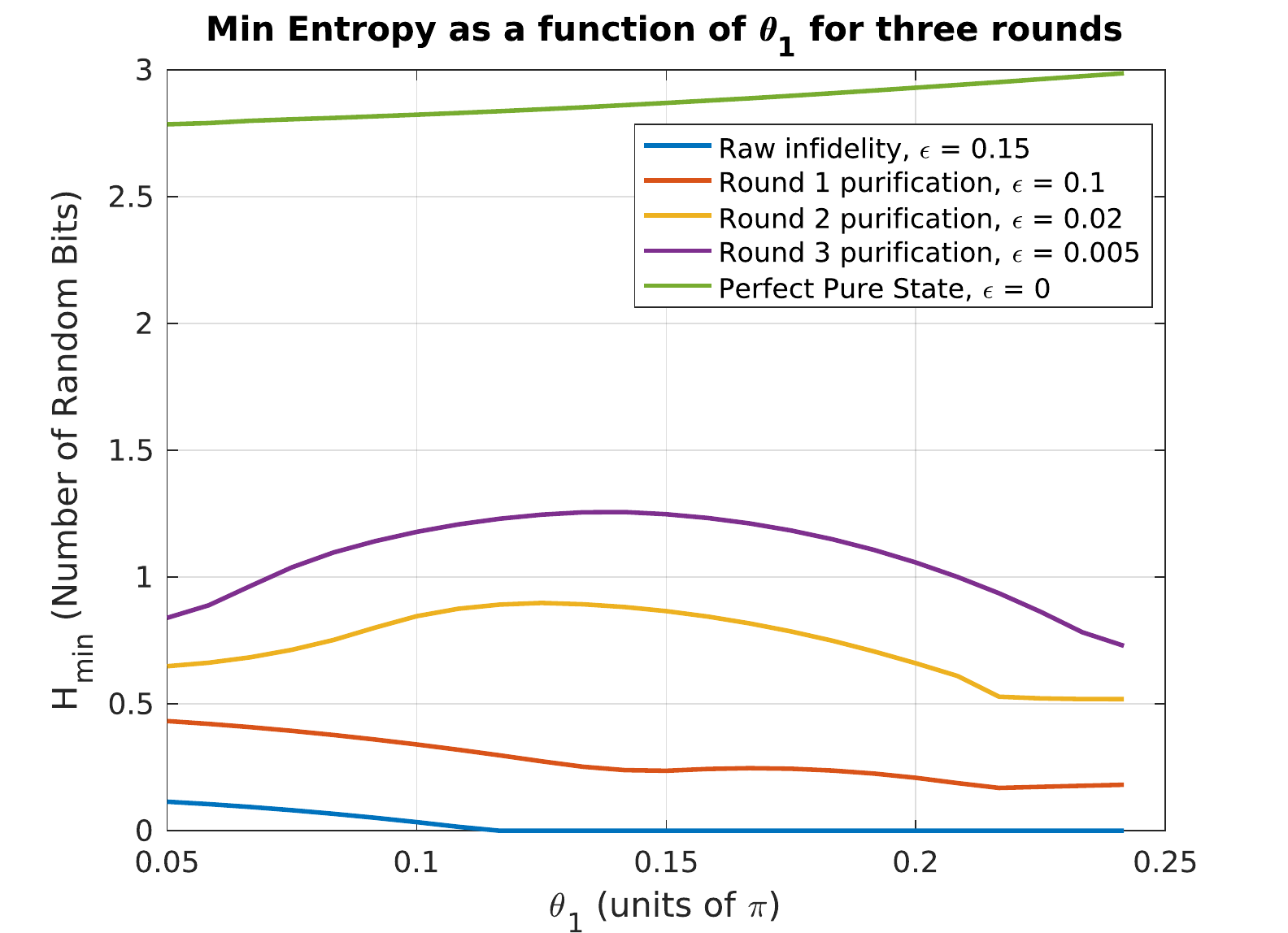}
   \caption{}
  \label{fig:threeroundspure1}
\end{subfigure}%
\begin{subfigure}{.5\textwidth}
  \centering
  \includegraphics[width=\linewidth]{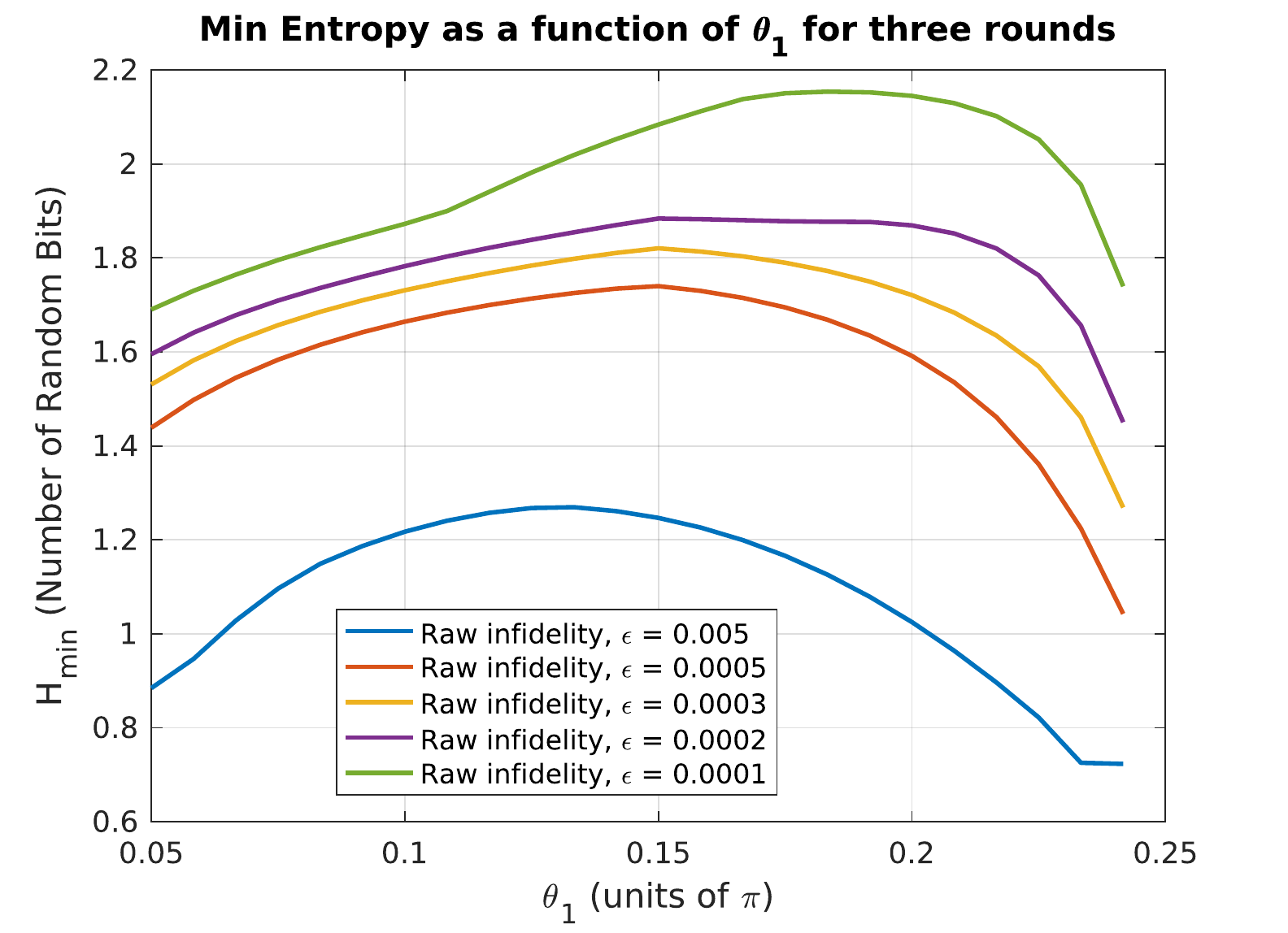}
   \caption{}
\label{fig:threeroundspure2}
\end{subfigure}
\caption{\textbf{(a)} Three rounds of measurement on the raw entangled state, (\ref{rawstate1}), ($\epsilon = 0.15$), the states produced after three rounds of the purification protocol, (\ref{1roundpure}, \ref{tworoundpure}, \ref{threeroundpure}), with $\epsilon = 0.1, 0.02, 0.005$ respectively and a perfect pure state with $\epsilon =0 $. \textbf{(b)} Three rounds of measurement on raw entangled states with infidelities $\epsilon = \{5\times 10^{-3}, 5\times 10^{-4}, 3\times 10^{-4}, 2\times 10^{-4}, 1\times 10^{-4}\}$}.
\label{fig:threeroundspure}
\end{figure}

Finally, \figref{fig:threeroundspure1} shows the results after three rounds of measurements, where the third, and final round of measurements are projective with $\theta_3 = \phi_3 = 0$. The second round of measurements is chosen in this case to be a noisy $Z$ measurement, with $\phi_2 = 0.08 \text{ rad}$.

Unfortunately, it can be seen that no extra randomness can be certified by implementing three measurements, than with two rounds. This is even the case for the purified states, (\ref{1roundpure}, \ref{tworoundpure}, \ref{threeroundpure}), so even these levels of purity are not sufficient to extract more randomness from a single state with three rounds of measurements. The perfect pure state, with $\epsilon = 0$ is also plotted for comparison.

Clearly, one would expect the existence of \textit{some} level at which the state becomes pure enough to be useful so \figref{fig:threeroundspure2}) shows the results of the protocol for very small infidelities, specifically:

\begin{align*}
\epsilon = \{5\times 10^{-3}, 5\times10^{-4}, 3\times 10^{-4}, 2\times 10^{-4}, 1\times 10^{-4}\}
\end{align*}

It can be seen that for an infidelity approximately in the interval,  $\epsilon \in (1\times 10^{-4}, 2\times 10^{-4})$, the state is pure enough to be able to certify more randomness with three rounds of measurement, than with two. This corresponds to being able to create pure entangled states experimentally with fidelities of greater than $99.98 \%$. This level could be reached by repeating the purification protocol more times but clearly this decreases the efficiency of the protocol as many more extra qubits would need to be introduced to implement this purification. Furthermore, for 4 and higher rounds of measurement, states which have an even higher level of purity would be required to make the protocol worthwhile, i.e. so that rounds of measurements on a single state would give better results than single measurements on new states each time.

\subsection{Atom-Photon Implementation}

We also examined a potential state arising from an atom-photon (AP) interaction. This case is even more applicable to the above 1sDI scenario as discussed in \secref{sec:intro}. In light of this, it makes sense to consider a situation where an entangled state is produced by some process between an atom, and a coherent photon state. As an example, we investigate the state produced in \cite{sangouard_loophole-free_2011}, which is the simplest for our purposes since it only involves single photon and vacuum states. However, an alternative method, using coherent photon states, such as the approach of \cite{teo_realistic_2013, teo_analysis_2013} could be studied. These scenarios are particularly relevant as the authors have the aim of performing a Bell test, and observing a violation of a Bell inequality.

It is possible to examine two possible cases in this scenario, since the setup is asymmetric. We can either consider noise introduced in either imperfections in the atom side, or on the photon side.

The ideal case considered in \cite{sangouard_loophole-free_2011} is given by (keeping our notation):

\begin{align}
    \ket{\Psi^{\zeta_1}} &= \cos(\zeta_1)\ket{0, g} + \sin(\zeta_1)\ket{1, s}
\end{align}

Where $\ket{g}, \ket{s}$ are two atomic states (held by Bob) and $\ket{0}, \ket{1}$ are the photon vacuum and single photon state respectively, held by Alice.

For simplicity, we will consider two of the cases presented in \cite{sangouard_loophole-free_2011} as sources of imperfections. The first error is introduced in the transmission efficiency, and we also consider the possibility that the photon was lost during the transmission. The transmission inefficiency is given by $\eta_t$, and if the photon is lost, we get an extra contribution to the overall state corresponding to $\ket{0, s}$, with a weight of $\sin^2(\zeta_1)(1-\eta_t)$, such that the final state is given by:

\begin{align}
    \rho_{\eta_t} = N\ket{\Psi^{\zeta_1}_{\eta_t}}\bra{\Psi^{\zeta_1}_{\eta_t}}  + \sin^2(\zeta_1)(1-\eta_t)\ket{s,0}\bra{s,0} \label{atomphotonstate}
\end{align}

where: 

\begin{align}
    \ket{\Psi^{\zeta_1}_{\eta_t}} &= \frac{1}{N}(\cos(\zeta_1)\ket{0, g} + \sin(\zeta_1)\sqrt{\eta_t}\ket{1, s}), \\
    N &= \cos^2(\zeta_1)+\sin^2(\zeta_1)\eta_t 
    \label{atomphotonpurestate}
\end{align}

Since both sets of atomic, $(A)$, and photonic, $(P)$, states are orthogonal to each other, we can make the translation to `logical' basis states: $\ket{g}\rightarrow \ket{0}_A, \ket{s}\rightarrow \ket{1}_A, \ket{0}\rightarrow \ket{0}_P, \ket{1}\rightarrow \ket{1}_P$ the state is given in the computational basis by:

\begin{align}
   \rho_{\eta_t} = \left(\begin{array}{cccc}
        c^2 & 0 & 0 &  \sqrt{\eta_t} cs  \\
         0 & s^2(1-\eta_t) & 0 & 0  \\
         0 & 0 & 0 & 0  \\
         \sqrt{\eta_t}cs & 0 & 0 & \eta_t s^2 
    \end{array}\right)\label{atomphotondensitymatrix}
\end{align}

\begin{figure}[H]
\centering
\begin{subfigure}{.34\textwidth}
  \centering
  \includegraphics[width=\linewidth, height=0.9\linewidth]{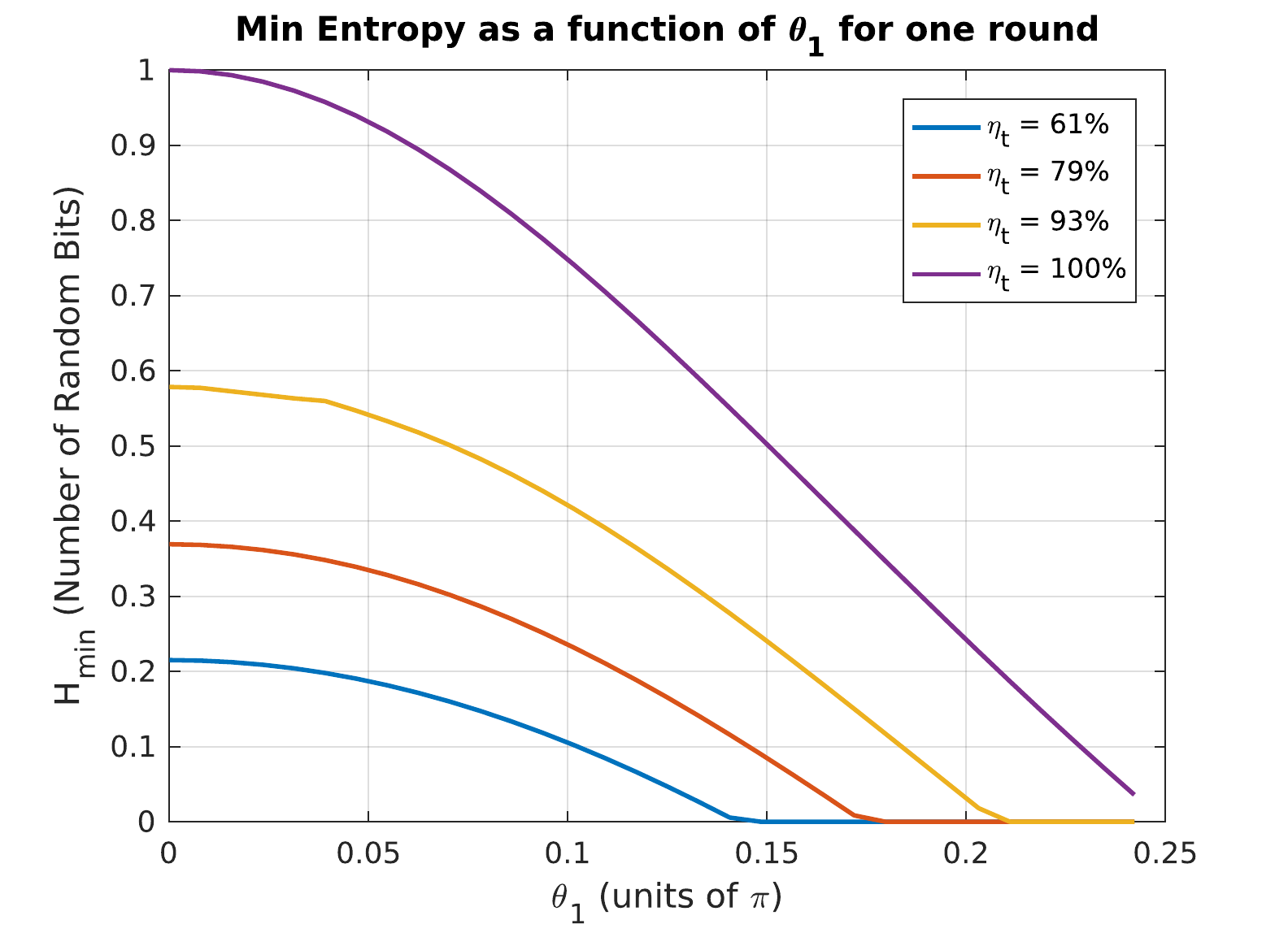}
   \caption{\ }
\label{fig:OneRoundAtomPhoton}
\end{subfigure}%
\begin{subfigure}{.34\textwidth}
  \centering
  \includegraphics[width=\linewidth, height=0.9\linewidth]{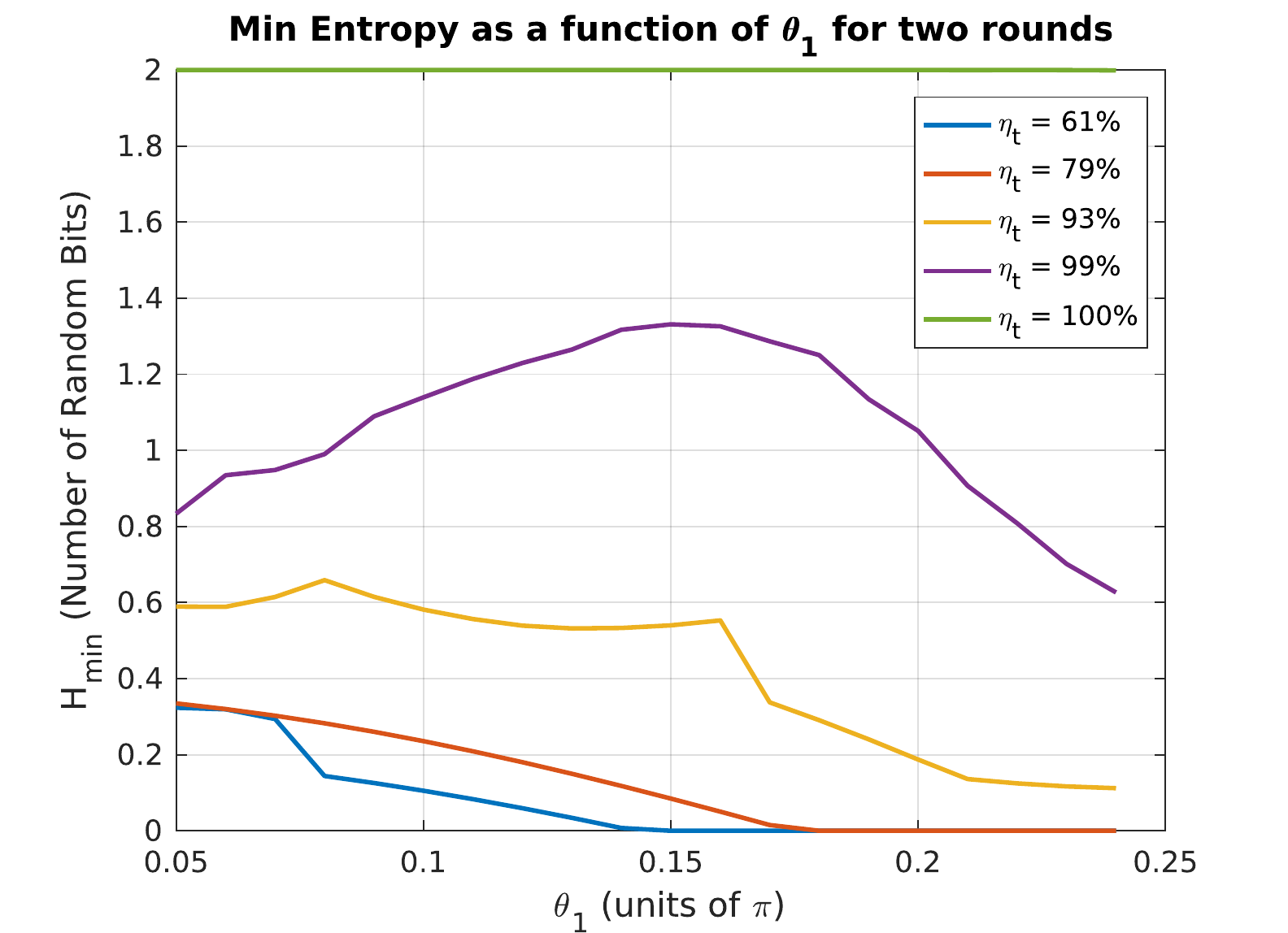}
   \caption{\ }
    \label{fig:TwoRoundsAtomPhoton}
\end{subfigure}%
\begin{subfigure}{.34\textwidth}
  \centering
  \includegraphics[width=\linewidth, height=0.9\linewidth]{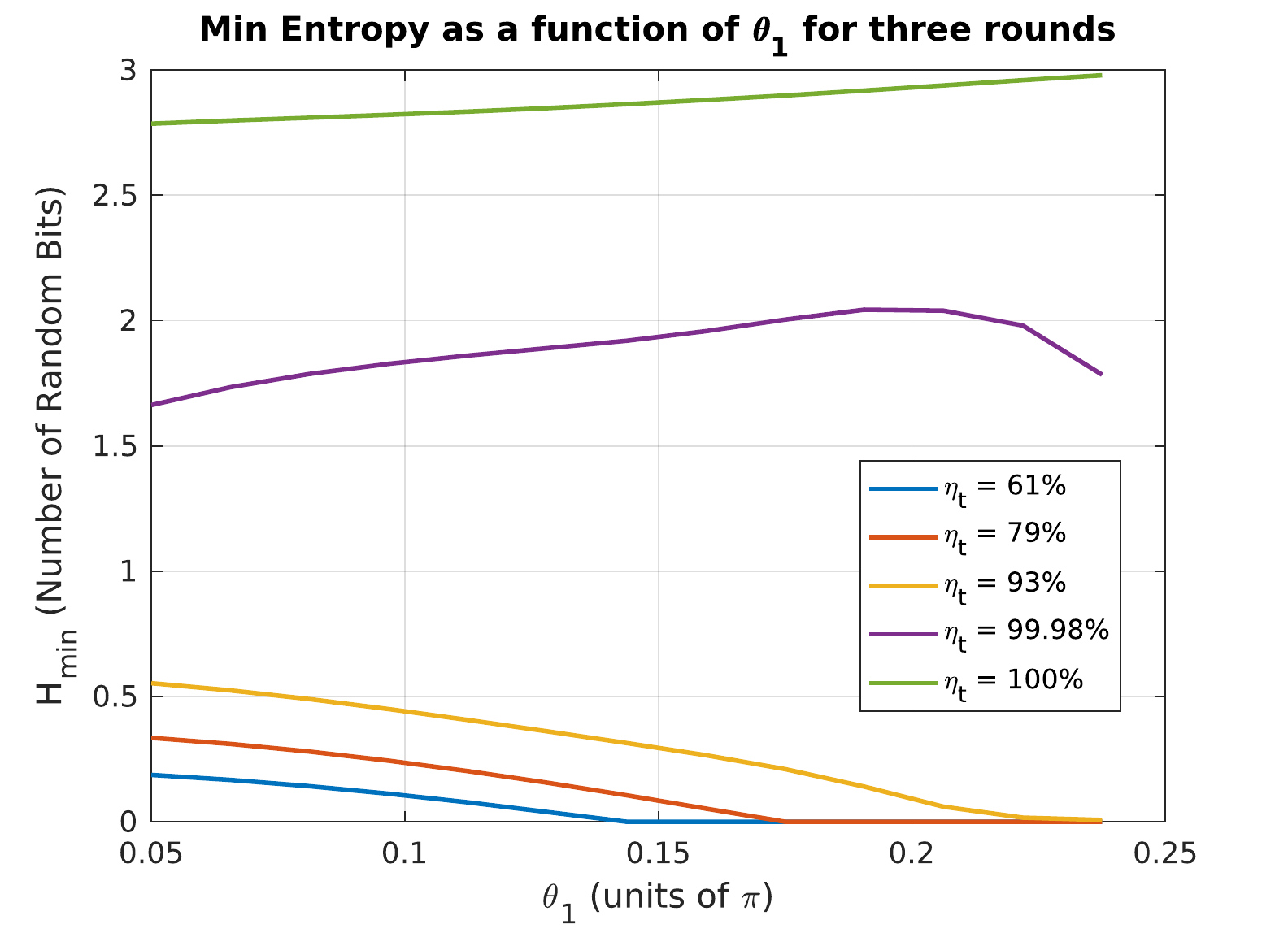}
   \caption{\ }
\label{fig:ThreeRoundsAtomPhoton}
\end{subfigure}
\caption{\textbf{(a)} $H_{min}$ for one round, for various levels of $\eta_t$. \textbf{(b)} $H_{min}$ for two rounds, for various levels of $\eta_t$. \textbf{(c)} $H_{min}$ for three rounds, for various levels of $\eta_t$. The second round measurement angle is $\phi_2 =0.08$ rad.}
\label{fig:AtomPhoton}
\end{figure}

Where we have defined $c = \cos(\zeta_1) ,\ s = \sin(\zeta_1) $
Figure~(\ref{fig:AtomPhoton}) illustrates the results after one, two and three measurement rounds for an atom-photon state (\ref{atomphotonstate}) with $\zeta_1 = \pi/4$. Values of the transmission efficiency, $\eta_t$ were chosen for interest to correspond with those described in \cite{teo_realistic_2013}. In that paper, the authors examine Bell inequality violations where Bob (Alice in our case) has access to the photonic system, and can make either homodyne measurements or photon counting to determine his Bell statistics. This intuitively corresponds in our case to his choice of measurement basis. Alternatively, Bob might not use photon counting, but instead choose between homodyne measurements in two different quadratures. Values of $\eta_t = \{61\%, 79\%\}$ are the required levels of efficiency to produce a Bell violation, if the measurements on the photonic side are either homodyne and photon counting, or both homodyne respectively. It should be noted that in our case, we would not need to distinguish between these two cases as we do not need to reproduce statistics for binary outcomes on Alice's side, since she is fully trusted, and needs only to do state tomography on the photonic mode. As such, the measurement scheme which allows her to more easily do tomography is the one that should be chosen in the actual implementation of our protocol. Also, a value of $\eta_t = 93\%$ was plotted as this is the level that would be required to close the locality loophole in the Bell violation, as stated in \cite{teo_realistic_2013}.

From Figure~(\ref{fig:OneRoundAtomPhoton}), it can be seen that for a value of $\eta_t = 61\%$, more randomness can be certified with a single measurement with the AP state, than with the one produced in two ion traps, with a fidelity of $\epsilon = 85\%$, as in the latter case, only about 0.15 random bits could be certified, but in the former over 0.2 random bits can be certified.

In this implementation, we see once again see the same general trends as with the ion trap apparatus. At some level of transmission efficiency, illustrated by $\eta_t = 99\%, \eta_t = 99.98\%$ in Figures~(\ref{fig:TwoRoundsAtomPhoton}), (\ref{fig:ThreeRoundsAtomPhoton}) respectively, the state becomes pure enough for a sequence to become worthwhile. In particular, for $\eta_t = 99\%$, two measurements on the state generates more certifiable randomness than is possible with one, and for $\eta_t = 99.98\%$, we can get more than 2 certifiable random bits. 

As a result, we can see that this particular atom photon model has more promise for randomness certification than the ion trap model, although it may be an unfair comparison to directly compare transmission efficiency in the former case to state fidelity in the latter. However, while the state in \cite{nigmatullin_minimally_2016} only takes into account a very simple depolarising noise model, which may be unrealistic in practice, the atom-photon state, (\ref{atomphotonstate}), of \cite{teo_realistic_2013} takes into account all coupling errors in the state preparation between Alice and Bob. Another interesting property to investigate would be the detection efficiency of the photons and how this effects the protocol.

\subsection{Nitrogen-Vacancy Center Implementation} \label{ssec:nv_center_implementation}
Next, we consider an entangled state produced between Alice and Bob using qubits based on electronic spins of nitrogen-vacancy defect centers in diamond. In particular, we examine the state used in the first loophole free Bell test, \cite{hensen_loophole-free_2015, pfaff_unconditional_2014}. This state is again relevant due to its use in the Bell test, and as mentioned in \cite{hensen_loophole-free_2015}, the setup could readily be used for randomness certification, albeit in a fully device independent scenario. The shared state between Alice and Bob in this experiment is given by the following density matrix:

\begin{align}
    \rho_{NV} = \frac{1}{2}\left(\begin{array}{cccc}
        1- F_z & 0 & 0 & 0  \\
        0 & F_z & -V & 0 \\
        0 & -V & F_z & 0 \\
        0 & 0 & 0 & 1-F_z \\
    \end{array}\right) \label{nvstate}
\end{align}

Where, $F_z = 1/2[(1-e_{\text{early}}^A)(1-e_{\text{late}}^B)+(1-e_{\text{early}}^B)(1-e_{\text{late}}^A)]$, and $V$ is the visibility which describes the indistinguishibility of the photons used to create entanglement. The residual errors, $e^{A/B}_{early/late}$, are due to the spin-photon coupling, as described in \cite{hensen_loophole-free_2015}. In this case, the ideal case is not particular Bell state we have assumed above, $\Phi^+$, instead it is another Bell state, $\Psi^- = \ket{\psi^-}\bra{\psi^-}, \ket{\psi^-} = \frac{1}{\sqrt{2}}(\ket{01}-\ket{10})$. The best estimate for the visibility is given to be $V = 0.873 \pm 0.060$, and the residual errors are found to be $e_{\text{early}}^A = 1.4 \pm 0.2\%, e_{\text{early}}^B = 1.6 \pm 0.2\%, e_{\text{late}}^A = 0.8 \pm 0.4\%, e_{\text{late}}^B = 0.7 \pm 0.4\%$. For these values, the fidelity of the state used in their Bell test is reported to be $\bra{\psi^-}\rho_{NV}\ket{\psi^-} = 0.92\pm 0.03$, and $F_z \approx 0.9775$. 

\begin{figure}[H]
\centering
\begin{subfigure}{.34\textwidth}
  \centering
  \includegraphics[width=\linewidth, height=0.9\linewidth]{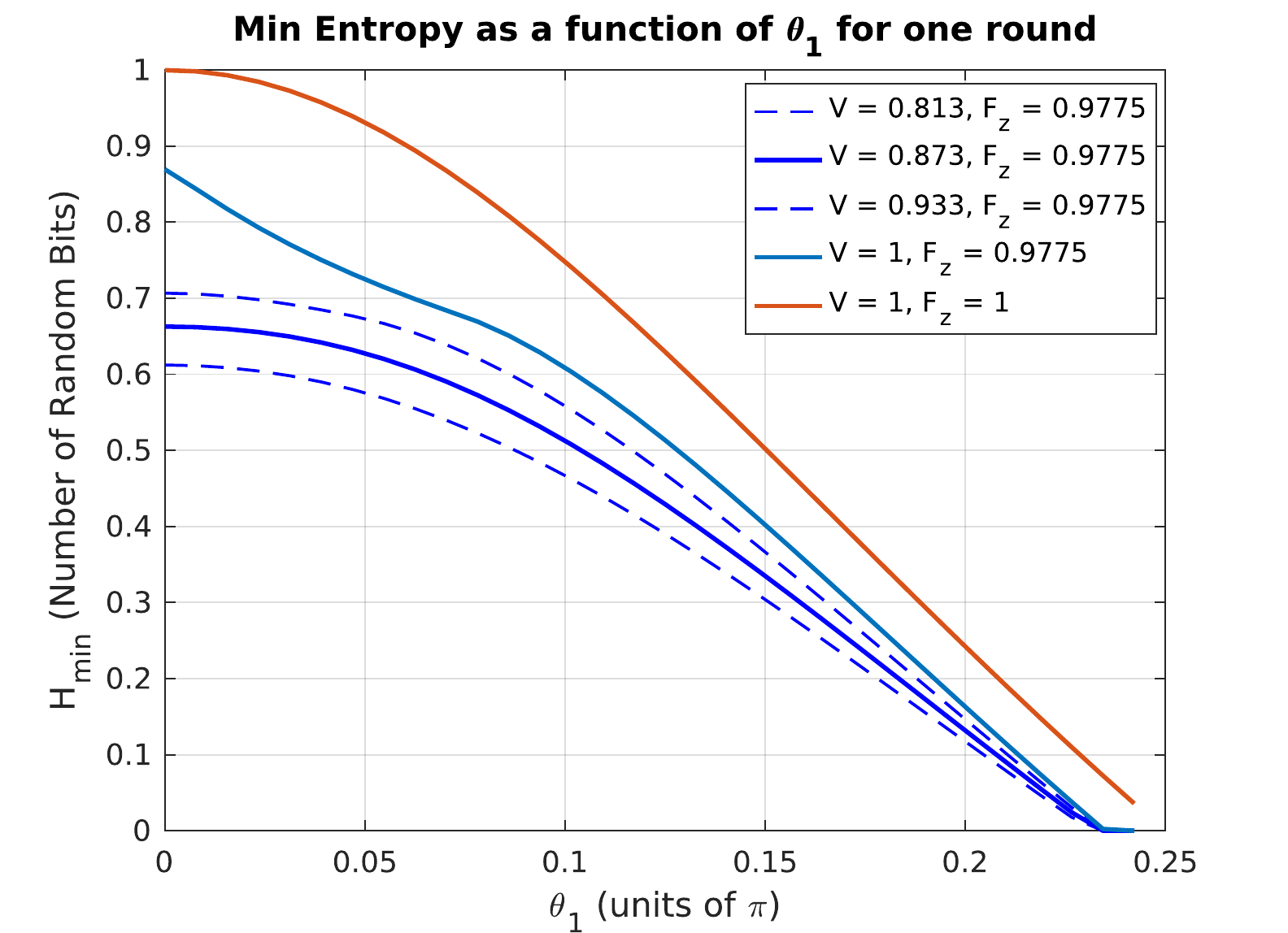}
   \caption{}
\label{fig:OneRoundNV}
\end{subfigure}%
\begin{subfigure}{.34\textwidth}
  \centering
  \includegraphics[width=\linewidth, height=0.9\linewidth]{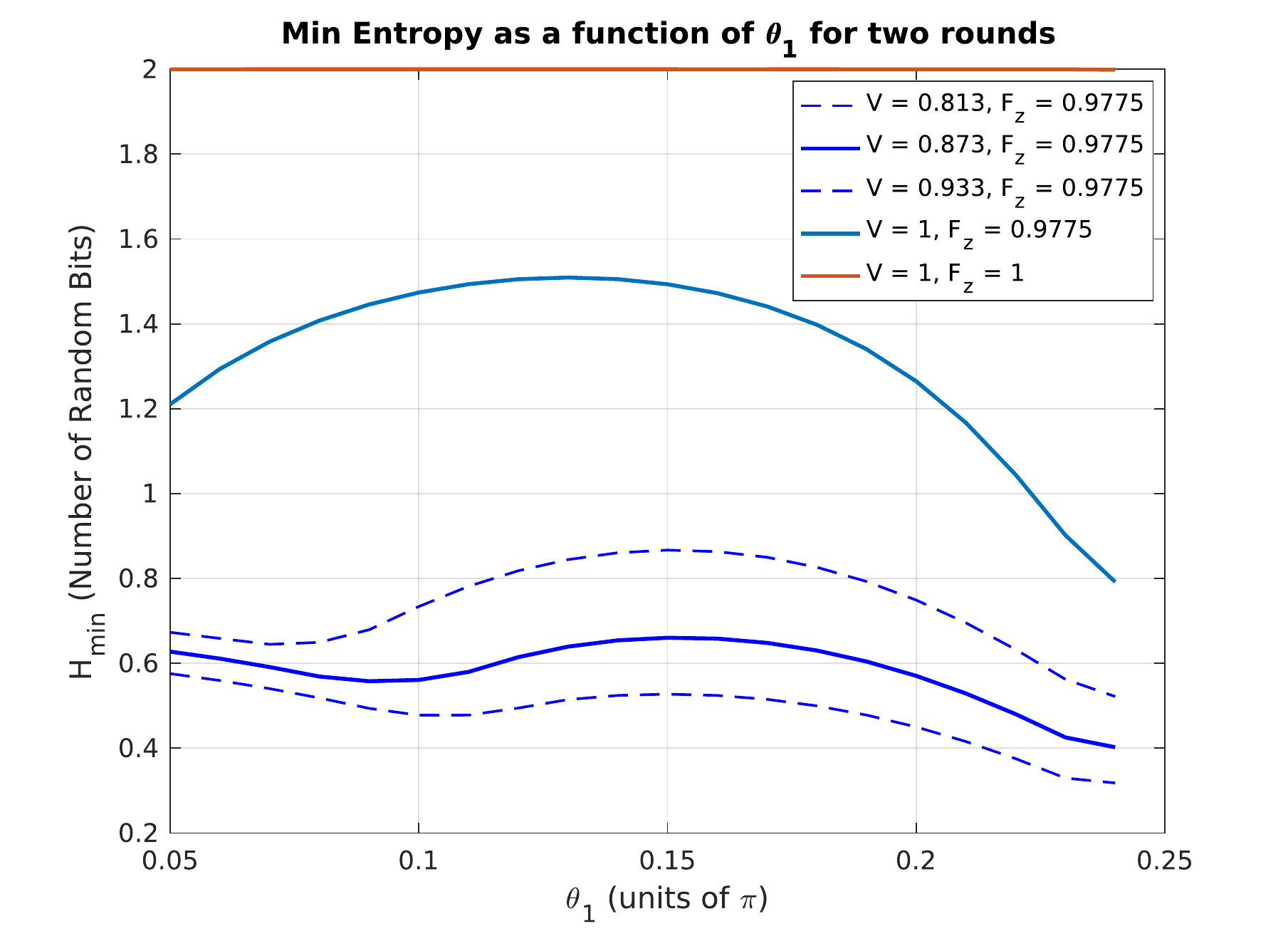}
   \caption{}
    \label{fig:TwoRoundsNV}
\end{subfigure}%
\begin{subfigure}{.34\textwidth}
  \centering
  \includegraphics[width=\linewidth, height=0.9\linewidth]{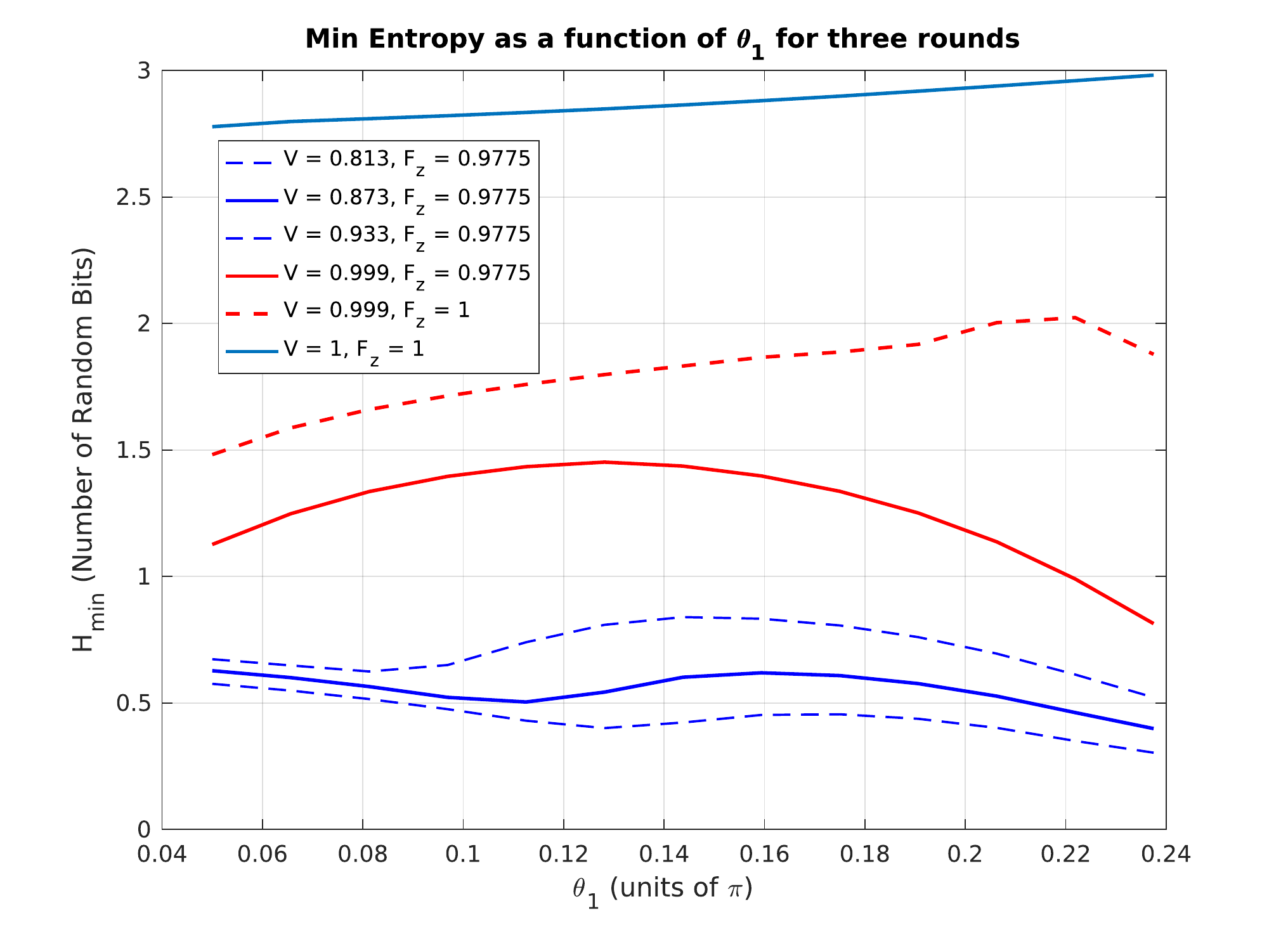}
   \caption{}
\label{fig:ThreeRoundsNV}
\end{subfigure}
\caption{$H_{min}$ using NV center state. The full blue line in each figure represents the best experimental estimate of \cite{hensen_loophole-free_2015} corresponding to $V = 0.873, F_z = 0.9775$. The dashed blue lines represent the effect of the error in the visibility, corresponding to states with $\Delta V = \pm 0.060$. Also, states with perfect visibility, $V=1$, but with residual errors, $F_z =0.9775$, along with a perfect pure Bell state, $\Psi^-$, is also plotted for comparison. Results for: \textbf{(a)} a single measurement, \textbf{(b)} two measurements and \textbf{(c)} for three measurement rounds, with a second round measurement angle of $\phi_2 = 0.08$ rad, all with various $\theta_1$ values.}
\label{fig:NV}
\end{figure}

Figure~(\ref{fig:NV}) shows the results of the protocol when the electronic spin state, (\ref{nvstate}), is used. In the experiment described in \cite{hensen_loophole-free_2015}, a very pure state was required to implement a reliable Bell test, and because of this, the state is substantially better for randomness certification than that available in the ion trap, or atom-photon implementation, with it being possible to certify $0.65 \pm 0.05$ random bits using electronic spins with a single measurement. Also, in both Figures~(\ref{fig:TwoRoundsNV}) and (\ref{fig:ThreeRoundsNV}), the effect of the residual errors can be seen to have a large consequence when it comes to randomness certification, and ultimately the state purity. For example, in (\ref{fig:TwoRoundsNV}) with a perfect visibility of $V=1$ and using a value of $F_z = 0.9775$ a maximum of 1.5 bits can be certified with two measurements, which is a substantially less than then maximal amount of 2 bits which can certified with a perfect pure state. A similar feature can be seen in (\ref{fig:ThreeRoundsNV}) for three measurement rounds. It would also be interesting to study the effect of the, $\Delta F_z$, derived from the statistical uncertainties on $e^{A/B}_{early/late}$, on the amount of randomness producible by the state. $\Delta F_z = 0$ was assumed in our numerical results for clarity.

The sensitivity of the randomness certification to errors is especially apparent in Figure~(\ref{fig:ThreeRoundsNV}). For a reduction in visibility, $V$, by only $0.1\%$, the amount of random bits drops by almost a full unit. Similarly, a reduction in $F_z$ by $0.023$ leads to a loss of $1/2$ a random bit, and even with this small drop, the situation changes from one in which a sequence of 3 measurements can do better than is \textit{ever} possible with two, to a scenario in which two measurement rounds produce a very similar amount of certifiable randomness, and the third measurement is almost unnecessary. 

\subsection{Implementation on Rigetti Forest Platform}
As a final example, we implement the protocol using Rigetti's Forest Platform, \cite{smith_practical_2016}. This is done in a proof of principle way using the following circuit:

\begin{align}
\Qcircuit @C=0.7em @R=0.6em {
\lstick{y_n}        &\cw    & \cw                                   & \cw                               & \cw               &\cw                                        & \cw       & \cw                                   & \cw               & \cctrlo{1} & \cctrl{1}\\
\lstick{\ket{0^n}}  &\qw    &\qw                                    &\qw                                & \qw               &  \qw                                      &  \qw      &  \qw                                  &  \qw              &\sgate{Z_{0}^{y_{n}\oplus 1}}{7}   &\sgate{X_{0}^{y_{n}}}{7}  &\measureD{Z^{(n)}}\\
\lstick{}           &\vdots &  \vdots                               & \vdots                            &  \vdots           &  \vdots                                   & \vdots    & \vdots                                & \vdots            &         &        & \\
\lstick{}           &       &                                       &                                   &                   &                                           &           &                                        &                   &          &             & \\
\lstick{y_{2}}      &       & \cw                                   &\cw                                & \cw               & \cctrlo{1}                                &   \cw        & \cctrl{1}                              &                   &         &           & \\ 
\lstick{\ket{0^2}}  &   \qw &\qw                                    &\qw                                & \qw               & \sgate{Z_{\phi_2}^{y_2\oplus 1}}{3}   &    \qw       &  \sgate{X_{\theta_2}^{y_2}}{3}      &\measureD{Z^{(2)}} &       &            & \\
\lstick{y_1}        &       &\cctrlo{1}                             & \cctrl{1}                         &                   &                                  &           &                              &                 &         &             & \\
\lstick{\ket{0^1}}  &       &\sgate{Z_{\phi_1}^{y_1\oplus 1}}{1} & \sgate{X_{\theta_1}^{y_1}}{1} &\measureD{Z^{(1)}} &                                  &           &          &                 &                       &            & \\
\lstick{\rho_B}     &       &\gate{Z}       & \gate{X}        & \qw               &  \gate{Z}        &\qw        &  \gate{X}           & \qw               & \gate{Z}         & \gate{X}           &\qw\\
    &       &       &         &            &           &       &         &                &          &           & \\
    &       &       &       & \ustick{\uparrow}  &      &  &  & \ustick{\uparrow}&&  &\ustick{\uparrow} & & \\
        &       &       &         &            &           &       &         &                &          &           & \\
    &       &       &         & \ustick{\text{Round 1 Measurement}}  &           &        &  &\ustick{\text{Round 2 Measurement}} & &  &\ustick{\text{Round n Measurement}}&&
} \label{circuitclassicalcontrol}
\end{align}

Where we have defined the following two qubit unitary gates, that effectively implement the non-projective measurements in the X and Z bases, denoted as $X^\theta$, (\ref{nonprojectivegates})  respectively.

\begin{small}
\begin{align}
\Qcircuit @C=0.7em @R=0.7em {
\lstick{\ket{0^i}} & \sgate{X_\theta}{1}& \qw &=& & \qw &\gate{R_{y}(2\theta)} & \gate{H} & \ctrl{1} & \gate{H}  & \qw &\qw\\
\lstick{\ket{\psi}}& \gate{X} & \qw& &&\qw & \qw & \qw& \gate{X} &\qw &\qw &\qw \gategroup{1}{7}{2}{10}{2em}{--} 
} 
\\
\Qcircuit @C=0.7em @R=0.7em {
\lstick{\ket{0^i}} & \sgate{Z_\phi}{1}& \qw &=& & \qw &\gate{R_{y}(2\phi)} & \gate{H} & \ctrl{1} & \gate{H}  & \qw &\qw\\
\lstick{\ket{\psi}}& \gate{Z} & \qw& &&\qw & \qw & \qw& \gate{Z} &\qw &\qw &\qw \gategroup{1}{7}{2}{10}{2em}{--} 
}\label{nonprojectivegates}
\end{align}
\end{small}

 The index on the ancilla represents the measurement round it is used in. The input string, $y$, for $n$ measurement rounds is used as classical input to the circuit, and conditioned on this input for each round, either the noisy X or noisy Z measurement is implemented. As mentioned above, it is the topmost ancilla that is used as a control qubit for each gate in the circuit. At the end of each round of the protocol, a single ancillas can be measured in the usual computational basis, where $Z^{(k)}$ represents the measurement done in round $k$. Clearly, if the input $y_k = 0$, the noisy Z measurement is implemented, $Z_{\phi_k}^{y_k\oplus 1}$, while if $y_k = 1$, the noisy X measurement is implemented, $X_{\theta_k}^{y_k}$, and the other is not. In this fashion, only one quantum gate acts on the state per measurement round. Also, the state $\ket{\psi}_B$ is only Bob's initial reduced state.

The circuit could be further improved since it is possible to only use a single ancilla. This ancilla would undergo multiple measurements, with the addition of a series of CNOT gates to the ancilla wire in order to reset the ancilla post measurement. These CNOT gates would return the ancilla to the usual $\ket{0}$ state conditioned on the previous measurement outcome. It is actually essential that the measurements occur in a sequential manner, i.e.\@ it is Bob's \textit{post-measurement} state which is rotated in the next round of the protocol. In this way, the measurements actually cannot be deferred to the end of the circuit since if this was done, there would be a cheating strategy for Eve. Causality is essential for the proper security of the protocol.

However, since the quantum hardware prohibits intermediate measurements in a quantum circuit, and instead it is necessary to defer all measurements to the end of the circuit. While this would not be sufficient for security against a malicious adversary, it is useful as a proof-of principle, assuming any deviation occurs from noise errors alone.

To implement the protocol, we proceed as described above and perform tomography on Alice's qubit, $\rho_A$ after the sequence of measurements on Bob's qubit, $\rho_B$ (deferred onto the ancillary qubits). We proceed using the simulator of the {\fontfamily{cmtt}\selectfont Aspen} quantum processing unit (QPU) with the sublattice {\fontfamily{cmtt}\selectfont Aspen-4-3Q-A}. For this scheme, we require a $2+n$ qubit chip to implement $n$ sequential measurements..

We perform Direct Inversion Tomography \cite{schmied_quantum_2016} by measuring the expectation values of the Pauli Observables, $X, Y, Z$ to reconstruct the state:

\begin{align}
\rho_A = \frac{1}{2}\left(\mathds{1} + r_xX  + r_yY + r_zZ\right)   \label{tomography} 
\end{align}

Direct inversion tomography is the simplest method of state tomography, and compensates for the fact that, due to measurement errors, the state which is estimated naively may lie outside the Bloch Sphere (i.e.\@ it has a norm greater than $1$). If this is the case, the vector, $\mathbf{r} = (r_x, r_y, r_z)$ are simply rescaled by its norm in the following way:

\begin{align}
    \hat{\mathbf{r}} = \begin{cases}
    \mathbf{r} \qquad &\text{if } ||\mathbf{r}||_2 \leq 1\\
    \mathbf{r}/ ||\mathbf{r}||_2 \qquad &\text{if } ||\mathbf{r}||_2 > 1\\
    \end{cases}
\end{align}

Where $||\mathbf{x}||_2 = \left(|x_1|^2 + |x_2|^2 + |x_3|^2 \right)^{1/2}$ is the $\ell_2$ norm. The original vector is estimated by approximating the expectation values, $(\tr(X\rho_A), \tr(Y\rho_A), \tr(Z\rho_A))$. This is achieved by counting the number of times the positive eigenvalue is observed, minus the number of times the negative eigenvalue is observed and normalising the answer, for each operator.

However, when implementing the protocol on the simulator, Alice can use her foreknowledge that Bob makes measurements only in the noisy $X/Z$ bases. In this case, the steered states, $\rho_A$, would have no $Y$ contribution so Alice would only be required to estimate  $(\tr(X\rho_A), \tr(Z\rho_A)$. However, if the protocol was to run on the physical hardware, it would be necessary to include measurements of the $Y$ observable also. To generate the full assemblage, this must be done for each of Bob's measurement choices and outcomes, $\sigma_{b|y}$.

The full protocol requires the assemblage after \textit{each} round, $\sigma_{b^i|y^i} \forall i \leq n$, but it is sufficient to compute these from the final round assemblages elements. This is due to the causality relationship $\sum_{b_i} \sigma_{b^i|y^i} = \sigma_{b^{i-1}|y^{i-1}}$.

\begin{figure}[H]
\centering
\begin{subfigure}{.34\textwidth}
  \centering
  \includegraphics[width=\linewidth, height=0.9\linewidth]{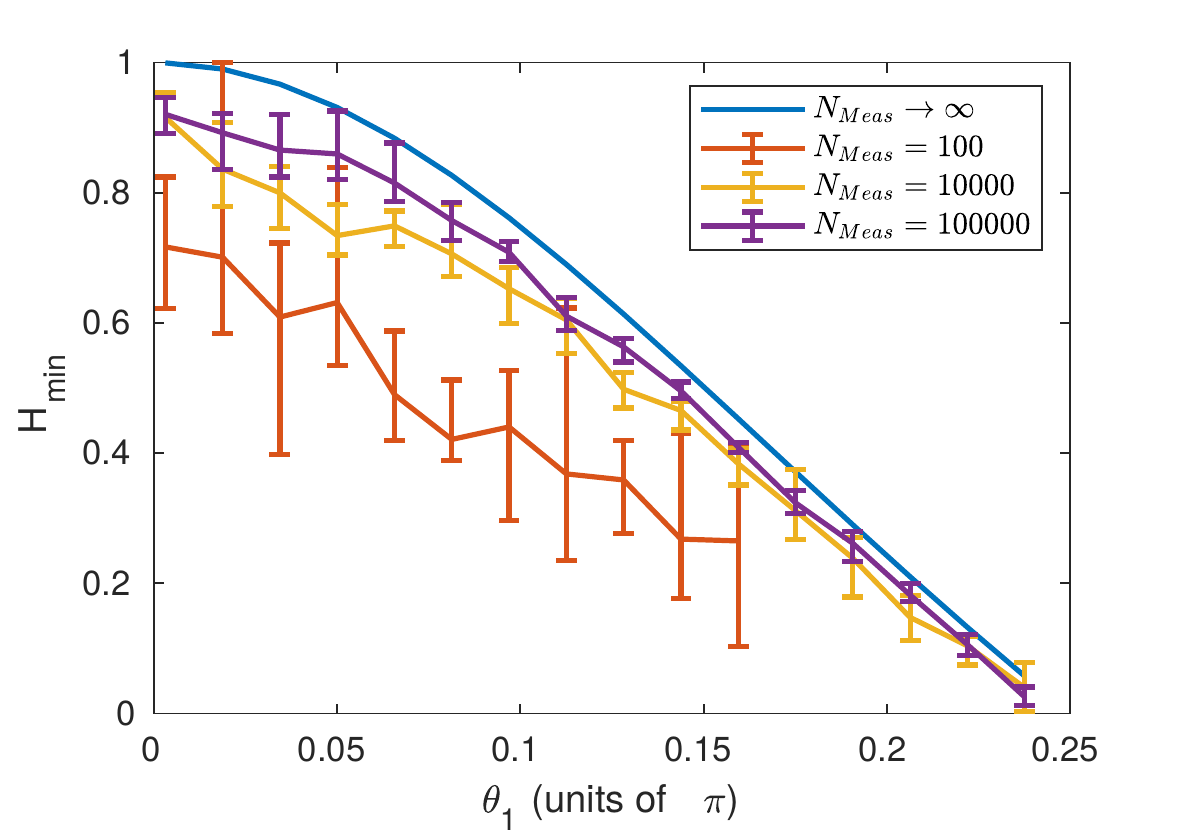}
   \caption{}
\label{subfig:OneRoundRigettiQVM}
\end{subfigure}%
\begin{subfigure}{.34\textwidth}
  \centering
  \includegraphics[width=\linewidth, height=0.9\linewidth]{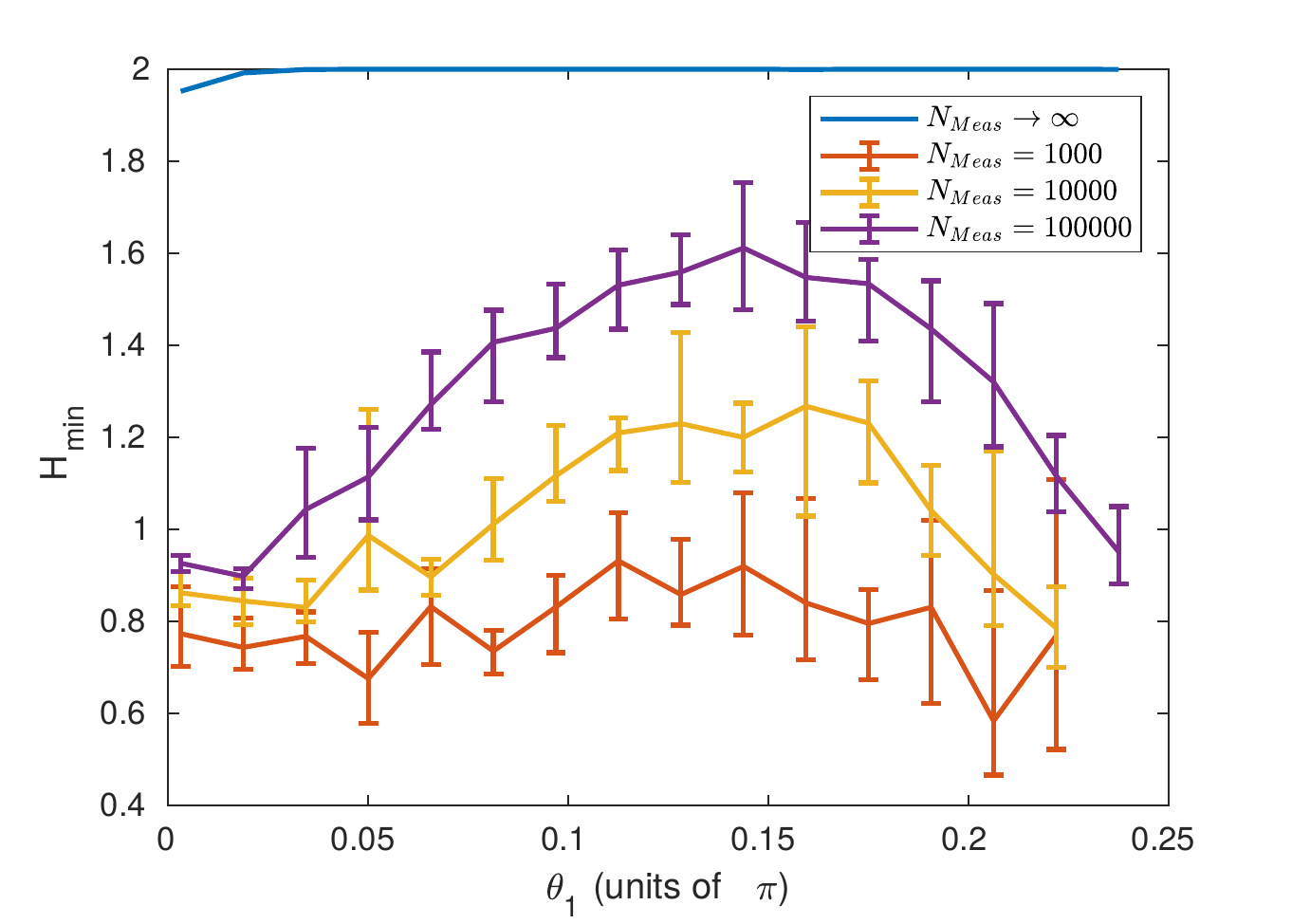}
   \caption{}
    \label{subfig:TwoRoundsRigettiQVM}
\end{subfigure}%
\begin{subfigure}{.34\textwidth}
  \centering
  \includegraphics[width=\linewidth, height=0.9\linewidth]{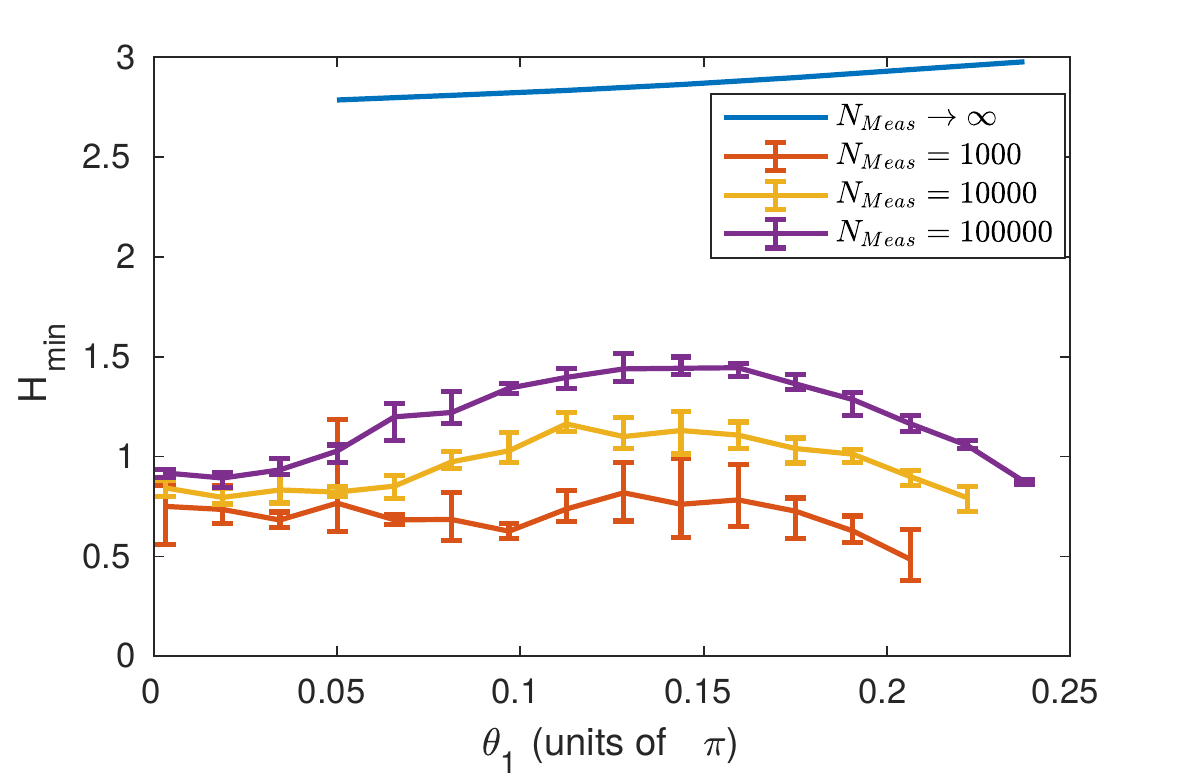}
   \caption{}
\label{subfig:ThreeRoundsRigettiQVM}
\end{subfigure}
\\[1.5ex]
\begin{subfigure}{.34\textwidth}
  \centering
  \includegraphics[width=0.8\linewidth, height=0.6\linewidth]{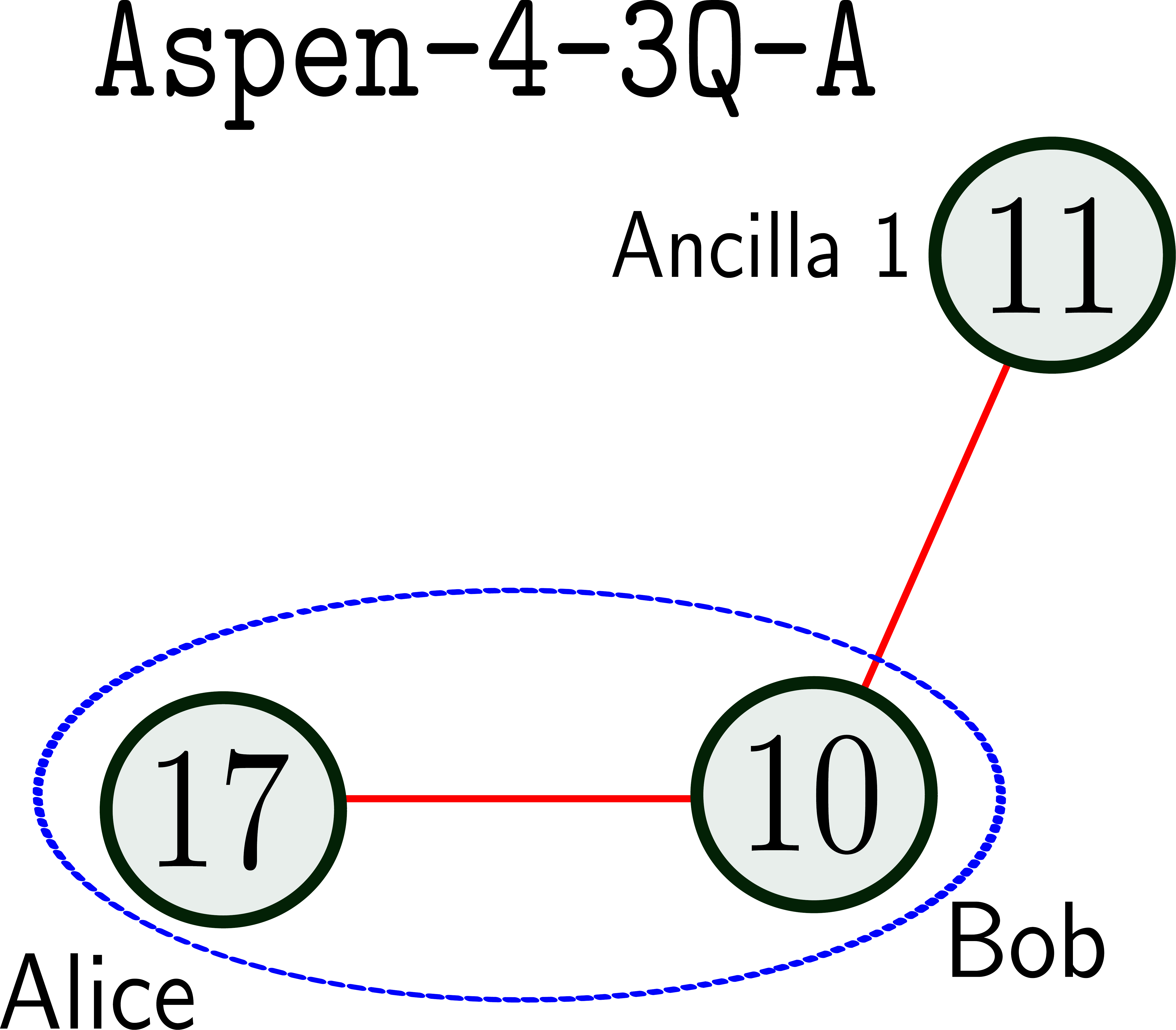}
   \caption{}
\label{subfig:3Q_qvm}
\end{subfigure}%
\begin{subfigure}{.34\textwidth}
  \centering
  \includegraphics[width=0.8\linewidth, height=0.8\linewidth]{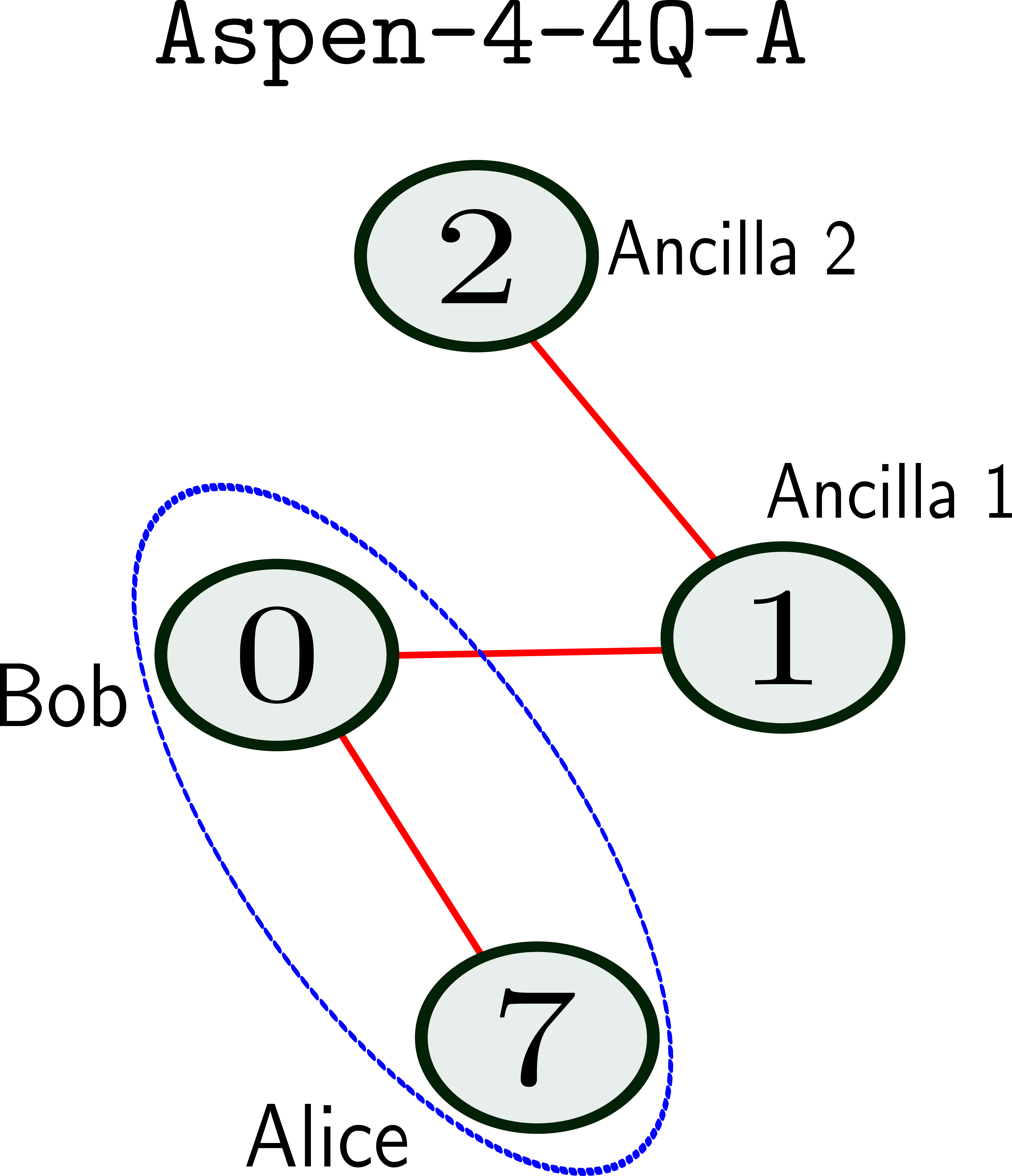}
   \caption{}
    \label{subfig:4Q_qvm}
\end{subfigure}%
\begin{subfigure}{.34\textwidth}
  \centering
  \includegraphics[width=0.8\linewidth, height=0.8\linewidth]{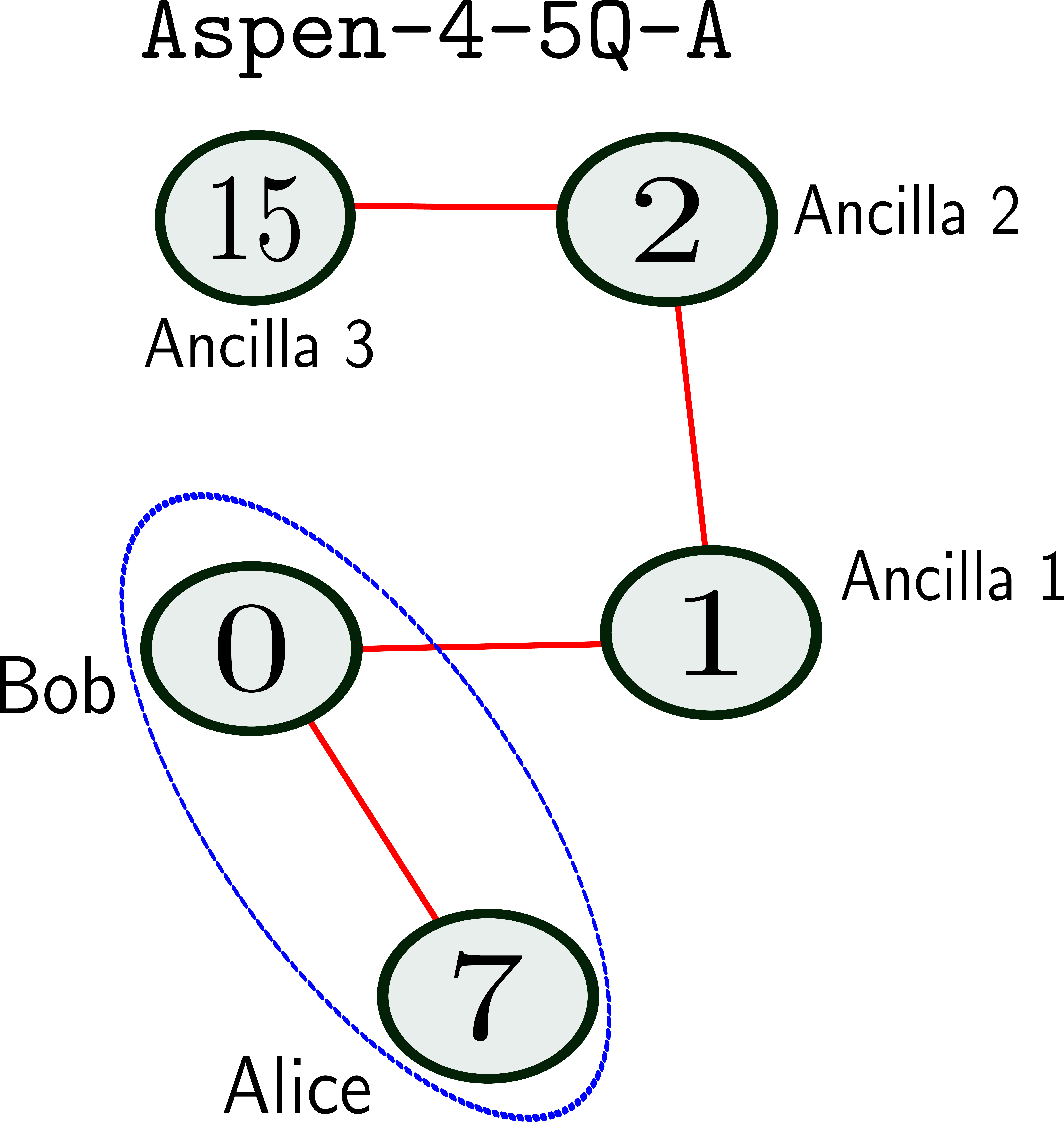}
   \caption{}
\label{subfig:5Q_qvm}
\end{subfigure}
\caption{$H_{min}$ for \textbf{(a)} one, \textbf{(b)} two and \textbf{(c)} three measurement rounds on Rigetti {\fontfamily{cmtt}\selectfont Aspen} Simulator using the pure entangled state \eqref{initialstate}, with $\zeta_1 = \pi/4$. $N_{meas}$ is the number of measurements taken when estimating each of the expectations values required to approximate \eqref{tomography}. Also plotted for comparison is the limit of infinite measurements, such that the ideal assemblage elements are obtained. In each case, the protocol is run five times, with the average $H_{min}$ plotted, and the error bars represent the maximum and minimum values obtained over the five runs. Figures \textbf{(d)}, \textbf{(e)}, \textbf{(f)} illustrate the chip simulator topology used for one, two and three measurements respectively. The qubits shared between Alice and Bob are indicated also.}
\label{fig:Rigetti}
\end{figure}

\figref{fig:Rigetti} illustrates the protocol using Rigetti's Simulator of sublattices of the QPU  containing three, four and five qubits to implement one, two and three measurement rounds in \figref{subfig:OneRoundRigettiQVM}, \figref{subfig:TwoRoundsRigettiQVM}, \figref{subfig:ThreeRoundsRigettiQVM} respectively. For a single measurement, the results are encouraging, but this is because 100,000 measurements allows a good characterisation of the 4 assemblage elements received by Alice. It is apparent that the exponential scaling quickly overtakes the number of measurements such that three measurement rounds does not increase the randomness certified over two, it actually reduces it. Unfortunately, we were not able to get sensible results when running the protocol on the QPU versions of the corresponding simulators in \figref{fig:Rigetti}, even for a single measurement, due to noise. Potentially, this could be mitigated by using more sophisticated tomography techniques, or some error mitigation scheme.  

\section{Discussion}\label{sec:discussion}
 We presented a novel scheme to certify an unbounded amount of random numbers from sequential measurements on one half of a quantum state shared by two parties, building on the work of \cite{curchod_unbounded_2017, passaro_optimal_2015, skrzypczyk_paulskrzypczyk/steeringreview_2018}. The `certificate' in this case can be a set of statistical criteria, or the states into which the other party is `steered' as a result of the sequence of well-chosen measurements. We studied the behaviour of the scheme both in the ideal setting, and in experimentally realistic settings, \cite{nigmatullin_minimally_2016}, including those which has actually been implemented \cite{hensen_loophole-free_2015, sangouard_loophole-free_2011}. We also demonstrated the feasibility of the scheme in being able to certify multiple random bits produced from a single quantum state, rather than multiple states each producing only one bit. This distinction is important given the valuable nature of controlled quantum systems, and hence represents an important step in resource reduction.
 
 Our scheme could be readily turned into a protocol for randomness expansion, especially now that we have improved upon the work of \cite{curchod_unbounded_2017} in reducing the number of measurements required. We leave this to future work.
 
 Interesting future work would be to investigate the reason behind the apparent anomaly in the steering scenario with two sequential measurements on a maximally entangled state, as discussed in \secref{sec:numericalresults}. Given our focus in this work on studying the behaviour of the protocol in realistic experimental implementations, it would be insightful to actually implement the protocol in a physical system, similar to those carried out in Bell testing, \cite{hensen_loophole-free_2015}.

\section{Materials and Methods}

All numerical results in this work were obtained using the Matlab convex optimisation package, \textit{cvx}, \cite{grant_graph_2008} and a package for managing quantum states, \textit{qetlab}, \cite{johnston_qetlab:_2016}. The resulting code required to produce all images in this work is available at \cite{briancoyle_briancoyle/tpmscproject2017_2017}.

\authorcontributions{B.C. and M.H. devised the scheme. B.C. wrote the code and produced the numerical results. M.H. derived the theorems. E.K. supervised the work.}

\funding{This work was supported by the Engineering and Physical Sciences Research Council (grant EP/L01503X/1), EPSRC Centre for Doctoral Training in Pervasive Parallelism at the University of Edinburgh, School of Informatics and Entrapping Machines, (grant FA9550-17-1-0055). }

\acknowledgments{B.C. thanks Atul Mantri and Niraj Kumar for useful discussions. We also thank Rigetti Computing for the use of their quantum compute resources, and views expressed in this paper are those of the authors and do not necessarily reflect the views or policies of Rigetti Computing.}

\conflictsofinterest{The authors declare no conflict of interest. The funders had no role in the design of the study; in the collection, analyses, or interpretation of data; in the writing of the manuscript, or in the decision to publish the results.}

\appendix

\section{Proof of Theorem 1}\label{app:min-entropy-proofs}


We will at first consider the case of a single round of the scheme outlined in this paper, that is Bob's first measurement is his only one. Recall we have that Eve has created a quantum state $\ket{\psi_{ABE}}$ with sub-systems denoted by $A$, $B$ and $E$, which are Alice, Bob and Eve's respective subsystems. 

We want to bound Eve's guessing probability of the outcome $b$ of Bob's measurement. In order to do this we will constrain what form $\ket{\psi_{ABE}}$ takes, and if the dichotomic observable corresponding to Bob's measurement is denoted $X_{B}$, we will also constrain the form of $\mathbb{I}_{A}\otimes X_{B}\otimes\mathbb{I}_{E}\ket{\psi_{ABE}}$. In particular, we will use the techniques of self-testing in the one-sided device-independent setting \cite{supic} to show that there exists a local isometry acting on Bob's systems that map the state $\ket{\psi_{ABE}}$ to the state $\ket{{\zeta}}_{AB'}\ket{\textrm{anc}}_{E'}$, where $B'$ denotes a virtual qubit system held by Bob respectively such that $\ket{{\zeta}}_{AB'}=\cos(\zeta)\ket{00}+\sin(\zeta)\ket{11}$, and $E'$ is an arbitrary virtual system held by Eve. We also show something analogous for $\mathbb{I}_{A}\otimes X_{B}\otimes\mathbb{I}_{E}\ket{\psi_{ABE}}$. 

To establish rigorously our self-testing results, we first need to establish the statistical conditions that need to be satisfied. Bob's aforementioned observable $X_{B}$ has two eigenvalues taking values $\pm 1$ (corresponding to the binary outcomes $b$). Without loss of generality, Bob's observable can be taken to be sharp with eigenvalues $+1$ and $-1$ being associated with projectors $\Sigma_{B}^{0}$ and $\Sigma_{B}^{1}$ respectively such that $\Sigma_{B}^{0}+\Sigma_{B}^{1}=\mathbb{I}_{B}$ and $\Sigma_{B}^{0}-\Sigma_{B}^{1}=X_{B}$. In addition to $X_{B}$, Bob has a second sharp observable $Z_{B}$ with eigenvalues $+1$ and $-1$ being associated with projectors $\Pi_{B}^{0}$ and $\Pi_{B}^{1}$ respectively such that $\Pi_{B}^{0}+\Pi_{B}^{1}=\mathbb{I}_{B}$ and $\Pi_{B}^{0}-\Pi_{B}^{1}=Z_{B}$. In addition to Bob's measurement, we denote Alice's Pauli X and Z matrices as $\tau_{X}$ and $\tau_{Z}$ respectively. Now we can state the statistical criteria that Alice and Bob need to satisfy:
\begin{eqnarray}\label{crit}
\left\vert\langle \tau^{A}_{Z}\otimes Z_{B}\rangle-1\right\vert&\leq&\epsilon_{1}\nonumber\\
\left\vert\langle \tau^{A}_{X}\otimes X_{B}\rangle-\sin(2\zeta)\right\vert&\leq&\epsilon_{2}\nonumber\\
\left\vert\langle \tau^{A}_{Z}\rangle-\cos(2\zeta)\right\vert&\leq&\epsilon_{1}
\end{eqnarray} 
with $\epsilon_{1}$ and $\epsilon_{2}$ being small error terms, i.e. a positive real number. We chose the errors to have this symmetry motivated by our original scheme, as will hopefully be clear.

We can now state the self-testing result we will need:

\begin{theorem_fix}\label{selftest}
If Alice and Bob's statistics satisfy the criteria in \eqref{crit} and $\zeta\in]0,\frac{\pi}{4}]$, then there exist a quantum state $\ket{\textrm{anc}}\in\mathcal{H}_{E'}$ in Hilbert space $\mathcal{H}_{E'}$ and local isometry $\Phi_{B}$ such that
\begin{eqnarray*}
\left\Vert \mathbb{I}_{A}\otimes\Phi_{B}\otimes\mathbb{I}_{E}(\ket{\psi_{ABE}})-\ket{{\zeta}}_{AB'}\ket{\textrm{anc}}_{E'} \right\Vert &\leq& \sqrt{\epsilon_{1}}(\sqrt{2}+1)+\sqrt{\epsilon_{1}+\epsilon_{2}}\\
\left\Vert \mathbb{I}_{A}\otimes\Phi_{B}\otimes\mathbb{I}_{E}(\mathbb{I}_{A}\otimes X_{B}\otimes\mathbb{I}_{E}\ket{\psi_{ABE}})-\mathbb{I}_{A}\otimes\tau^{B'}_{X}\ket{{\zeta}}_{AB'}\ket{\textrm{anc}}_{E'}\right\Vert&\leq& \sqrt{\epsilon_{1}}\left(2\sqrt{2}+1+\frac{5}{2\sin(\zeta)}\right)\\
&+&\sqrt{\epsilon_{1}+\epsilon_{2}}\left(\frac{3}{\sqrt{2}\sin(\zeta)}+2\right).
\end{eqnarray*}
\end{theorem_fix}

To prove this result we need to state a few lemmas. Before doing this, we will simplify notation to have to have $\ket{\psi}:=\ket{\psi_{ABE}}$ and we will suspend denoting tensor products and identities when it is clear from context. The lemmas we need now follow.
\begin{lemma_fix}\label{lemm0}
If \eqref{crit} is satisfied and $\zeta\in]0,\frac{\pi}{4}]$ the following is true:
\begin{equation*}
\left\Vert \left(\ket{x}\langle x\vert_{A}-\Pi^{x}_{B}\right)\ket{\psi}\right\Vert  \leq \sqrt{\epsilon_{1}/2},
\end{equation*}
for $x\in\{0,1\}$ and $\{\ket{x}\}_{x}$ being the computational basis, i.e.\@ the eigenstates of $\tau_{Z}$.
\end{lemma_fix}

\begin{proof}
We will proof the case where $x=0$, but the proof for $x=1$ is essentially the same. Note that by definition:
\begin{eqnarray*}
\Vert \left(\ket{0}\langle 0\vert_{A}-\Pi^{0}_{B}\right)\ket{\psi}\Vert&=&\sqrt{|\langle\psi \vert 0\rangle_{A}|^{2} + \langle \Pi^{0}_{B}\rangle -2\langle\psi\ket{0}\langle 0\vert_{A}\Pi^{0}_{B}\ket{\psi}}\\
&=&\sqrt{\frac{1}{2}-\frac{1}{2}\left(\bra{\psi}(2\ket{0}\bra{0}_{A}-\mathbb{I}_{A})\otimes(2\Pi^{0}_{B}-\mathbb{I}_{B})\ket{\psi}\right)}\\
&=&\sqrt{\frac{1}{2}-\frac{1}{2}\langle \tau^{A}_{Z}Z_{B}\rangle}\\
&\leq&\sqrt{\frac{\epsilon_{1}}{2}},
\end{eqnarray*}
where the first line of \eqref{crit} was used in the final inequality.
\end{proof}

\begin{lemma_fix}
If \eqref{crit} is satisfied and $\zeta\in]0,\frac{\pi}{4}]$ the following is true:
\begin{eqnarray}
\Vert \left(\sin(\zeta)\ket{1}\bra{0}_{A}-\cos(\zeta)\ket{1}\bra{1}_{A}X_{B}\right)\ket{\phi}\Vert&\leq&\sqrt{(\epsilon_{1}+\epsilon_{2}) / 2} \nonumber\\
\Vert \left(\cos(\zeta)\ket{0}\bra{1}_{A}-\sin(\zeta)\ket{0}\bra{0}_{A}X_{B}\right)\ket{\phi}\Vert&\leq&\sqrt{(\epsilon_{1}+\epsilon_{2}) / 2}\nonumber.
\end{eqnarray}
\end{lemma_fix}
\begin{proof}
We will prove the first case as the second case has essentially the same proof. By definition we have
\begin{align*}
&\Vert \left(\sin(\zeta)\ket{1}\bra{0}_{A}-\cos(\zeta)\ket{1}\bra{1}_{A}X_{B}\right)\ket{\phi}\Vert=\sqrt{\sin^{2}(\zeta)\vert\bra{0}\psi\rangle_{A}\vert^{2}+\cos^{2}(\zeta)\vert\bra{1}\psi\rangle_{A}\vert^{2}-\sin(\zeta)\cos(\zeta)\bra{\psi}\tau^{A}_{X}X_{B}\ket{\psi}}\\
&\leq\sqrt{\sin^{2}(\zeta)(\cos^{2}(\zeta)+\frac{\epsilon_{1}}{2})+\cos^{2}(\zeta)(\sin^{2}(\zeta)+\frac{\epsilon_{1}}{2})-\sin(\zeta)\cos(\zeta)(\sin(2\zeta)-\epsilon_{2})}\\
&\leq\sqrt{\frac{\epsilon_{1}+\sin(2\zeta)\epsilon_{2}}{2}}\leq\sqrt{\frac{\epsilon_{1}+\epsilon_{2}}{2}},
\end{align*}
where in the first inequality the second and third line of \eqref{crit} were utilised. In particular, we utilised the fact that
\begin{equation}
\langle\psi\vert Z_{A}\vert\psi\rangle=2\vert\bra{0}\psi\rangle_{A}\vert^{2}-1=1-2\vert\bra{1}\psi\rangle_{A}\vert^{2},
\end{equation}
which concludes the proof.
\end{proof} 

\begin{lemma_fix}\label{lemm1}
If \eqref{crit} is satisfied and $\zeta\in]0,\frac{\pi}{4}]$ the following is true:
\begin{eqnarray}
\left\Vert \left(\sin(\zeta)\ket{1}\bra{0}_{A}-\cos(\zeta)X_{B}\Pi^{1}_{B}\right)\ket{\phi}\right\Vert&\leq& \sqrt{\frac{\epsilon}{2}}+\cos(\zeta)\sqrt{\frac{\epsilon_{1}}{2}}\nonumber\\
\left\Vert \left(\cos(\zeta)\ket{0}\bra{1}_{A}-\sin(\zeta)X_{B}\Pi^{0}_{B}\right)\ket{\phi}\right\Vert&\leq& \sqrt{\frac{\epsilon}{2}}+\sin(\zeta)\sqrt{\frac{\epsilon_{1}}{2}}\nonumber,
\end{eqnarray}
for $\epsilon:=\epsilon_{1}+\epsilon_{2}$, and thus
\begin{eqnarray}
\left\Vert \left(X_{B}-\frac{\sin(\zeta)}{\cos(\zeta)}\ket{1}\bra{0}_{A}-\frac{\cos(\zeta)}{\sin(\zeta)}\ket{0}\bra{1}_{A}\right)\ket{\phi}\right\Vert&\leq&\sqrt{2\epsilon_{1}}+\sqrt{\frac{\epsilon}{2}}\left(\frac{1}{\sin(\zeta)}+\frac{1}{\cos(\zeta)}\right)\nonumber.
\end{eqnarray}
\end{lemma_fix}

\begin{proof}We address the first line in the lemma as the proof of the second line is essentially the same:
\begin{align*}
&\left\Vert \left(\sin(\zeta)\ket{1}\bra{0}_{A}-\cos(\zeta)X_{B}\Pi^{1}_{B}\right)\ket{\phi}\right\Vert\\
= &\left\Vert \left(\sin(\zeta)\ket{1}\bra{0}_{A}-\cos(\zeta)\ket{1}\bra{1}_{A}X_{B}+\cos(\zeta)\ket{1}\bra{1}_{A}X_{B}-\cos(\zeta)X_{B}\Pi^{1}_{B}\right)\ket{\phi}\right\Vert\\
\leq&\sqrt{\frac{\epsilon_{1}+\epsilon_{2}}{2}}+\cos(\zeta)\sqrt{\frac{\epsilon_{1}}{2}},
\end{align*}

where the inequality uses the triangle inequality and the results from the previous two lemmata. For the second part of the proof, we have 
\begin{align*}
&\left\Vert \left(X_{B}-\frac{\sin(\zeta)}{\cos(\zeta)}\ket{1}\bra{0}_{A}-\frac{\cos(\zeta)}{\sin(\zeta)}\ket{0}\bra{1}_{A}\right)\ket{\phi}\right\Vert \\
&\left\Vert \left(X_{B}(\Pi^{0}_{B}+\Pi^{1}_{B})-\frac{\sin(\zeta)}{\cos(\zeta)}\ket{1}\bra{0}_{A}-\frac{\cos(\zeta)}{\sin(\zeta)}\ket{0}\bra{1}_{A}\right)\ket{\phi}\right\Vert\\
\leq &\sqrt{2\epsilon_{1}}+\sqrt{\frac{\epsilon}{2}}\left(\frac{1}{\sin(\zeta)}+\frac{1}{\cos(\zeta)}\right),
\end{align*}

thus completing the proof.
\end{proof}

\begin{lemma_fix}\label{lemm2}
If \eqref{crit} and $\zeta\in]0,\frac{\pi}{4}]$ is satisfied the following is true:
\begin{eqnarray*}
\left\Vert \left(\frac{\sin(\zeta)}{\cos(\zeta)}\ket{1}\bra{0}_{A}-\Pi^{0}_{B}X_{B}\right)\ket{\phi}\right\Vert&\leq& \sqrt{2\epsilon_{1}}\left(1+\frac{1}{\sin(2\zeta)}\right)+\sqrt{\frac{\epsilon}{2}}\left(\frac{1}{\sin(\zeta)}+\frac{1}{\cos(\zeta)}\right)\\
\left\Vert \left(\frac{\cos(\zeta)}{\sin(\zeta)}\ket{0}\bra{1}_{A}-\Pi^{1}_{B}X_{B}\right)\ket{\phi}\right\Vert&\leq& \sqrt{2\epsilon_{1}}\left(1+\frac{1}{\sin(2\zeta)}\right)+\sqrt{\frac{\epsilon}{2}}\left(\frac{1}{\sin(\zeta)}+\frac{1}{\cos(\zeta)}\right),
\end{eqnarray*}

for $\epsilon:=\epsilon_{1}+\epsilon_{2}$, and thus
\begin{eqnarray*}
\left\Vert \left(\ket{0}\bra{0}_{A}-X_{B}\Pi^{1}_{B}X_{B}\right)\ket{\phi}\right\Vert&\leq& \sqrt{\frac{\epsilon_{1}}{2}}\left(\frac{\cos(\zeta)}{\sin(\zeta)}+2\left(1+\frac{1}{\sin(2\zeta)}\right)\right)+\sqrt{\frac{\epsilon}{2}}\left(\frac{2}{\sin(\zeta)}+\frac{1}{\cos(\zeta)}\right).
\end{eqnarray*}

\end{lemma_fix}

\begin{proof}
We address the first line in the lemma as the proof of the second line is essentially the same:
\begin{multline*}
\left\Vert \left(\frac{\sin(\zeta)}{\cos(\zeta)}\ket{1}\bra{0}_{A}-\Pi^{0}_{B}X_{B}\right)\ket{\phi}\right\Vert = \left\Vert \left[\frac{\sin(\zeta)}{\cos(\zeta)}\ket{1}\bra{0}_{A}-\left(\frac{\sin(\zeta)}{\cos(\zeta)}\ket{1}\bra{0}_{A}+\frac{\cos(\zeta)}{\sin(\zeta)}\ket{0}\bra{1}_{A}\right)\Pi_{B}^{0}+ \right. \right. \\
\left. \left.\left(\frac{\sin(\zeta)}{\cos(\zeta)}\ket{1}\bra{0}_{A}+\frac{\cos(\zeta)}{\sin(\zeta)}\ket{0}\bra{1}_{A}\right)\Pi_{B}^{0}-\Pi_{B}^{0}X_{B}\right]\ket{\psi}\right\Vert \\
\leq \left\Vert\left(\frac{\sin(\zeta)}{\cos(\zeta)}\ket{1}\bra{0}_{A}-\left(\frac{\sin(\zeta)}{\cos(\zeta)}\ket{1}\bra{0}_{A}+\frac{\cos(\zeta)}{\sin(\zeta)}\ket{0}\bra{1}_{A}\right)\Pi_{B}^{0}\right)\ket{\psi}\right\Vert\\
+\sqrt{2\epsilon_{1}}+\sqrt{\frac{\epsilon}{2}}\left(\frac{1}{\sin(\zeta)}+\frac{1}{\cos(\zeta)}\right),
\end{multline*}


where lemma \ref{lemm1} was used in the inequality along with the fact that $\left\Vert\Pi_{B}^{0}\right\Vert_{\infty}\leq 1$. We now need to provide a bound on the remaining norm, which we now do:
\begin{eqnarray*}
&&\left\Vert\left(\frac{\sin(\zeta)}{\cos(\zeta)}\ket{1}\bra{0}_{A}-\left(\frac{\sin(\zeta)}{\cos(\zeta)}\ket{1}\bra{0}_{A}+\frac{\cos(\zeta)}{\sin(\zeta)}\ket{0}\bra{1}_{A}\right)\Pi_{B}^{0}\right)\ket{\psi}\right\Vert\\
&\leq &\frac{\sin(\zeta)}{\cos(\zeta)}\left\Vert\left(\left(\ket{1}\bra{0}_{A}\right)\ket{0}\bra{0}_{A}-\left(\ket{1}\bra{0}_{A}\right)\Pi^{0}_{B}\right)\ket{\psi}\right\Vert+\frac{\cos(\zeta)}{\sin(\zeta)}\left\Vert\left(\left(\ket{1}\bra{0}_{A}\right)\ket{0}\bra{0}_{A}-\left(\ket{1}\bra{0}_{A}\right)\Pi^{0}_{B}\right)\ket{\psi}\right\Vert\\
&\leq&\left(\frac{\sin(\zeta)}{\cos(\zeta)}+\frac{\cos(\zeta)}{\sin(\zeta)}\right)\sqrt{\frac{\epsilon_{1}}{2}}=\frac{1}{\sin(2\zeta)}\sqrt{2\epsilon_{1}},
\end{eqnarray*}
thus proving the first claim. 
To prove the final claim we first observe that $X_{B}\Pi^{1}_{B}X_{B}=X_{B}\Pi^{1}_{B}\Pi^{1}_{B}X_{B}$ and thus
\begin{eqnarray}
&&\Vert\left(\ket{0}\bra{0}_{A}-\frac{\cos(\zeta)}{\sin(\zeta)}\ket{0}\bra{1}_{A}X_{B}\Pi_{B}^{1}+\frac{\cos(\zeta)}{\sin(\zeta)}\ket{0}\bra{1}_{A}X_{B}\Pi_{B}^{1}-X_{B}\Pi_{B}^{1}\Pi_{B}^{1}X_{B}\right)\ket{\psi}\Vert\nonumber\\
&\leq&\Vert\left(\ket{0}\bra{0}_{A}-\frac{\cos(\zeta)}{\sin(\zeta)}\ket{0}\bra{1}_{A}X_{B}\Pi_{B}^{1}\right)\ket{\psi}\Vert+\sqrt{2\epsilon_{1}}\left(1+\frac{1}{\sin(2\zeta)}\right)+\sqrt{\frac{\epsilon}{2}}\left(\frac{1}{\sin(\zeta)}+\frac{1}{\cos(\zeta)}\right),\nonumber
\end{eqnarray}
where the inequality utilises the first part of the lemma and the fact that $\Vert X_{B}\Pi^{1}_{B}\Vert_{\infty}\leq 1$, and we have that
\begin{eqnarray*}
&&\left\Vert\left(\ket{0}\bra{0}_{A}-\frac{\cos(\zeta)}{\sin(\zeta)}\ket{0}\bra{1}_{A}X_{B}\Pi_{B}^{1}\right)\ket{\psi}\right\Vert\\
&=&\left\Vert\left(\ket{0}\bra{0}_{A}+\left(\ket{0}\bra{1}_{A}\ket{1}\bra{0}_{A}-\ket{0}\bra{1}_{A}\ket{1}\bra{0}_{A}\right)
-\frac{\cos(\zeta)}{\sin(\zeta)}\ket{0}\bra{1}_{A}X_{B}\Pi_{B}^{1}\right)\ket{\psi}\right\Vert\\
&=&\left\Vert\left(\left(\ket{0}\bra{1}_{A}\right)\ket{1}\bra{0}_{A}-\frac{\cos(\zeta)}{\sin(\zeta)}\ket{0}\bra{1}_{A}X_{B}\Pi_{B}^{1}\right)\ket{\psi}\right\Vert\\
&\leq& \frac{1}{\sin(\zeta)}\left(\sqrt{\frac{\epsilon}{2}}+\cos(\zeta)\sqrt{\frac{\epsilon_{1}}{2}}\right),
\end{eqnarray*}

where we used lemma \ref{lemm1} and $\Vert \ket{0}\bra{1}_{A}\Vert_{\infty}\leq 1$ in the last inequality, thus concluding the proof. 
\end{proof}

Now we have the useful lemmas, we are in a position to prove Theorem \ref{selftest}. To outline the structure of the proof, we will explicitly construct an isometry $\Phi_{B}$ acting on Bob's system, which introduces an auxiliary qubit system $B'$, and then is a unitary acting on $B$ and $B'$. 
\\

\noindent
\textit{Proof of Theorem \ref{selftest}} We first will explicitly construct the isometry $\Phi_{B}$ that acts upon Bob's system in the statement of the theorem. The isometry introduces an auxiliary qubit system $B'$ in the state $\ket{+}_{B'}=\frac{1}{\sqrt{2}}(\ket{0}_{B'}+\ket{1}_{B'})$ and then applies a unitary $U_{BB'}$ jointly to Bob's system $B$ and $B'$. The unitary is constructed as $U_{BB'}=\Lambda^{X}_{BB'}(\mathbb{I}_{B}\otimes H_{B'})\Lambda^{Z}_{BB'}$ where $\Lambda^{P}_{BB'}=\ket{0}\bra{0}_{B'}\otimes\mathbb{I}_{B}+\ket{1}\bra{1}_{B'}\otimes P_{B}$ for $P\in\{X,Z\}$ and $H_{B'}$ is the Hadamard applied to qubit $B'$. Since $X_{B}$ and $Z_{B}$ can be shown to be unitaries, one can show that $U_{BB'}$ is unitary. We now take the state $\ket{\psi}$ and apply $\Phi_{BB'}$ to get
\begin{eqnarray}
U_{BB'}\ket{\psi}\ket{+}_{B'}&=&\frac{1}{2}\left[\ket{\psi}\ket{0}_{B'}+X_{B}\ket{\psi}\ket{1}_{B'}+Z_{B}\ket{\psi}\ket{0}_{B'}-X_{B}Z_{B}\ket{\psi}\ket{1}_{B'}\right]\nonumber\\
&=&\Pi^{0}_{B}\ket{\psi}\ket{0}_{B'}+X_{B}\Pi^{1}_{B}\ket{\psi}\ket{1}_{B'}\nonumber,
\end{eqnarray}
using the fact that $\Pi^{x}_{B}=\frac{1}{2}\left(\mathbb{I}_{B}+(-1)^{x}Z_{B}\right)$ for $x\in\{0,1\}$. Therefore, in the first line of the statement of the theorem, we need to bound 
\begin{equation}\label{eq1}
\Vert \Phi_{B}(\ket{\psi})-\ket{{\zeta}}_{AB'}\ket{\textrm{anc}}_{E'} \Vert=\Vert \Pi^{0}_{B}\ket{\psi}\ket{0}_{B'}+X_{B}\Pi^{1}_{B}\ket{\psi}\ket{1}_{B'} - \ket{{\zeta}}_{AB'}\ket{\textrm{anc}}_{E'}\Vert,
\end{equation}
and from lemmata \ref{lemm0} and \ref{lemm1}, we have the following useful identity:
\begin{equation*}
\Vert(\Pi^{0}_{B}-\ket{0}\bra{0}_{A})\ket{\psi}\ket{0}_{B'}+(X_{B}\Pi_{B}^{1}-\tan(\zeta)\ket{1}\bra{0}_{A})\ket{\psi}\ket{1}_{B'}\Vert\leq\sqrt{2\epsilon_{1}}+\frac{1}{\cos(\zeta)}\sqrt{\frac{\epsilon}{2}}.
\end{equation*}
We can use this identity to give the bound on \eqref{eq1} of
\begin{eqnarray*}
&&\Vert (\Pi^{0}_{B}\ket{0}_{B'}+X_{B}\Pi^{1}_{B}\ket{1}_{B'})\ket{\psi} - \ket{{\zeta}}_{AB'}\ket{\textrm{anc}}_{E'}\Vert\nonumber\\
&\leq &\sqrt{2\epsilon_{1}}+\frac{1}{\cos(\zeta)}\sqrt{\frac{\epsilon}{2}}+\Vert(\ket{0}\bra{0}_{A}\ket{\psi}\ket{0}_{B'}+(\tan(\zeta)\ket{1}\bra{0}_{A})\ket{\psi}\ket{1}_{B'}-\ket{{\zeta}}_{AB'}\ket{\textrm{anc}}_{E'}\Vert\nonumber,
\end{eqnarray*}
and by the Schmidt decomposition we also have $\ket{\psi}=\alpha\ket{0}_{A}\ket{\psi_{0}}_{BE}+\beta\ket{1}_{A}\ket{\psi_{1}}_{BE}$ for $\alpha$ and $\beta$ being real positive numbers and $\ket{\psi_{0}}$ being orthogonal to $\ket{\psi_{1}}$. Utilising this fact we have 
\begin{eqnarray*}
\ket{0}\bra{0}_{A}\ket{\psi}\ket{0}_{B'}+(\tan(\zeta)\ket{1}\bra{0}_{A})\ket{\psi}\ket{1}_{B'}&=&\alpha\ket{0}_{A}\ket{\psi_{0}}_{B}\ket{0}_{B'}+\alpha\tan(\zeta)\ket{1}_{A}\ket{\psi_{0}}_{B}\ket{1}_{B'}\\
&=&\left(\ket{00}+\tan(\zeta)\ket{11}\right)_{AB'}(\alpha\ket{\psi_{0}})\\
&:=&\ket{\zeta'}(\alpha\ket{\psi_{0}}),
\end{eqnarray*}
where $\bra{\zeta'}\zeta\rangle=1+\tan^{2}(\zeta)$ and $\bra{\zeta}\zeta'\rangle=\bra{\zeta'}\zeta\rangle=\cos(\zeta)+\sin(\zeta)\tan(\zeta)$. Therefore, if we set $\ket{\textrm{anc}}_{E'}\ket{\psi_{0}}_{BE}$ then
\begin{eqnarray*}
&&\Vert(\ket{0}\bra{0}_{A}\ket{\psi}\ket{0}_{B'}+(\tan(\zeta)\ket{1}\bra{0}_{A})\ket{\psi}\ket{1}_{B'}-\ket{{\zeta}}_{AB'}\ket{\textrm{anc}}_{E'}\Vert\\
&=&\sqrt{1+\alpha^{2}(1+\tan^{2}(\zeta))-2\alpha(\cos(\zeta)+\sin(\zeta)\tan(\zeta))}\\
&=&1-\alpha\sec(\zeta)\\
&\leq&1-\sec(\zeta)\sqrt{\cos^{2}(\zeta)-\frac{\epsilon_{1}}{2}}\\
&=&1-\sqrt{1-\frac{\epsilon_{1}}{2\cos^{2}(\zeta)}}\\
&\leq&\frac{\sqrt{\epsilon_{1}}}{\sqrt{2}\cos(\zeta)},
\end{eqnarray*}
where in the second line we use the fact that $\langle\tau_{Z}^{A}\rangle=2\langle\psi\vert(\ket{0}\bra{0}_{A})\vert\psi\rangle-1=2\alpha^{2}-1$ and thus by virtue of \eqref{crit} being satisfied we have $\vert\alpha^{2}-\cos^{2}(\zeta)\vert\leq\frac{\epsilon_{1}}{2}$. In principle this allows the system $BE$ to be isomorphic to the system $E'$ under the action of this isometry. Putting all of this together we have
\begin{equation*}
\Vert \Phi_{B}(\ket{\psi})-\ket{{\zeta}}_{AB'}\ket{\textrm{anc}}_{E'} \Vert\leq \sqrt{\epsilon_{1}}\left(\sqrt{2}+\frac{1}{\sqrt{2}\cos(\zeta)}\right)+\frac{1}{\cos(\zeta)}\sqrt{\frac{\epsilon}{2}}\leq\sqrt{\epsilon_{1}}(\sqrt{2}+1)+\sqrt{\epsilon},
\end{equation*}
where in the final inequality we used the fact that $\cos(\zeta)\geq\frac{1}{\sqrt{2}}$ for $\zeta\in]0,\frac{\pi}{4}]$, thus concluding the proof for the first line of Theorem \ref{selftest}.

For the second line of the proof for Theorem \ref{selftest}, thus we need to bound the following expression:
\begin{equation*}
\Vert \Phi_{B}(X_{B}\ket{\psi})-\ket{{\zeta}}_{AB'}\ket{\textrm{anc}}_{E'} \Vert=\Vert \Pi^{0}_{B}X_{B}\ket{\psi}\ket{0}_{B'}+X_{B}\Pi^{1}_{B}X_{B}\ket{\psi}\ket{1}_{B'} - \tau_{X}^{B'}\ket{{\zeta}}_{AB'}\ket{\textrm{anc}}_{E'}\Vert.
\end{equation*}
To provide an upper bound to the right-hand-side we have that
\begin{eqnarray*}
&&\Vert \Pi^{0}_{B}X_{B}\ket{\psi}\ket{0}_{B'}+X_{B}\Pi^{1}_{B}X_{B}\ket{\psi}\ket{1}_{B'} - \tau_{X}^{B'}\ket{{\zeta}}_{AB'}\ket{\textrm{anc}}_{E'}\Vert\\
&\leq &\Vert\left(\Pi^{0}X_{B}-\frac{\sin(\zeta)}{\cos(\zeta)}\ket{1}\bra{0}_{A}\right)\ket{\psi}\ket{0}_{B'}\Vert+\Vert\left(X_{B}\Pi^{1}_{B}X_{B}-\ket{0}\bra{0}_{A}\right)\ket{\psi}\ket{1}_{B'}\Vert\\
&+&\Vert\frac{\sin(\zeta)}{\cos(\zeta)}\ket{1}\bra{0}_{A}\ket{\psi}\ket{0}_{B'}+\ket{0}\bra{0}_{A}\ket{\psi}\ket{1}_{B'}-\tau_{X}^{B'}\ket{{\zeta}}_{AB'}\ket{\textrm{anc}}_{E'}\Vert\\
&\leq&\sqrt{\epsilon_{1}}\left[2\sqrt{2}\left(1+\frac{1}{\sin(2\zeta)}\right)+\frac{\cos(\zeta)}{2\sin(\zeta)}\right]+\sqrt{\frac{\epsilon}{2}}\left[\frac{3}{\sin(\zeta)}+\frac{2}{\cos(\zeta)}\right]\\
&+&\Vert\frac{\sin(\zeta)}{\cos(\zeta)}\ket{1}\bra{0}_{A}\ket{\psi}\ket{0}_{B'}+\ket{0}\bra{0}_{A}\ket{\psi}\ket{1}_{B'}-\tau_{X}^{B'}\ket{{\zeta}}_{AB'}\ket{\textrm{anc}}_{E'}\Vert,
\end{eqnarray*}
where the first inequality is just resulting from the triangle inequality and the second inequality is an application of lemma \ref{lemm2}. The final term we need to bound is
\begin{eqnarray*}
&&\Vert\frac{\sin(\zeta)}{\cos(\zeta)}\ket{1}\bra{0}_{A}\ket{\psi}\ket{0}_{B'}+\ket{0}\bra{0}_{A}\ket{\psi}\ket{1}_{B'}-\tau_{X}^{B'}\ket{{\zeta}}_{AB'}\ket{\textrm{anc}}_{E'}\Vert\\
&=&\Vert\tau_{X}^{B'}\left(\alpha(\tan(\zeta)\ket{11}_{AB'}+\ket{00}_{AB'})\ket{\psi_{0}}_{B}-\ket{\zeta}_{AB'}\ket{\textrm{anc}}_{E'}\right)\Vert\\
&=&\Vert\left[(\tan(\zeta)\ket{11}_{AB'}+\ket{00}_{AB'})(\alpha\ket{\psi_{0}}_{B})-\ket{\zeta}_{AB'}\ket{\textrm{anc}}_{E'}\right]\Vert\\
&\leq&\frac{\sqrt{\epsilon_{1}}}{\sqrt{2}\cos(\zeta)},
\end{eqnarray*}
where the first inequality is a consequence of the aforementioned Schmidt decomposition and the second equality results from the fact that $\tau_{X}^{B'}$ is a unitary, and the final inequality is just the bound derived earlier for the same norm. Putting all of this together we then get the bound
\begin{eqnarray*}
&&\Vert \Phi_{B}(\otimes X_{B}\ket{\psi})-\tau^{B'}_{X}\ket{{\zeta}}_{AB'}\ket{\textrm{anc}}_{E'}\Vert\\
&\leq& \sqrt{\epsilon_{1}}\left[2\sqrt{2}\left(1+\frac{1}{\sin(2\zeta)}\right)+\frac{\cos(\zeta)}{2\sin(\zeta)}+\frac{1}{\sqrt{2}\cos(\zeta)}\right]+\sqrt{\frac{\epsilon}{2}}\left[\frac{3}{\sin(\zeta)}+\frac{2}{\cos(\zeta)}\right]\\
&\leq&\sqrt{\epsilon_{1}}\left(2\sqrt{2}+1+\frac{5}{2\sin(\zeta)}\right)+\sqrt{\epsilon}\left(\frac{3}{\sqrt{2}\sin(\zeta)}+2\right),
\end{eqnarray*}
where the last inequality utilised the fact that $1>\cos(\zeta)\geq\frac{1}{\sqrt{2}}$ as outlined earlier. $\square$

\begin{Corollary}\label{corr1}
For Alice, Bob and Eve sharing the state $\ket{\psi}$ then for Bob making the measurement associated with the observable $X_{B}$ and for Eve making a dichotomic measurement $\{M_{z},\mathbb{I}_{E}-M_{z}\}$ with $M_{z}$ being the POVM associated with Eve's guess $z$ of Bob's measurement outcome, then Eve's maximum probability of guessing Bob's outcome is
\begin{equation*}
p_{\textrm{guess}}=\underset{\{\ket{\psi},M_{z},X_{B}\}}{\text{max}}\sum_{z\in\{0,1\}}\frac{1}{2}\langle\psi\vert(1+(-1)^{z}X_{B})M_{z}\vert\psi\rangle,
\end{equation*}
and if Alice and Bob's statistics satisfy the criteria in \eqref{crit} and $\zeta\in]0,\frac{\pi}{4}]$, then 
\begin{equation*}
p_{\textrm{guess}}\leq\frac{1}{2}+\sqrt{\epsilon_{1}}\left(3\sqrt{2}+2+\frac{5}{2\sin(\zeta)}\right)+3\sqrt{\epsilon_{1}+\epsilon_{2}}\left(\frac{1}{\sqrt{2}\sin(\zeta)}+1\right).
\end{equation*}
\end{Corollary}

\begin{proof}
Starting with the definition of $p_{\textrm{guess}}$ and $\sum_{z}M_{z}=\mathbb{I}_{E}$ in the statement we have that 
\begin{equation*}
p_{\textrm{guess}}=\frac{1}{2}+\frac{1}{2}\underset{\{\ket{\psi},M_{z},X_{B}\}}{\text{max}}\bra{\psi}X_{B}\otimes(M_{0}-M_{1})\ket{\psi}.
\end{equation*}
Since the statistics of Alice and Bob satisfy \eqref{crit} then Theorem \ref{selftest} implies that there exists an isometry that (up to some error) maps $\ket{\psi}$ to $\ket{\zeta}\ket{\textrm{anc}}$ and $X_{B}\ket{\psi}$ to $\tau_{X}\ket{\zeta}\ket{\textrm{anc}}$. Isometries do not change probabilities so we have that
\begin{eqnarray*}
p_{\textrm{guess}}&=&\frac{1}{2}+\frac{1}{2}\underset{\{\ket{\psi},M_{z},X_{B}\}}{\text{max}}\Phi_{B}^{\dagger}(\bra{\psi})\otimes(M_{0}-M_{1})\Phi_{B}(X_{B}\ket{\psi})\\
&\leq&\frac{1}{2}+\sqrt{\epsilon_{1}}\left(3\sqrt{2}+2+\frac{5}{2\sin(\zeta)}\right)+3\sqrt{\epsilon_{1}+\epsilon_{2}}\left(\frac{1}{\sqrt{2}\sin(\zeta)}+1\right)\\
&+&\bra{\zeta_{AB'}}\tau_{X}^{B'}\ket{\zeta_{AB'}}\underset{\{\ket{\psi},M_{z}\}}{\text{max}}\bra{\textrm{anc}}(M_{0}-M_{1})\ket{\textrm{anc}}\\
&\leq&\frac{1}{2}+\sqrt{\epsilon_{1}}\left(3\sqrt{2}+2+\frac{5}{2\sin(\zeta)}\right)+3\sqrt{\epsilon_{1}+\epsilon_{2}}\left(\frac{1}{\sqrt{2}\sin(\zeta)}+1\right),
\end{eqnarray*}
where the inequality in the second line is through applying (via the triangle inequality) both of the self-testing results in Theorem \ref{selftest} along with the fact that $\Vert(M_{0}-M_{1})\Vert_{\infty}\leq 1$, and the second inequality comes from the fact that $\Vert(M_{0}-M_{1})\Vert_{\infty}\leq 1$ and that $\bra{\zeta_{AB'}}\tau_{X}^{B'}\ket{\zeta_{AB'}}=0$. 
\end{proof}

\noindent
If we consider one round of our particular scheme then (up to local unitaries) Alice and Bob share the state $\ket{\zeta}=\cos(\zeta)\ket{00}+\sin(\zeta)\ket{11}$ and Bob makes a measurement from two possible measurements with observables $Z_{B}$ and $X_{B}$, where $Z=\tau_{Z}$ and $X$ is associated with POVM elements $\{\cos^{2}(\xi)\ket{+}\bra{+}+\sin^{2}(\xi)\ket{-}\bra{-},\cos^{2}(\xi)\ket{-}\bra{-}+\sin^{2}(\xi)\ket{+}\bra{+}\}$ such that Bob's observable is $X=\cos(2\xi)\tau_{X}$. Given this scheme, we have the following correlations:
\begin{eqnarray}
\langle \tau^{A}_{Z}\otimes Z_{B}\rangle&=&1\nonumber\\
\langle \tau^{A}_{X}\otimes X_{B}\rangle&=&\sin(2\zeta)\cos(2\xi)\nonumber\\
\langle \tau^{A}_{Z}\rangle&=&\cos(2\zeta).\nonumber
\end{eqnarray} 
Thus this scheme will pass the statistical criteria in \eqref{crit} with $\epsilon_{1}=0$ and $\epsilon_{2}=2\sin^{2}(\xi)$, and a guessing probability of
\begin{equation*}
p_{\textrm{guess}}\leq\frac{1}{2}+3\sin(\xi)\left(\frac{1}{\sin(\zeta)}+\sqrt{2}\right).
\end{equation*}
We will return to this observation later on, but first we need to now consider the case of Bob making a sequence of measurements. Recall that in our scheme described in Section \ref{sec:unbounded_randomness}, after each $i$th measurement made by Bob, the state shared between Alice and Bob is (after Bob's post-measurement unitary corrections)
\begin{equation}
\ket{\psi_{b^{i}|y^{i}}}=U_{A}^{b^{i}|y^{i}}\otimes\mathbb{I}_{B}\left(\cos(\zeta_{b^{i}|y^{i}})\ket{00}+\sin(\zeta_{b^{i}|y^{i}})\ket{11}\right),
\end{equation}
where $b^{i}|y^{i}$ is the bit-string of outcomes of Bob's sequence of measurements from round $1$ to round $i$, also $U_{A}^{b^{i}|y^{i}}$ and $\zeta_{b^{i}|y^{i}}$ depend on the initial state shared by Alice and Bob and Bob's sequence of measurement outcomes (and the type of measurement). By convention, we have that $b^{0}$ is the empty string. The important thing is that we know in an honest implementation of our scheme what the values of $U_{A}^{b^{i}|y^{i}}$ and $\zeta_{b^{i}|y^{i}}$ will be. If we want to certify randomness from the sequence of measurements we need statistical criteria for Alice and Bob such that if they pass, we are guaranteed randomness. This statistical criteria will be that in round $i+1$ of the sequence of measurements, Alice and Bob's statistics need to satisfy:
\begin{eqnarray}\label{critnew}
\vert\langle U_{A}^{b^{i}|y^{i}}\tau^{A}_{Z}\left(U_{A}^{b^{i}|y^{i}}\right)^{\dagger}\otimes Z_{B}^{b^i|y^i}\rangle-1\vert&\leq&\epsilon_{1}^{i+1}\nonumber\\
\vert\langle U_{A}^{b^{i}|y^{i}}\tau^{A}_{X}\left(U_{A}^{b^{i}|y^{i}}\right)^{\dagger}\otimes X_{B}^{b^i|y^i}\rangle-\sin(2\zeta_{b^{i}|y^{i}})\vert&\leq&\epsilon_{2}^{i+1}\nonumber\\
\vert\langle U_{A}^{b^{i}|y^{i}}\tau^{A}_{Z}\left(U_{A}^{b^{i}|y^{i}}\right)^{\dagger}\rangle-\cos(2\zeta_{b^{i}|y^{i}})\vert&\leq&\epsilon_{1}^{i+1}.
\end{eqnarray} 
We can now state a useful corollary of Theorem \ref{selftest} (and Corollary \ref{corr1}).
\begin{Corollary}
After Bob makes $i$ measurements yielding the outcomes $b^{i}|y^{i}$, Alice, Bob and Eve share some quantum state $\ket{\phi_{b^{i}|y^{i}}}$, then for Bob making the measurement associated with the observable $X_{B}$ and for Eve making a dichotomic measurement $\{M_{z},\mathbb{I}_{E}-M_{z}\}$ with $M_{z}$ being the POVM associated with Eve's guess $z$ of Bob's measurement outcome, then Eve's probability of guessing Bob's outcome is
\begin{equation*}
p_{\textrm{guess}}=\underset{\{\ket{\phi_{b^{i}|y^{i}}},M_{z},X_{B}\}}{\text{max}}\sum_{z\in\{0,1\}}\frac{1}{2}\bra{\phi_{b^{i}|y^{i}}}(1+(-1)^{z}X_{B})M_{z}\ket{\phi_{b^{i}|y^{i}}},
\end{equation*}
and if Alice and Bob's statistics satisfy the criteria in \eqref{critnew}, then 
\begin{equation*}
p_{\textrm{guess}}\leq\frac{1}{2}+\sqrt{\epsilon_{1}^{i+1}}\left(3\sqrt{2}+2+\frac{5}{2\sin(\zeta_{b^{i}|y^{i}})}\right)+3\sqrt{\epsilon^{i+1}_{1}+\epsilon^{i+1}_{2}}\left(\frac{1}{\sqrt{2}\sin(\zeta_{b^{i}|y^{i}})}+1\right).
\end{equation*}
\end{Corollary}
\textit{Proof.} First notice that the statistical criteria in \eqref{critnew} is equivalent to \eqref{crit} up to a local unitary $U_{A}^{b^{i}|y^{i}}$ on Alice's system. This local unitary can be simulated by applying the inverse of this operation to Alice's part of the state that Alice and Bob share. The self-testing results in Theorem \ref{selftest} will now hold for the state  $U_{A}^{b^{i}|y^{i}}\otimes\mathbb{I}_{B'}\ket{\zeta}_{AB'}$ instead of $\ket{\zeta}_{AB'}$ since the norms are invariant under the action of unitaries. Given this, one can apply the result from Corollary \ref{corr1} to complete the proof. $\square$

This result quantifies the randomness of each round of our scheme. Now, if we consider the scheme in general, there is a sequence of measurements made by Bob, and we want to bound Eve's probability to guess the total string $b$ of Bob's outcomes. The following result gives us a bound on this probability.
\begin{Corollary}\label{corr2}
For Bob making a sequence of measurements yielding the outcome bit-string $b$, if Alice, Bob and Eve share some initial state $\ket{\psi}$, Bob's measurement in round $i$ is associated with observable $X_{B}^{b^{i}|y^{i}}$, and Eve makes a measurement associated with operators $\{M_{z}\}_{z}$, where $z$ is Eve's guess of Bob's outcome $b$, the probability of guessing it correctly is
\begin{eqnarray*}
p_{\textrm{guess}}&=&\underset{\{\ket{\psi},M_{z},\{X_{B}^{b^{i}|y^{i}}\}\}}{\text{max}}\sum_{b,z}\delta_{z}^{b}P_{\ket{\psi}}(z,b|\{X_{B}^{b^{i}|y^{i}}\}_{i}),
\end{eqnarray*}
and if for each $i$, there exist unitaries $U_{A}^{b^{i}|y^{i}}$ and angles $\zeta_{b^{i}|y^{i}}$ such that the statistical criteria in \eqref{critnew} is satisfied, then 
\begin{equation}
p_{\textrm{guess}}\leq \prod_{i=1}^{n}\left(\frac{1}{2}+\sqrt{\epsilon^{i}_{1}}\left(3\sqrt{2}+2+\frac{5}{2\sin(\zeta_{b^{i-1}|y^{i-1}})}\right)+3\sqrt{\epsilon^{i}_{1}+\epsilon^{i}_{2}}\left(\frac{1}{\sqrt{2}\sin(\zeta_{b^{i-1}|y^{i-1}})}+1\right)\right),
\end{equation}
and if $\epsilon^{i}_{1}=0$ for all $i$, then
\begin{equation}
p_{\textrm{guess}}\leq \prod_{i=1}^{n}\left(\frac{1}{2}+3\sqrt{\epsilon^{i}_{2}}\left(\frac{1}{\sqrt{2}\sin(\zeta_{b^{i-1}|y^{i-1}})}+1\right)\right).
\end{equation}
\end{Corollary}
\textit{Proof.} First we can rewrite $p_{\textrm{guess}}$ as
\begin{eqnarray*}
\underset{\{\ket{\psi},M_{z},\{X_{B}^{i}\}\}}{\text{max}}\sum_{b_{1},z}\delta^{b_{1}}_{z_{1}}P_{\ket{\psi}}(b_{1},z|X_{B}^{1})\sum_{b_{2}}\delta^{b_{2}}_{z_{2}}P_{\ket{\psi}}(b_{2},z|\{X_{B}^{i}\}_{i\leq 2},b_{1})...\sum_{b_{n}}\delta^{b_{n}}_{z_{n}}P_{\ket{\psi}}(b_{n},z|\{X_{B}^{i}\}_{i\leq n},b^{n-1}),
\end{eqnarray*}
by applying Bayes' theorem and then the constraints from causality; roughly that outcome $b_{i}$ cannot be affected by outcome $b_{j}$ if $j>i$. Using this structure we have that 
\begin{eqnarray*}
p_{\textrm{guess}}&\leq&\prod_{j=1}^{n}\left(\underset{\{\ket{\psi},M_{z},\{X_{B}^{i}\}\}}{\text{max}}\sum_{b_{j},z}\delta^{b_{j}}_{z_{j}}P_{\ket{\psi}}(b_{j},z|\{X_{B}^{i}\}_{i\leq j},b^{j-1})\right),
\end{eqnarray*}
where there are now multiple maximizations, essentially for each summand of $p_{\textrm{guess}}$. Now notice that the sum over strings $z$ along with the probability in the summand can be interpreted as a coarse-grained measurement by Eve. To wit, Eve makes a measurement with outcomes corresponding to each string $z$, and for each $b_{j}$, Eve will produce a guess of this based on the value of $z_{j}$. For example, if Eve generates a string $z$ such that $z_{j}=0$ then Eve's guess of $b_{j}$ is $0$. This then reduces each maximisation to the guessing probability for each round, and thus from Corollary \ref{corr2} we have the statement. $\square$

This final corollary gives us the proof of Theorem 1 in the main body of the paper.

\section{Alternative Quantum Circuit for Sequences of Measurements \label{app:quantumcircuits}}

The following circuit can be used to implement the sequence of non-projective measurements in Bob's device, in which the measurement choices are encoded in quantum states in the computational basis, as an alternative to the classically controlled version given in the main text. This circuit only implement Bob's half of the protocol on his quantum state denoted by $\rho_B$, which is prepared for him by Eve. Alice's part in the protocol is simply to do state tomography on her steered state so this is excluded from the circuit. These circuits, \eqref{quantumcircuit1} and \eqref{quantumcircuit2} measure the state for one, and two rounds respectively with a generalisation to higher rounds straightforward.  
\begin{align}
\Qcircuit @C=0.5em @R=0.4em {
\lstick{\ket{y}}& \ctrlo{4} & \qw& \qw & \qw & \qw & \qw & \qw & \qw &\\
\lstick{\ket{0}} &\qw & \qw &\qw & \gate{R_{y}(2\theta_1)} & \gate{H} & \ctrl{3}& \gate{H} &\measureD{Z^{(1)}}& \rstick{X_{\theta_1}} \\
\lstick{\ket{0}}& \qw &\gate{R_{y}(2\phi_1)}& \gate{H} & \ctrl{1} & \gate{H} & \qw  \qw & \qw &\measureD{Z^{(1)}} &\rstick{Z_{\phi_1}}\\
\lstick{\ket{0}} & \qswap & \qw&\qw & \gate{Z} & \qw & \qw & \qw & \qw &  \\
\lstick{\ket{\psi}_B}& \qswap & \qw &\qw & \qw& \qw & \gate{X} & \qw &\qw & \\
}\label{quantumcircuit1}
\end{align}
\begin{align}
\Qcircuit @C=0.5em @R=0.4em {
& &\lstick{\ket{y_1}} & \qw& \qw&\qw &\qw &\qw& \qw &\qw&\ctrlo{8} & \qw &\qw &\qw & \qw &\qw &\qw & \qw & \qw & &  \\
 & &\lstick{\ket{y_2}} & \ctrlo{7}&\qw & \qw & \qw & \qw & \qw & \qw & \qw & \qw  & \qw &\qw & \qw &\qw  & \qw & \qw &\qw & \\
&\lstick{\ket{0}} && \qw &\qw & \qw& \qw  &\qw &\qw &\qw      & \qw  &\qw  &\qw & \gate{R_{y}(2\theta_2)}  &\gate{H} &\ctrl{6} &\gate{H}& \qw &\measureD{Z^{(2)}} & \rstick{X_{\theta_2}}\\
&\lstick{\ket{0}} && \qw&\qw & \qw& \gate{R_{y}(2\theta_1)} &\gate{H} &\ctrl{5} &\gate{H} & \qw &\qw&\qw &\qw &\qw &\qw & \qw &\qw       &\measureD{Z^{(1)}} &\rstick{X_{\theta_1}}\\
&\lstick{\ket{0}}&& \qw  &\qw& \qw &\qw &\qw  &\qw &\qw &\qw & \gate{R_{y}(2\phi_2)} &\gate{H}& \ctrl{2} & \gate{H} &\qw &\qw &\qw & \measureD{Z^{(2)}} & \rstick{Z_{\phi_2}}\\
&\lstick{\ket{0}} && \qw & \gate{R_{y}(2\phi_1)}& \gate{H} & \ctrl{2} &\gate{H}     &\qw &\qw & \qw & \qw &  \qw&\qw & \qw & \qw & \qw & \qw & \measureD{Z^{(1)}} & \rstick{Z_{\phi_1}} \\
&\lstick{\ket{0}} && \qw & \qw& \qw &\qw &\qw &\qw &\qw &\qswap &\qw& \qw & \gate{Z} & \qw &\qw &\qw & \qw & \qw &     &  \\
&\lstick{\ket{0}} && \qswap &\qw &\qw & \gate{Z} \qw & \qw  &  \qw &\qw & \qw  & \qw & \qw &\qw &\qw & \qw&\qw & \qw & \qw &  &\\
&\lstick{\ket{\psi}_B}&& \qswap &\qw&\qw &\qw & \qw & \gate{X} \qw &\qw &\qswap & \qw &\qw& \qw & \qw  &\gate{X} & \qw &\qw &\qw& \\
} \label{quantumcircuit2}
\end{align}


\end{document}